    \documentclass[pra,aps,twocolumn,10pt,floatfix,nofootinbib,superscriptaddress,longbibliography,nobibnotes]{revtex4-1}
    
    \usepackage{amsmath}
    \usepackage{amssymb}
    \usepackage{stmaryrd}
    \usepackage{txfonts}
    \usepackage{siunitx}
    \usepackage{suffix}
    
    \usepackage{amsfonts}
    \usepackage{dsfont}                 % double stroke font
    \usepackage[utf8]{inputenc}
    \usepackage{bbold}                  % \mathbb{0}
    
    \usepackage[usenames,dvipsnames,table]{xcolor}
    \definecolor{verylightgray}{RGB}{230,230,230}
    \usepackage{graphicx}
    \graphicspath{{Images/}} % Setting the graphics path
    
    \usepackage[version=4]{mhchem}      % chemical formulae
    \usepackage{physics}                % physics notation
    \usepackage{csquotes}               % quotations
    \usepackage{placeins}

    \usepackage[colorlinks=true,citecolor=cyan,linkcolor=magenta,filecolor=magenta,linktoc=all]{hyperref}
    \usepackage[capitalize]{cleveref}
    
    \usepackage{enumerate}
    \usepackage[shortlabels]{enumitem}
    \usepackage{titlesec}
    \usepackage{listings}
    
    \usepackage{orcidlink}
    
    \usepackage[globalcitecopy]{bibunits}
    \defaultbibliography{bibliography}
    \defaultbibliographystyle{apsrev4-1}
    
    \newcommand{\e}{\mathrm{e}}         % Euler number / exponential function
    \renewcommand{\i}{\mathrm{i}}       % imaginary unit

    \renewcommand{\vec}[1]{{\vb{#1}}}  % vector
    \newcommand{\subg}<
    
    \newcommand{\Mathematica}{\textsc{Mathematica}}
    
    \DeclareMathAlphabet{\mathbbold}{U}{bbold}{m}{n}
    \def\abs#1{\left|{#1}\right|}      	% absolute value, \abs{x} gives |x|
    \def\bs#1{\boldsymbol{#1}}			% shorted version of "boldsymbol"
    \def\imi{\mathrm{i}}				% imaginary i
    \def\de{\mathrm{d}}
    
    \def\mcH{\mathcal{H}}					% Hamiltonian
    \def\mcT{\mathcal{T}}					% Time-reversal symmetry operator
    \def\mcK{\mathcal{K}}					% Complex conjugation operator

    \def\ztwo{\mathbbold{Z}_2}				% parity group
    \DeclareMathOperator{\arccosh}{arcosh}
    \DeclareMathOperator{\arctanh}{artanh}

    \newcommand{\cip}[2]{\left\langle{#1},{#2}\right\rangle}
    \WithSuffix\newcommand\cip*[2]{\langle{#1},{#2}\rangle}
    
    \newcommand{\cconj}[1]{{#1}^*}
    
    \definecolor{TB}{rgb}{1,0.5,0}
    \definecolor{DU}{rgb}{0,0,1}
    \definecolor{PL}{rgb}{0,0.8,0.4}
\begin{document}
    \begin{bibunit}
    
    \title{Hyperbolic Topological Band Insulators}
    
    \author{David M. Urwyler}
    \affiliation{Department of Physics, University of Zurich, Winterthurerstrasse 190, 8057 Zurich, Switzerland}
    
    \author{Patrick M. Lenggenhager\,\orcidlink{0000-0001-6746-1387}
    }
    \affiliation{Department of Physics, University of Zurich, Winterthurerstrasse 190, 8057 Zurich, Switzerland}
    \affiliation{Condensed Matter Theory Group, Paul Scherrer Institute, 5232 Villigen PSI, Switzerland}
    \affiliation{Institute for Theoretical Physics, ETH Zurich, 8093 Zurich, Switzerland}
    
    \author{Igor Boettcher\,\orcidlink{0000-0002-1634-4022}
    }
    \affiliation{Department of Physics, University of Alberta, Edmonton, Alberta T6G 2E1, Canada}
    \affiliation{Theoretical Physics Institute, University of Alberta, Edmonton, Alberta T6G 2E1, Canada}
    
    \author{Ronny Thomale\,\orcidlink{0000-0002-3979-8836}
    }\affiliation{Institut für Theoretische Physik und Astrophysik, Universität Würzburg, 97074 Würzburg, Germany}
    
    \author{Titus Neupert\,\orcidlink{0000-0003-0604-041X}
    }
    \affiliation{Department of Physics, University of Zurich, Winterthurerstrasse 190, 8057 Zurich, Switzerland}
    
    \author{Tom\'{a}\v{s} Bzdu\v{s}ek\,\orcidlink{0000-0001-6904-5264}
    }\email[corresponding author: ]{tomas.bzdusek@psi.ch}
    \affiliation{Condensed Matter Theory Group, Paul Scherrer Institute, 5232 Villigen PSI, Switzerland}
    \affiliation{Department of Physics, University of Zurich, Winterthurerstrasse 190, 8057 Zurich, Switzerland}
    
    \date{\today}
    
    \begin{abstract}
    Recently, hyperbolic lattices that tile the negatively curved hyperbolic plane emerged as a new paradigm of synthetic matter, and their energy levels were characterized by a band structure in a four- (or higher-)dimensional momentum space. 
    To explore the uncharted topological aspects arising in hyperbolic band theory, we here introduce elementary models of hyperbolic topological band insulators: the hyperbolic Haldane model and the hyperbolic Kane-Mele model; both obtained by replacing the hexagonal cells of their Euclidean counterparts by octagons. 
    Their non-trivial topology is revealed by computing topological invariants in both position and momentum space. The bulk-boundary correspondence is evidenced by comparing bulk and boundary density of states, by modelling propagation of edge excitations, and by their robustness against disorder.
    \end{abstract}
    
    \maketitle
    
    \end{bibunit}
    
    %%%%%%%%%%%%%%%%%%%%%%%%
    %%%    MAIN TEXT     %%%
    %%%%%%%%%%%%%%%%%%%%%%%%
    
    \emph{Introduction.}---The interplay between the crystal structure of materials and their electronic band-structure topology is pivotal to modern condensed matter physics, with major recent developments in areas such as topological quantum chemistry~\cite{Bradlyn:2017,Po:2017,Vergniory:2022} and moir\'{e} materials~\cite{Andrei:2021}. 
    With the ground-breaking experimental realization of hyperbolic lattices in coupled waveguide resonators~\cite{Kollar:2019} and electric-circuit networks~\cite{Lenggenhager:2021d}, such exotic lattices have been elevated from purely mathematical objects~\cite{BookMagnus,BookCoxeter} to promising tabletop platforms for simulating quantum many-body physics in curved space.
    These experimental achievements have also inspired numerous theoretical studies of hyperbolic lattices.
    Notably, hyperbolic band theory (HBT) has been formulated~\cite{Maciejko:2021}, enabling the characterization of their energy spectra via band structures in momentum space.
    The range of recently investigated physical phenomena further includes the effects of magnetic fields~\cite{Yu:2020,Ikeda:2021,Stegmaier:2021}, continuum approximation~\cite{Boettcher:2020},
    periodic boundary conditions~\cite{Maciejko:2022,Zhu:2021,Breuckmann:2016}, hyperbolic crystallography~\cite{Boettcher:2021}, photon bound states~\cite{Bienias:2021}, exact trace formulas~\cite{Attar2022}, Bose-Hubbard model~\cite{Zhu:2021}, elastic vibrations~\cite{Ruzzene:2021}, and flat bands~\cite{Kollar:2019b,Saa:2021,Bzdusek:2022,Mosseri:2022}. 
    Notably, two very recent works proposed concrete models of hyperbolic topological insulators~\cite{Zhang:2022,Liu:2022}; however, a systematic investigation of
    topological quantum numbers on hyperbolic lattices remains largely unexplored.
            
    Among the multitude of hyperbolic lattices, which are tessellations of the two-dimensional (2D) hyperbolic plane of negative curvature~\cite{Katok:1992}, the so-called $\{8 , 3\}$ lattice presents a unique opportunity for a first systematic study of band topology in toy models with topological ground states. The graph of this lattice consists of regular octagons with three lines meeting at each vertex. Hence it derives from the honeycomb lattice (denoted $\{6 , 3\}$ in this context) through replacing hexagons by octagons. 
    Importantly, HBT predicts that the Brillouin zone (BZ) of this lattice is four-dimensional (4D)~\cite{Boettcher:2021}, with crystal momentum $\bs{k} = (k_1 , k_2 , k_3 , k_4)$, separating the dimensions of position and momentum space as a genuine property of hyperbolic lattices. This enhanced dimensionality suggests~\cite{Kitaev:2009,Ryu:2010} that hyperbolic models may host larger families of strong and weak topological band insulators than their Euclidean counterparts.     
    
    In this Letter, we introduce two elemental models of hyperbolic topological band insulators, the hyperbolic Haldane and hyperbolic Kane-Mele (KM) models on the $\{8 , 3\}$ lattice, which generalize the quintessential namesake Euclidean models formulated on the $\{6 , 3\}$ lattice \cite{Haldane:1988,Kane:2005,Khanikaev:2013,Jotzu:2014,Ding:2019}. These models could be implemented experimentally using the platforms of Refs.~\onlinecite{Kollar:2019,Lenggenhager:2021d,Imhof:2018,Lee:2018,Hofmann:2019}.
    Importantly, due to the applicability of both HBT and real-space topological markers~\cite{Bellissard:1994,Kitaev:2006,Bianco:2011,Prodan:2011,Huang:2018} on the $\{8 , 3\}$ lattice, we are able to study band-topological properties in both position and momentum space. 
    This dual point of view allows us to compare topological invariants of hyperbolic topological insulators in momentum and position space, and to study their associated bulk-boundary correspondence.
    Our models and analysis surpass the study of
    hyperbolic Hofstadter and Haldane-like models in Refs.~\onlinecite{Yu:2020,Zhang:2022,Liu:2022}
    as they do not utilize the complementary momentum-space picture.
    
    \emph{Tight-binding models.}---We consider models     on the hyperbolic $\{8 , 3\}$ lattice, which is comprised of octagonal faces with three lines meeting at each vertex, see Fig.~\ref{fig:Bolza-cell}\textbf{a}. 
    This lattice consists of a 16-site unit cell that is repeated infinitely many times according to a hyperbolic Bravais lattice, which is the $\{8 , 8\}$ lattice in this case \cite{Boettcher:2021}, comprised of octagonal plaquettes with coordination number eight.
    We refer to the unit cell as the \emph{Bolza cell} -- a name inspired by the fact that this cell covers the Bolza surface (the most symmetric genus-two 
    Riemann surface~\cite{kazaryan2019}).
    The Bravais lattice is generated by four non-commuting hyperbolic translations, denoted $\gamma_1,\dots,\gamma_4$. 
    
    To obtain the energy bands for tight-binding model with nearest-neighbor (NN) hopping on the $\{8 , 3\}$ lattice from HBT, the Bolza cell is equipped with twisted boundary conditions~\cite{Maciejko:2021}, defined by four phase factors, $e^{{\rm i} k_1},\dots,e^{{\rm i} k_4}$, along the directions of the four generators: each bond crossing one of the eight sides of the Bolza cell acquires a phase factor.
    This yields a $16\times 16$ \emph{hyperbolic Bloch Hamiltonian} $\mcH_{\rm Bloch}(\bs{k})$, whose eigenvalues comprise $16$ energy bands of the $\{8 , 3\}$ lattice in 4D momentum space. 
    We henceforth set the NN hopping parameter to unity.
    
    The density of states (DoS) of tight-binding models, $\rho(E)$, can be obtained either (1)~through exact diagonalization (ED) on finite hyperbolic graphs, or (2)~via HBT by sampling $\bs{k}$ over the 4D BZ. 
    We refer to finite hyperbolic graphs with open boundary as \emph{flakes}. 
    To remove the contribution of boundary states, we define in ED calculations the \emph{bulk-DoS} as the sum of local-DoS on the 16 sites in the innermost Bolza cell~\cite{supp}.
    Whether the two just-defined DoS functions should match for large systems remains at present an open problem, since HBT only identifies eigenstates transforming %according to 
    in Abelian representations of the non-commutative translation group~\cite{Cheng2022}.
    Nevertheless, the results for the NN model, compared in Fig.~\ref{fig:Bolza-cell}\textbf{b}, indicate an auspicious level of agreement, with the deviations partly attributable to residual boundary effects~\cite{Chen2022}. 
    \begin{figure}[t!]
    \centering
        \includegraphics[width=\linewidth]{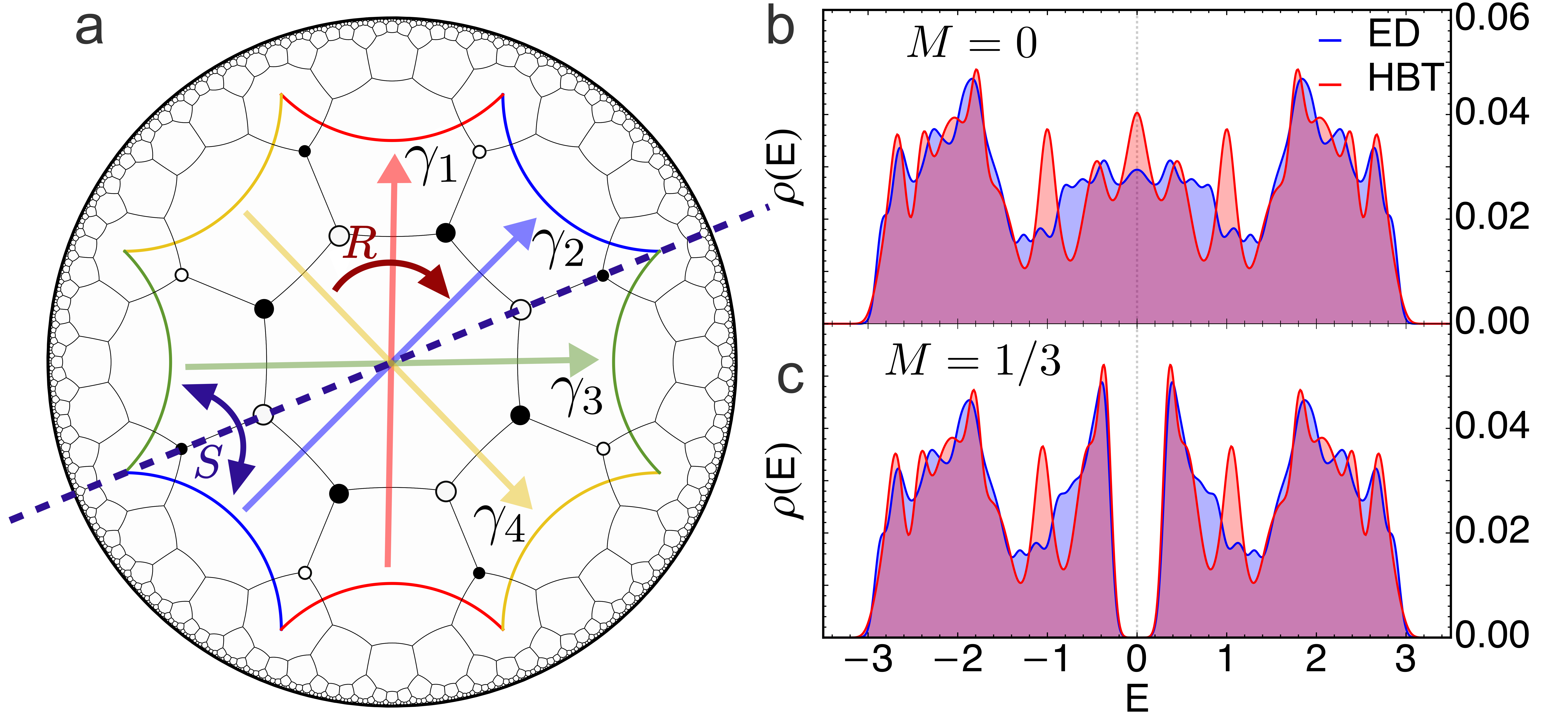} \caption[]{
        \textbf{Nearest-neighbor model on the $\boldsymbol{\{8 , 3\}}$ lattice.}
        \textbf{a.} The lattice consists of octagons with coordination number three. The Bolza cell (multi-colored octagon), the fundamental tile of the hyperbolic Bravais lattice, contains six elementary octagons and 16 sites (black/white dots).
        The colored arrows labelled  ${\gamma_1,\dots,\gamma_4}$ are the generators of the hyperbolic Bravais lattice.
        Rotation $R$ (dark red) and reflection $S$ (dark blue) are symmetries of the model.
        \textbf{b,~c.}~Bulk density of states $\rho(E)$ for the nearest-neighbor model on the $\{8 , 3\}$-lattice, extracted from hyperbolic band theory (HBT, red) vs.~exact diagonalization (ED, blue) in the absence~(\textbf{b}) vs.~presence~(\textbf{c}) of a sublattice potential $M$.
        }
    \label{fig:Bolza-cell}
    \end{figure}
    
    We next consider the inclusion of an on-site potential $\pm M$ with opposite sign on the two sublattices of the $\{8 , 3\}$ lattice, marked with white/black in Fig.~\ref{fig:Bolza-cell}\textbf{a}. 
    In the absence of a sublattice potential, the DoS is gapless at $E = 0$, whereas we observe a gap $\Delta E = 2M$ for $M {\neq} 0$. 
    This feature is reproduced both with ED and HBT.
    In fact, for all tight-binding models studied in this work, whenever HBT predicts a gap in the DoS at certain energies, then a bulk gap is also found in this energy range with ED on flakes. 
    Whether this behavior generalizes to all hyperbolic lattice models constitutes a formidable question for future investigations.
            
    \emph{Topological band insulators.}---We introduce the \emph{hyperbolic Haldane model} on the $\{8 , 3\}$ lattice by including complex-valued next-to-nearest-neighbor hopping terms, $t_2 \e^{ \pm  \imi\Phi}$, to the tight-binding Hamiltonian of the previous section.
    The positive (negative) sign is chosen in the exponent for hopping in the clockwise (counter-clockwise) direction within an octagon. 
    This model describes spinless fermions coupled orbitally to staggered magnetic fluxes, see Fig.~\ref{fig:Haldane}\textbf{a}. 
    The associated Bloch Hamiltonian $\mathcal{H}_{\rm H}(\bs{k})$ with crystal momentum $\bs{k} = (k_1 , k_2 , k_3 , k_4)$ is constructed using HBT~\cite{supp}. 
    The magnetic field breaks time-reversal symmetry, locating the model in Altland-Zirnbauer class~$\textrm{A}$~\cite{Kitaev:2009,Altland:1997}; we therefore anticipate that its topology is encoded by the Chern class~\cite{Ryu:2010,Nakahara:1990}. 
        
    We investigate in Fig.~\ref{fig:Haldane}\textbf{b} %the spectrum of 
    the model in terms of the DoS function     $\rho^\textrm{HBT}(E;\Phi)$ as $\Phi$ is varied for $t_2 = M/2 = 1/6$. We identify extended gapped regions at filling fractions $f  = \tfrac{5}{16} , \tfrac{8}{16} , \tfrac{11}{16}$, corresponding respectively to chemical potentials $\mu=-1.3 , 0 , 1.3$ at $\Phi = \pi/2$. (In contrast to the Euclidean case, non-vanishing $M {\neq} 0$ is necessary to open an energy gap at half-filling for the hyperbolic Haldane model~\cite{supp}.)
    In Fig.~\ref{fig:Haldane}\textbf{c} we observe that the bulk gaps obtained from ED on flakes resp.~from HBT again agree. 
    Two of the three gaps are special in that they are filled by boundary states, as is inferred from ED by integrating the local-DoS over boundary sites~\cite{supp}. 
    Below, we reveal that these energy gaps are associated with non-trivial Chern topology and chiral edge states.
    
    \begin{figure}[t!]
    \centering
        \includegraphics[width=\linewidth]{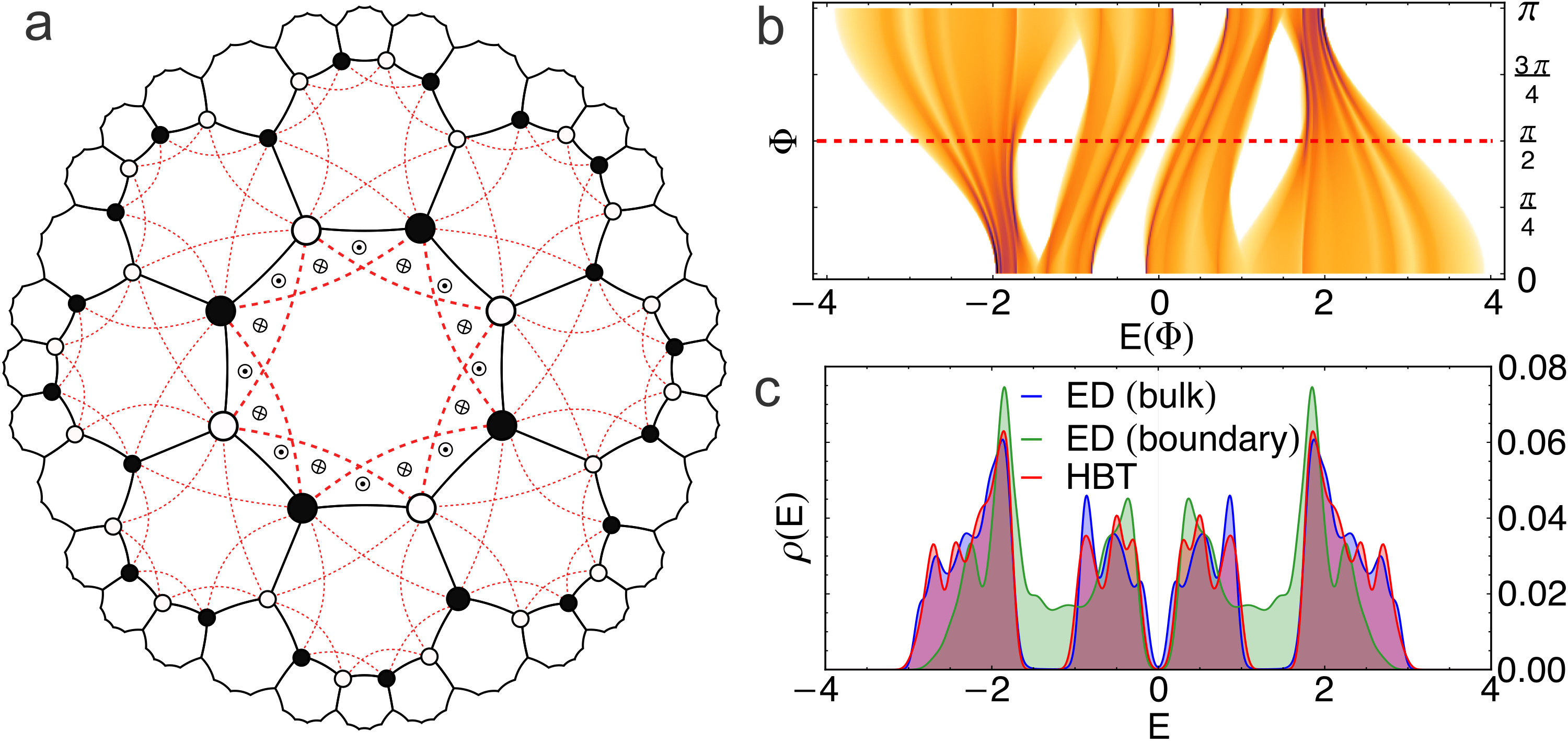} 
        \caption[]{
        \textbf{Hyperbolic Haldane model.} 
        \textbf{a.} Schematic depiction of the model.
        Red dashed lines indicate next-to-nearest neighbor hopping with amplitude $t_2 \e^{\pm\imi \Phi}$. 
        The phases $ \pm \Phi$ arise due to alternating magnetic fluxes (symbols $\odot$ resp.~$\otimes$ in the innermost octagon) through the system. 
        \textbf{b.} Density of states (DoS) for $t_2 = \tfrac{M}{2} = \tfrac{1}{6}$ computed from HBT, revealing three energy gaps.
        \textbf{c.} Bulk-DoS functions $\rho^\textrm{HBT}(E)$ and $\rho^\textrm{ED}_\textrm{bulk}(E)$ for $\Phi = \tfrac{\pi}{2}$ (red dashed line in \textbf{b}) computed using HBT (red) and ED (blue), and the boundary-DoS $\rho^\textrm{ED}_\textrm{boundary}$ (green). 
        }
    \label{fig:Haldane}
    \end{figure}
    
    Next we introduce the \emph{hyperbolic Kane-Mele (KM) model} of spin-$\tfrac{1}{2}$ fermions on the $\{8 , 3\}$ lattice as a time-reversal-symmetric topological model in Altland-Zirnbauer class~$\textrm{AII}$. 
    The model can be interpreted as a ``doubled'' version of the Haldane model, graded with a spin degree of freedom, in the following sense~\cite{Kane:2005,Urwyler:2021,*Urwyler:2022b}: denoting the Hamiltonian of the Haldane model by $\mathcal{H}_{\rm H}$, the Kane-Mele Hamiltonian for the spin-up (spin-down) fermions reads $\mathcal{H}_{\rm H}$ ($\mathcal{H}_{\rm H}^*$), and is supplied with spin-mixing Rashba term with amplitude $\lambda_\textrm{R}$.
    Since the hyperbolic curvature induces non-trivial holonomy of the spin along closed loops of the lattice, constructing a symmetry-compatible Rashba term is challenging. 
    We therefore simplify the model here by assuming a non-constant curvature that is concentrated at the corners of the Bolza cell, while the curvature is flat everywhere else (for detailed construction see Supplementary Fig.~\ref{fig:reduced-KM-geometry}~\cite{supp}).
    We call this simplified model the \emph{reduced hyperbolic KM model}.
    
    The reduced KM model is expected to exhibit energy gaps at the same filling fractions as the Haldane model as long as  $\lambda_\textrm{R}$ is sufficiently small. This is verified by the plot of $\rho^\textrm{HBT}(E;\lambda_\textrm{R})$ in Fig.~\ref{fig:Kane-Mele}\textbf{a}. 
    The obtained data motivate us to fix $\lambda_\textrm{R} = -1/6$, in which case all three gaps are still present. 
    The comparison of the resulting $\rho^\textrm{HBT}(E)$ and $\rho^\textrm{ED}_\textrm{bulk}(E)$ is shown in Fig.~\ref{fig:Kane-Mele}\textbf{b}. 
    The same panel also displays the corresponding $\rho^\textrm{ED}_\textrm{boundary}(E)$, which reveals filling of the two outer energy gaps by edge states, portending a non-trivial Kane-Mele topology.

    \begin{table}[t]
    \caption{Values of topological invariants for the considered hyperbolic models, with the three energy gaps labelled by their filling fraction $(f)$ and chemical potential $(\mu)$. 
    }
    \begin{ruledtabular}
    \begin{tabular}{cccccccccccc}
    	\multicolumn{2}{c}{} &
    	$\;\;$ &
    	\multicolumn{4}{c}{Haldane} &
    	$\;\;$ &
    	\multicolumn{4}{c}{Kane-Mele}  \\ \hline
    	$f$ & $\mu$ & $\;\;$ & $\mathcal{C}_a$ & $\mathcal{C}_b$ & $\mathcal{C}_c$ & $\mathcal{C}_\textrm{RS}$ & $\;\;$ & $\nu_a$ & $\nu_b$ & $\nu_c$ & $\nu_\textrm{RS}$  \\ \hline
    	$5/16$  & $-1.3$ & $\quad$ &$-1$ & $+1$ & $-1$ & $-0.986$ & $\;\;$ & $1$ & $1$ & $1$ & $-0.971$	\\
    	$8/16$  & $0$ & $\;\;$ &$0$ & $0$ & $0$ & $0$ & $\;\;$ & $0$ & $0$ & $0$ & $0$	\\ 
    	$11/16$  & $+1.3$ & $\quad$ &$-1$ & $+1$ & $-1$ & $-0.986$ & $\;\;$ & $1$ & $1$ & $1$ & $-0.971$	\\
    \end{tabular}
    \end{ruledtabular}
    \label{tab:topology}
    \end{table}
    
    \emph{Topological invariants.}---We compute topological invariants in momentum and position space for the band gaps of both constructed models.
    In $\bs{k}$-space, we compute the first Chern numbers of the Haldane model from the Bloch Hamiltonian $\mcH_{\rm H}(\bs{k})$ in the six planes spanned by pairs $(k_i,k_j)$ of momentum components, $i , j = 1,\ldots, 4$.
    The model exhibits $(\pi/2)$-rotation symmetry $R$ around the center of the Bolza cell [dark red arrow in Fig.~\ref{fig:Bolza-cell}(a)], which transforms the group generators as $(\gamma_1 , \gamma_2 , \gamma_3 , \gamma_4)\mapsto(\gamma_3 , \gamma_4 , \gamma_1^{-1} , \gamma_2^{-1})$; therefore, the Hamiltonians $\mcH_\textrm{H}(k_1 , k_2 , k_3 , k_4)$ and $\mcH_\textrm{H}(k_3 ,  k_4 , -k_1 , - k_2)$ are related by a unitary transformation. 
    Consequently, one can relate Chern numbers $\mathcal{C}_{12} = \mathcal{C}_{34} =: \mathcal{C}_a$ and $\mathcal{C}_{14} = \mathcal{C}_{23} =: \mathcal{C}_c$.
    In addition, while reflection $S$ [dashed blue line in Fig.~\ref{fig:Bolza-cell}(a)] flips the magnetic fluxes $\pm\Phi$, its composition with time-reversal constitutes an antiunitary symmetry of the Haldane model, transforming $(k_1 , k_2 , k_3 , k_4)\mapsto ( - k_4 ,  - k_3 ,  - k_2 ,  - k_1)$ and relating $\mathcal{C}_{13}=\mathcal{C}_{24}=:\mathcal{C}_b$.
    We compute $\mathcal{C}_{a,b,c}$ using Wilson loops~\cite{Soluyanov:2011,Gresch:2017}, and find that energy gaps at $f = \tfrac{5}{16} , \tfrac{11}{16}$ have non-trivial Chern number $ \pm  1$ in all planes, while the gap at $f = \tfrac{8}{16}$ does not exhibit Chern topology, see Table~\ref{tab:topology}. 
    Similarly, after utilizing the $R$ and $S$ symmetry of $\mathcal{H}_{\rm KM}(\bs{k})$, we identify three independent $\mathbb{Z}_2$-topological invariants $\nu_{ij}$, namely $\nu_a := \nu_{12} = \nu_{34}$, $\nu_b := \nu_{13} = \nu_{24}$, and $\nu_c := \nu_{14} = \nu_{23}$.
    We find that $\nu_{a,b,c}$ are all non-trivial for the two outer band gaps, while they are trivial for the inner gap, see Table~\ref{tab:topology}. 
    [We discuss in the Supplementary Material~\cite{supp} that if $M = 0$ then an additional $(\pi/4)$-rotation symmetry further implies $\mathcal{C}_a = \mathcal{C}_c$ and $\nu_a = \nu_c$].
    
    We also computed higher-dimensional topological invariants for $\mathcal{H}_{\rm H}(\bs{k})$ and $\mathcal{H}_{\rm KM}(\bs{k})$, namely the $\ztwo$-valued Fu-Kane-Mele~\cite{Fu:2007} invariant on 3D subspaces (for KM) and the second Chern number~\cite{Nakahara:1990,Zhang:2001} in 4D BZ (for both). 
    These, however, are all trivial. 
    Nevertheless, hyperbolic toy-models with non-vanishing values of such topological invariants could be constructed through reverse engineering: starting from a 4D Euclidean Bloch-Hamiltonian, $\mathcal{H}_{\rm Euc}(\bs{k})$, which features such topological invariants, one constructs a hyperbolic tight-binding model where each component $k_i$ of $\bs{k}$ is replaced by a generator $\gamma_i$ of the hyperbolic Bravais lattice. 
    We leave this promising route for designing topological hyperbolic Hamiltonians for future research. 
    
    \begin{figure}[t!]
    \centering
        \includegraphics[width=\linewidth]{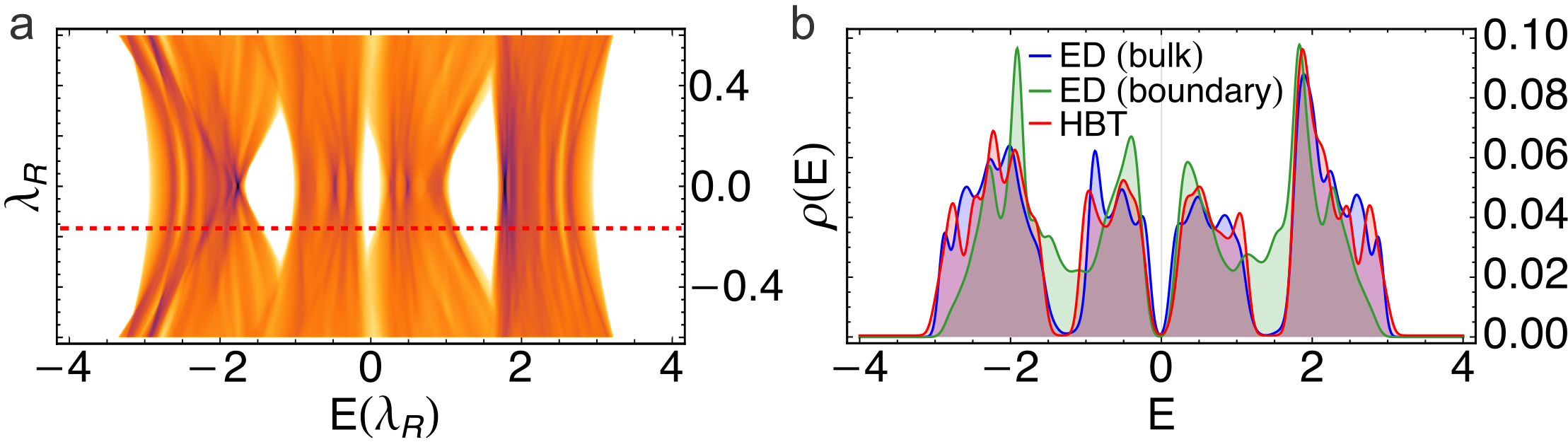} 
        \caption[]{
        \textbf{Hyperbolic Kane-Mele model.}
        \textbf{a.} Density of states (DoS) for $t_2 = \tfrac{M}{2} = \tfrac{1}{6}$ computed from HBT, which reveals three energy gaps at small Rashba coupling $\lambda_\textrm{R}$. 
        \textbf{b.} Bulk-DoS functions for $\lambda_\textrm{R} = {-}\tfrac{1}{6}$ (red dashed line in~\textbf{a}) computed using HBT (red) and ED (blue), and the boundary-DoS function (green).
        }
    \label{fig:Kane-Mele}
    \end{figure}
    
    We complement the momentum-space discussion of the band topology of the two models     with real-space topological markers~\cite{Bellissard:1994,Kitaev:2006,Bianco:2011,Prodan:2011,Huang:2018}. 
    Importantly, these computations include states transforming in \emph{all} representations of the non-commutative translation group, hence going beyond HBT. 
    For the Haldane model, we compute the \emph{real-space Chern number} $\mathcal{C}_\textrm{RS}$ as introduced in Ref.~\onlinecite{Kitaev:2006} (detailed in Supplementary Material~\cite{supp}).
    This algorithm does not warrant quantized results, but we confirm that integers are approached as the summation regions are enlarged. 
    Our position-space analysis confirms that the energy gaps at $\mu  =  {\pm} 1.3$ are topological, while the one at $\mu  =  0$ is trivial, see Table~\ref{tab:topology}. 
    We observe $ \mathcal{C}_{\rm RS}= \mathcal{C}_a = - \mathcal{C}_b = \mathcal{C}_c$ for all phases with gapped bulk. 
    Recall here that for 2D Euclidean lattices we have $ \mathcal{C}_\textrm{RS}=\mathcal{C}_{12}$~\cite{Bellissard:1994,Kitaev:2006}, whereas no exact relation is currently known for hyperbolic lattices.
        
    We further adapt the techniques of Refs.~\onlinecite{Kitaev:2006,Bianco:2011,Prodan:2011,Huang:2018} to compute the \emph{real-space spin Chern number} $\nu_\textrm{RS} \in \mathbb{Z}$ of the reduced KM model for each bulk gap. This invariant is integer-valued as long as spin-mixing terms in the Hamiltonian are sufficiently weak, and for Euclidean lattice models it obeys $\nu_\textrm{RS}=\nu_{12}\;(\textrm{mod}\,2) $~\cite{Prodan:2011,Huang:2018}. We observe that the extracted $\ztwo$-invariants obey $\nu_\textrm{RS} = \nu_a = \nu_b =\nu_c$ for all cases analyzed. Whether there exists a simple \emph{universal} relation between the parities of $\nu_{a,b,c,\textrm{RS}}$ constitutes another open question.
    
    \begin{figure}[t!]
    \centering
        \includegraphics[width=\linewidth]{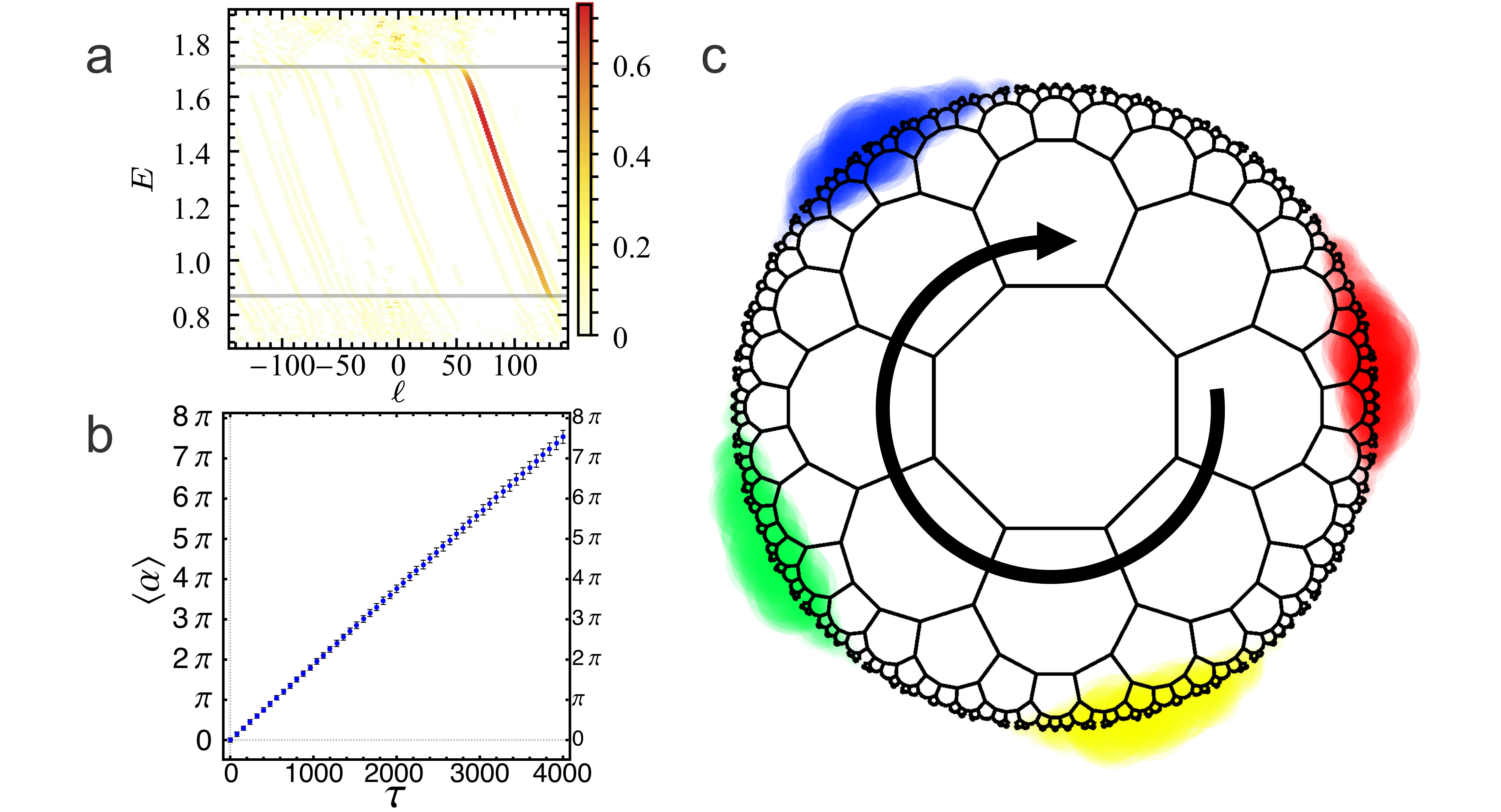}
        \caption[]{
        \textbf{Topological edge states.}
        \textbf{a.} Edge-dispersion $E$ as a function of angular momentum $\ell$ for chiral edge states in the upper energy gap of the hyperbolic Haldane model. 
        \textbf{b.} Propagation of a Gaussian wave packet  along the flake boundary; $\langle \alpha \rangle$ is the angular displacement in time $\tau$, and error bars indicate the width of the wave packet. 
        \textbf{c.} Snapshots of the wave packet at $\tau = 0 , 240 , 480 , 720$ (colored respectively red/yellow/green/blue). 
        The area of a disk centered at a given site encodes the probability to find the particle at that site.
        }
    \label{fig:edge-states}
    \end{figure}
    
    \emph{Bulk-boundary correspondence.}---We finally investigate whether the non-trivial band topology identified in the bulk for both models is reflected in topological edge states on the  boundary. For this, we (\emph{i})~extract the edge-state dispersion, (\emph{ii})~showcase the propagation of an edge-localized wave packet around the flake boundary, and (\emph{iii})~investigate their robustness against disorder. 
    
    (\emph{i})~When computing for a circular flake the edge-state dispersion ($E$) against angular momentum ($\ell$), note that the latter is only defined modulo~4 due to the fourfold symmetry $R$. 
    To obtain $E(\ell)$ for \emph{unbounded} $\ell \in \mathbb{Z}$, we decompose the lattice eigenstates $|{\phi_j}\rangle$ into eigenmodes $|{\psi_{n,\ell}}\rangle$ of the Laplace-Beltrami operator defined in the continuum~\cite{Boettcher:2020,Lenggenhager:2021d} and select the number $\ell$ with the largest contribution (for details see Supplementary Material~\cite{supp}). 
    We plot $E(\ell)$ for the edge states in the hyperbolic Haldane model for the energy gap at $\mu = 1.3$ in Fig.~\ref{fig:edge-states}\textbf{a}. 
    We observe a single dispersive branch for positive $\ell$ only, implying chiral edge state at the flake boundary, in agreement with $\abs{\mathcal{C}_\textrm{RS}} = 1$.  
    An analogous analysis for the reduced Kane-Mele model at the same filling reveals a pair of counter-propagating \emph{helical} branches, compatible with $\abs{\nu_\textrm{RS}} = 1$ (see Supplementary Fig.~\ref{fig:edgemode-dispersion}~\cite{supp}).
    
    (\emph{ii})~To study the propagation of a \emph{wave packet} along the edge, we construct a boundary-localized Gaussian state with energy near $\mu$ and with energy width $\sigma$~\cite{supp}.
    We plot in Fig.~\ref{fig:edge-states}\textbf{b,c} the time-evolution of a wave packet initialized with parameters $(\mu,\sigma) = (1.3 , 0.025)$ in the hyperbolic Haldane model. 
    We find the center of the packet smoothly propagates along the boundary in time.
    The angular group velocity of the wave packet extracted from the time-evolution matches the edge-dispersion from (\emph{i})~via $\omega_\textrm{group}  =  \mathrm{d} E/ \mathrm{d} \ell$. 
    We similarly analyzed the edge-state propagation for the reduced KM model~\cite{supp} and confirmed their anticipated helical \mbox{character}.
        
    \begin{figure}[t!]
    \centering
        \includegraphics[width=\linewidth]{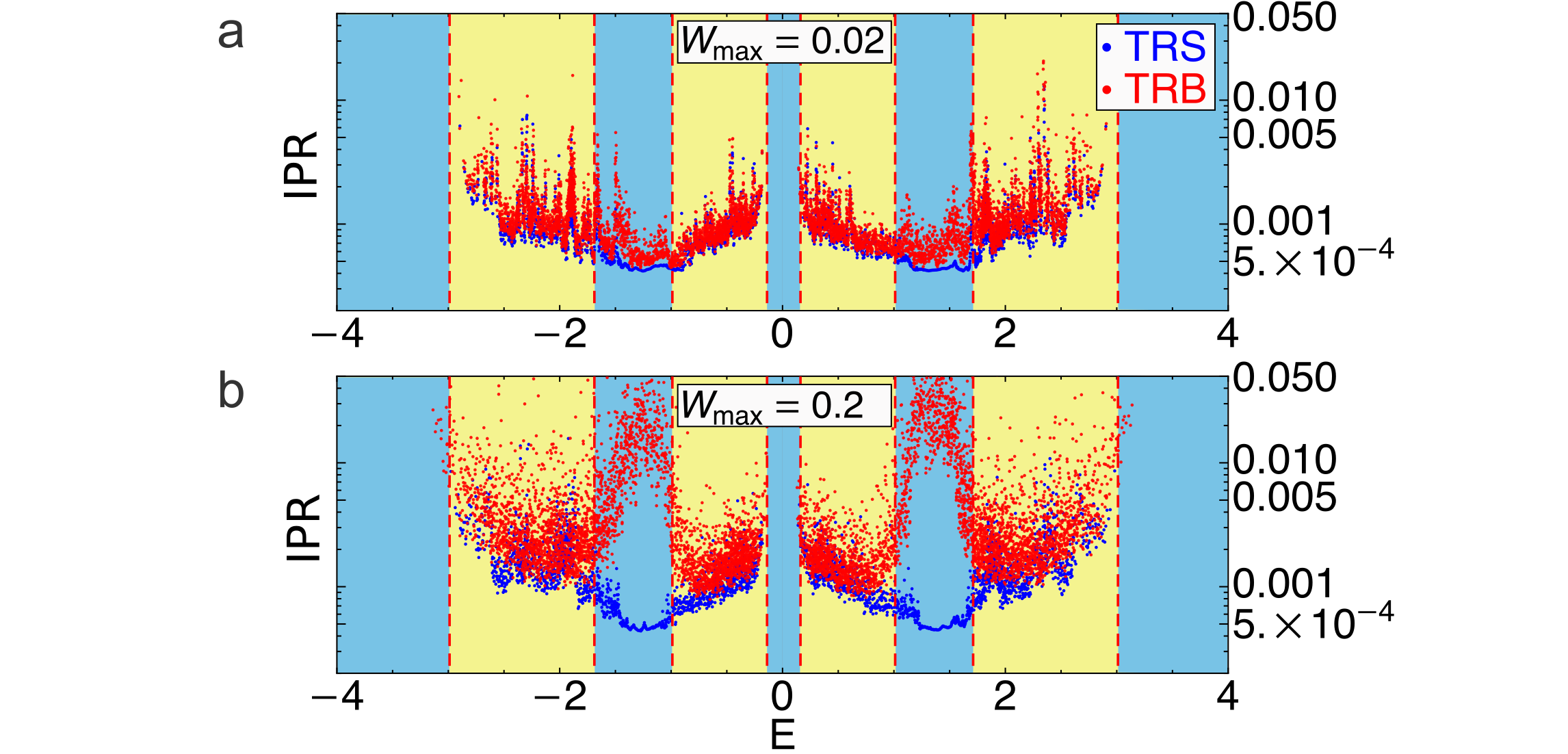} 
        \caption[]{
        \textbf{Robustness against Anderson localization.}
        For the reduced hyperbolic KM model, we consider inclusion of random spin-mixing terms that preserve (TRS, blue) or break (TRB, red) time-reversal symmetry.
        A dot with coordinates ($E_j, \textrm{IPR}_j$) represents the energy and the inverse participation ratio of an eigenstate $|{\phi_j}\rangle$ in the flake geometry.
        The random terms are drawn from a uniform distribution with 
        $W_\textrm{max}  =  0.02$ (\textbf{a}) and
        $W_\textrm{max}  =  0.2$ (\textbf{b}). 
        The blue (yellow) backgrounds indicate energy ranges corresponding to bulk gap (bulk band) in the absence of disorder.
        }
    \label{fig:Anderson}
    \end{figure}
    
    (\emph{iii})~To quantify the robustness of the edge states against Anderson localization when subject to disorder, we show that they retain a small \emph{inverse participation ratio} (IPR). 
    Here $0 <\textrm{IPR} \leq 1$ is defined such that an eigenstate $|{\phi_j}\rangle$ characterized by value $\textrm{IPR}_j$ has most support over approximately $1{/}\textrm{IPR}_j$ sites~\cite{supp}. 
    For the reduced KM model, we add random spin-mixing terms that either preserve (TRS) or break (TRB) time-reversal symmetry of the model. 
    These can be interpreted as \emph{random Rashba terms} resp.~\emph{random magnetic fields}, with an amplitude $W \in [-W_{\rm max} , W_{\rm max}]$ drawn from a uniform distribution. 
    The results in Fig.~\ref{fig:Anderson} indicate that disorder with TRB leads to localization of the edge states, whereas increasing $W_{\rm max}$ in the presence of TRS does not change their IPR values significantly, as expected for topological edge states. 
    We similarly verified for the hyperbolic Haldane model that inclusion of random on-site potential drawn from a uniform distribution $W \in [-W_{\rm max} , W_{\rm max}]$  has little effect on the IPR of the edge states. 
    In addition, we confirmed that the chiral propagation of the edge states without back-scattering is preserved under the addition of the random terms (see Supplementary Fig.~\ref{fig:HaldaneAnderson}~\cite{supp}).
    This provides further evidence of the non-trivial topology of the constructed model.
    
    \emph{Outlook.}---Our work constitutes an essential step towards designing topological hyperbolic Hamiltonians and exploring the interplay of geometry and topology in such systems.
    It is natural to wonder if similar constructions of topological insulators generalize to other hyperbolic $\{p , q\}$ lattices. 
    Indeed, the Haldane model on the $\{6 , 4\}$ lattice has very recently been implemented as an electric-circuit network by Ref.~\onlinecite{Zhang:2022}, while another model of Chern insulator on the $\{8 , 3\}$ lattice, although lacking translation symmetry, has been considered by Ref.~\onlinecite{Liu:2022}. 
    However, these works did not apply HBT to characterize the models, thus lacking the momentum-space language.
    
    A key difference between hyperbolic and Euclidean lattices is the extensive scaling of the boundary in the hyperbolic case, implying a finite fraction of boundary sites \emph{irrespective} of the system size~\cite{Boettcher:2020,Saa:2021}.
    Consequently, a \emph{macroscopic fraction of all states} contributes to topological edge modes, in stark contrast to Euclidean topological models.  
    It will be intriguing to investigate features of non-Hermitian topology in this context, as the non-Hermitian skin effect likewise affects a macroscopic fraction of the spectrum~\cite{Borgnia:2020,Okuma:2020,Weidemann:2020,Hofmann:2020,Helbig:2020,Lv:2021}.
    The extensive hyperbolic boundary is also anticipated to give access to novel one-dimensional many-body models affected by the bulk design through the bulk-boundary correspondence.     
    
    \emph{Code and data availability.}---The Wolfram Language code and the generated data used to arrive at the conclusions presented in this work are publicly available in the following data repository: \href{https://doi.org/10.5281/zenodo.6380568}{https://doi.org/10.5281/zenodo.6380568}~\cite{Urwyler:2022:SDC}. 
    The manuscript is based on the master's thesis of one of the authors~\cite{Urwyler:2021,*Urwyler:2022b}.
        
    % \begin{acknowledgments}
    \emph{Acknolwedgments.}---We would like to express our thanks to M.~Brzezi\'{n}ska, A.~Chen, B.~Lapierre, J.~Maciejko, A.~Stegmaier, T.~Tummuru, and L.~K.~Upreti for helpful discussions. 
    P.~M.~L.~and T.~B.~were supported by the Ambizione grant No.~185806 by the Swiss National Science Foundation.
    I.~B.~acknowledges support from the University of Alberta startup fund UOFAB Startup Boettcher and the Natural Sciences and Engineering Research Council of Canada (NSERC) Discovery Grants RGPIN-2021-02534 and DGECR2021-00043.
    R.~T.~acknowledges funding by the Deutsche Forschungsgemeinschaft (DFG, German Research Foundation) through Project-ID 258499086 - SFB 1170, and through the W\"urzburg-Dresden Cluster of Excellence on Complexity and Topology in Quantum Matter -- \textit{ct.qmat} Project-ID 39085490 - EXC 2147.
    T.~N.~acknowledges support from the European Research Council (ERC) under the European Union’s Horizon 2020 research and innovation programm (ERC-StG-Neupert-757867-PARATOP).
    % \end{acknowledgments}
    
    %\bibliography{bibliography}
    
    \let\oldaddcontentsline\addcontentsline
    \renewcommand{\addcontentsline}[3]{}
    \bibliography{bibliography}

\begin{thebibliography}{61}%
\makeatletter
\providecommand \@ifxundefined [1]{%
 \@ifx{#1\undefined}
}%
\providecommand \@ifnum [1]{%
 \ifnum #1\expandafter \@firstoftwo
 \else \expandafter \@secondoftwo
 \fi
}%
\providecommand \@ifx [1]{%
 \ifx #1\expandafter \@firstoftwo
 \else \expandafter \@secondoftwo
 \fi
}%
\providecommand \natexlab [1]{#1}%
\providecommand \enquote  [1]{``#1''}%
\providecommand \bibnamefont  [1]{#1}%
\providecommand \bibfnamefont [1]{#1}%
\providecommand \citenamefont [1]{#1}%
\providecommand \href@noop [0]{\@secondoftwo}%
\providecommand \href [0]{\begingroup \@sanitize@url \@href}%
\providecommand \@href[1]{\@@startlink{#1}\@@href}%
\providecommand \@@href[1]{\endgroup#1\@@endlink}%
\providecommand \@sanitize@url [0]{\catcode `\\12\catcode `\$12\catcode
  `\&12\catcode `\#12\catcode `\^12\catcode `\_12\catcode `\%12\relax}%
\providecommand \@@startlink[1]{}%
\providecommand \@@endlink[0]{}%
\providecommand \url  [0]{\begingroup\@sanitize@url \@url }%
\providecommand \@url [1]{\endgroup\@href {#1}{\urlprefix }}%
\providecommand \urlprefix  [0]{URL }%
\providecommand \Eprint [0]{\href }%
\providecommand \doibase [0]{http://dx.doi.org/}%
\providecommand \selectlanguage [0]{\@gobble}%
\providecommand \bibinfo  [0]{\@secondoftwo}%
\providecommand \bibfield  [0]{\@secondoftwo}%
\providecommand \translation [1]{[#1]}%
\providecommand \BibitemOpen [0]{}%
\providecommand \bibitemStop [0]{}%
\providecommand \bibitemNoStop [0]{.\EOS\space}%
\providecommand \EOS [0]{\spacefactor3000\relax}%
\providecommand \BibitemShut  [1]{\csname bibitem#1\endcsname}%
\let\auto@bib@innerbib\@empty
%</preamble>
\bibitem [{\citenamefont {Bradlyn}\ \emph {et~al.}(2017)\citenamefont
  {Bradlyn}, \citenamefont {Elcoro}, \citenamefont {Cano}, \citenamefont
  {Vergniory}, \citenamefont {Wang}, \citenamefont {Felser}, \citenamefont
  {Aroyo},\ and\ \citenamefont {Bernevig}}]{Bradlyn:2017}%
  \BibitemOpen
  \bibfield  {author} {\bibinfo {author} {\bibfnamefont {Barry}\ \bibnamefont
  {Bradlyn}}, \bibinfo {author} {\bibfnamefont {L.}~\bibnamefont {Elcoro}},
  \bibinfo {author} {\bibfnamefont {Jennifer}\ \bibnamefont {Cano}}, \bibinfo
  {author} {\bibfnamefont {M.~G.}\ \bibnamefont {Vergniory}}, \bibinfo {author}
  {\bibfnamefont {Zhijun}\ \bibnamefont {Wang}}, \bibinfo {author}
  {\bibfnamefont {C.}~\bibnamefont {Felser}}, \bibinfo {author} {\bibfnamefont
  {M.~I.}\ \bibnamefont {Aroyo}}, \ and\ \bibinfo {author} {\bibfnamefont
  {B.~Andrei}\ \bibnamefont {Bernevig}},\ }\bibfield  {title} {\enquote
  {\bibinfo {title} {Topological quantum chemistry},}\ }\href {\doibase
  10.1038/nature23268} {\bibfield  {journal} {\bibinfo  {journal} {Nature}\
  }\textbf {\bibinfo {volume} {547}},\ \bibinfo {pages} {298--305} (\bibinfo
  {year} {2017})}\BibitemShut {NoStop}%
\bibitem [{\citenamefont {Po}\ \emph {et~al.}(2017)\citenamefont {Po},
  \citenamefont {Vishwanath},\ and\ \citenamefont {Watanabe}}]{Po:2017}%
  \BibitemOpen
  \bibfield  {author} {\bibinfo {author} {\bibfnamefont {Hoi~Chun}\
  \bibnamefont {Po}}, \bibinfo {author} {\bibfnamefont {Ashvin}\ \bibnamefont
  {Vishwanath}}, \ and\ \bibinfo {author} {\bibfnamefont {Haruki}\ \bibnamefont
  {Watanabe}},\ }\bibfield  {title} {\enquote {\bibinfo {title} {Symmetry-based
  indicators of band topology in the 230 space groups},}\ }\href {\doibase
  10.1038/s41467-017-00133-2} {\bibfield  {journal} {\bibinfo  {journal} {Nat.
  Commun.}\ }\textbf {\bibinfo {volume} {8}},\ \bibinfo {pages} {50} (\bibinfo
  {year} {2017})}\BibitemShut {NoStop}%
\bibitem [{\citenamefont {Vergniory}\ \emph {et~al.}(2022)\citenamefont
  {Vergniory}, \citenamefont {Wieder}, \citenamefont {Elcoro}, \citenamefont
  {Parkin}, \citenamefont {Felse}, \citenamefont {Bernevig},\ and\
  \citenamefont {Regnault}}]{Vergniory:2022}%
  \BibitemOpen
  \bibfield  {author} {\bibinfo {author} {\bibfnamefont {Maia~G.}\ \bibnamefont
  {Vergniory}}, \bibinfo {author} {\bibfnamefont {Benjamin~J.}\ \bibnamefont
  {Wieder}}, \bibinfo {author} {\bibfnamefont {Luis}\ \bibnamefont {Elcoro}},
  \bibinfo {author} {\bibfnamefont {Stuart S.~P.}\ \bibnamefont {Parkin}},
  \bibinfo {author} {\bibfnamefont {Claudia}\ \bibnamefont {Felse}}, \bibinfo
  {author} {\bibfnamefont {B.~Andrei}\ \bibnamefont {Bernevig}}, \ and\
  \bibinfo {author} {\bibfnamefont {Nicolas}\ \bibnamefont {Regnault}},\
  }\bibfield  {title} {\enquote {\bibinfo {title} {{All topological bands of
  all nonmagnetic stoichiometric materials}},}\ }\href {\doibase
  10.1126/science.abg9094} {\bibfield  {journal} {\bibinfo  {journal}
  {Science}\ }\textbf {\bibinfo {volume} {376}},\ \bibinfo {pages} {abg9094}
  (\bibinfo {year} {2022})}\BibitemShut {NoStop}%
\bibitem [{\citenamefont {Andrei}\ \emph {et~al.}(2021)\citenamefont {Andrei},
  \citenamefont {Efetov}, \citenamefont {Jarillo-Herrero}, \citenamefont
  {MacDonald}, \citenamefont {Mak}, \citenamefont {Senthil}, \citenamefont
  {Tutuc}, \citenamefont {Yazdani},\ and\ \citenamefont {Young}}]{Andrei:2021}%
  \BibitemOpen
  \bibfield  {author} {\bibinfo {author} {\bibfnamefont {Eva~Y.}\ \bibnamefont
  {Andrei}}, \bibinfo {author} {\bibfnamefont {Dmitri~K.}\ \bibnamefont
  {Efetov}}, \bibinfo {author} {\bibfnamefont {Pablo}\ \bibnamefont
  {Jarillo-Herrero}}, \bibinfo {author} {\bibfnamefont {Allan~H.}\ \bibnamefont
  {MacDonald}}, \bibinfo {author} {\bibfnamefont {Kin~Fai}\ \bibnamefont
  {Mak}}, \bibinfo {author} {\bibfnamefont {T.}~\bibnamefont {Senthil}},
  \bibinfo {author} {\bibfnamefont {Emanuel}\ \bibnamefont {Tutuc}}, \bibinfo
  {author} {\bibfnamefont {Ali}\ \bibnamefont {Yazdani}}, \ and\ \bibinfo
  {author} {\bibfnamefont {Andrea~F.}\ \bibnamefont {Young}},\ }\bibfield
  {title} {\enquote {\bibinfo {title} {The marvels of moir\'{e} materials},}\
  }\href {\doibase 10.1038/s41578-021-00284-1} {\bibfield  {journal} {\bibinfo
  {journal} {Nat. Rev. Mater.}\ }\textbf {\bibinfo {volume} {6}},\ \bibinfo
  {pages} {201--206} (\bibinfo {year} {2021})}\BibitemShut {NoStop}%
\bibitem [{\citenamefont {Koll{\'{a}}r}\ \emph {et~al.}(2019)\citenamefont
  {Koll{\'{a}}r}, \citenamefont {Fitzpatrick},\ and\ \citenamefont
  {Houck}}]{Kollar:2019}%
  \BibitemOpen
  \bibfield  {author} {\bibinfo {author} {\bibfnamefont {Alicia~J.}\
  \bibnamefont {Koll{\'{a}}r}}, \bibinfo {author} {\bibfnamefont {Mattias}\
  \bibnamefont {Fitzpatrick}}, \ and\ \bibinfo {author} {\bibfnamefont
  {Andrew~A.}\ \bibnamefont {Houck}},\ }\bibfield  {title} {\enquote {\bibinfo
  {title} {{Hyperbolic lattices in circuit quantum electrodynamics}},}\ }\href
  {\doibase 10.1038/s41586-019-1348-3} {\bibfield  {journal} {\bibinfo
  {journal} {Nature}\ }\textbf {\bibinfo {volume} {571}},\ \bibinfo {pages}
  {45} (\bibinfo {year} {2019})}\BibitemShut {NoStop}%
\bibitem [{\citenamefont {Lenggenhager}\ \emph {et~al.}(2022)\citenamefont
  {Lenggenhager}, \citenamefont {Stegmaier}, \citenamefont {Upreti},
  \citenamefont {Hofmann}, \citenamefont {Helbig}, \citenamefont {Vollhardt},
  \citenamefont {Greiter}, \citenamefont {Lee}, \citenamefont {Imhof},
  \citenamefont {Brand}, \citenamefont {Kießling}, \citenamefont {Boettcher},
  \citenamefont {Neupert}, \citenamefont {Thomale},\ and\ \citenamefont
  {Bzdušek}}]{Lenggenhager:2021d}%
  \BibitemOpen
  \bibfield  {author} {\bibinfo {author} {\bibfnamefont {Patrick~M.}\
  \bibnamefont {Lenggenhager}}, \bibinfo {author} {\bibfnamefont {Alexander}\
  \bibnamefont {Stegmaier}}, \bibinfo {author} {\bibfnamefont {Lavi~K.}\
  \bibnamefont {Upreti}}, \bibinfo {author} {\bibfnamefont {Tobias}\
  \bibnamefont {Hofmann}}, \bibinfo {author} {\bibfnamefont {Tobias}\
  \bibnamefont {Helbig}}, \bibinfo {author} {\bibfnamefont {Achim}\
  \bibnamefont {Vollhardt}}, \bibinfo {author} {\bibfnamefont {Martin}\
  \bibnamefont {Greiter}}, \bibinfo {author} {\bibfnamefont {Ching~Hua}\
  \bibnamefont {Lee}}, \bibinfo {author} {\bibfnamefont {Stefan}\ \bibnamefont
  {Imhof}}, \bibinfo {author} {\bibfnamefont {Hauke}\ \bibnamefont {Brand}},
  \bibinfo {author} {\bibfnamefont {Tobias}\ \bibnamefont {Kießling}},
  \bibinfo {author} {\bibfnamefont {Igor}\ \bibnamefont {Boettcher}}, \bibinfo
  {author} {\bibfnamefont {Titus}\ \bibnamefont {Neupert}}, \bibinfo {author}
  {\bibfnamefont {Ronny}\ \bibnamefont {Thomale}}, \ and\ \bibinfo {author}
  {\bibfnamefont {Tomáš}\ \bibnamefont {Bzdušek}},\ }\bibfield  {title}
  {\enquote {\bibinfo {title} {Simulating hyperbolic space on a circuit
  board},}\ }\href {\doibase 10.1038/s41467-022-32042-4} {\bibfield  {journal}
  {\bibinfo  {journal} {Nat. Commun.}\ }\textbf {\bibinfo {volume} {13}},\
  \bibinfo {pages} {4373} (\bibinfo {year} {2022})}\BibitemShut {NoStop}%
\bibitem [{\citenamefont {Magnus}(1974)}]{BookMagnus}%
  \BibitemOpen
  \bibfield  {author} {\bibinfo {author} {\bibfnamefont {W.}~\bibnamefont
  {Magnus}},\ }\href@noop {} {\emph {\bibinfo {title} {{Noneuclidean
  tesselations and their groups}}}}\ (\bibinfo  {publisher} {Academic Press,
  New York},\ \bibinfo {year} {1974})\BibitemShut {NoStop}%
\bibitem [{\citenamefont {Coxeter}\ and\ \citenamefont
  {Moser}(1980)}]{BookCoxeter}%
  \BibitemOpen
  \bibfield  {author} {\bibinfo {author} {\bibfnamefont {H.~S.~M.}\
  \bibnamefont {Coxeter}}\ and\ \bibinfo {author} {\bibfnamefont {W.~O.~J.}\
  \bibnamefont {Moser}},\ }\href@noop {} {\emph {\bibinfo {title} {{Generators
  and Relations for Discrete Groups}}}}\ (\bibinfo  {publisher} {Springer
  Berlin Heidelberg, Berlin, Heidelberg},\ \bibinfo {year} {1980})\BibitemShut
  {NoStop}%
\bibitem [{\citenamefont {Maciejko}\ and\ \citenamefont
  {Rayan}(2021)}]{Maciejko:2021}%
  \BibitemOpen
  \bibfield  {author} {\bibinfo {author} {\bibfnamefont {Joseph}\ \bibnamefont
  {Maciejko}}\ and\ \bibinfo {author} {\bibfnamefont {Steven}\ \bibnamefont
  {Rayan}},\ }\bibfield  {title} {\enquote {\bibinfo {title} {{Hyperbolic band
  theory}},}\ }\href {\doibase 10.1126/sciadv.abe9170} {\bibfield  {journal}
  {\bibinfo  {journal} {Sci. Adv.}\ }\textbf {\bibinfo {volume} {7}},\ \bibinfo
  {pages} {eabe9170} (\bibinfo {year} {2021})}\BibitemShut {NoStop}%
\bibitem [{\citenamefont {Yu}\ \emph {et~al.}(2020)\citenamefont {Yu},
  \citenamefont {Piao},\ and\ \citenamefont {Park}}]{Yu:2020}%
  \BibitemOpen
  \bibfield  {author} {\bibinfo {author} {\bibfnamefont {Sunkyu}\ \bibnamefont
  {Yu}}, \bibinfo {author} {\bibfnamefont {Xianji}\ \bibnamefont {Piao}}, \
  and\ \bibinfo {author} {\bibfnamefont {Namkyoo}\ \bibnamefont {Park}},\
  }\bibfield  {title} {\enquote {\bibinfo {title} {{Topological Hyperbolic
  Lattices}},}\ }\href {\doibase 10.1103/PhysRevLett.125.053901} {\bibfield
  {journal} {\bibinfo  {journal} {Phys. Rev. Lett.}\ }\textbf {\bibinfo
  {volume} {125}},\ \bibinfo {pages} {053901} (\bibinfo {year}
  {2020})}\BibitemShut {NoStop}%
\bibitem [{\citenamefont {Ikeda}\ \emph {et~al.}(2021)\citenamefont {Ikeda},
  \citenamefont {Aoki},\ and\ \citenamefont {Matsuki}}]{Ikeda:2021}%
  \BibitemOpen
  \bibfield  {author} {\bibinfo {author} {\bibfnamefont {Kazuki}\ \bibnamefont
  {Ikeda}}, \bibinfo {author} {\bibfnamefont {Shoto}\ \bibnamefont {Aoki}}, \
  and\ \bibinfo {author} {\bibfnamefont {Yoshiyuki}\ \bibnamefont {Matsuki}},\
  }\bibfield  {title} {\enquote {\bibinfo {title} {Hyperbolic band theory under
  magnetic field and {D}irac cones on a higher genus surface},}\ }\href
  {\doibase 10.1088/1361-648x/ac24c4} {\bibfield  {journal} {\bibinfo
  {journal} {J. Phys. Condens. Matter}\ }\textbf {\bibinfo {volume} {33}},\
  \bibinfo {pages} {485602} (\bibinfo {year} {2021})}\BibitemShut {NoStop}%
\bibitem [{\citenamefont {Stegmaier}\ \emph {et~al.}(2022)\citenamefont
  {Stegmaier}, \citenamefont {Upreti}, \citenamefont {Thomale},\ and\
  \citenamefont {Boettcher}}]{Stegmaier:2021}%
  \BibitemOpen
  \bibfield  {author} {\bibinfo {author} {\bibfnamefont {Alexander}\
  \bibnamefont {Stegmaier}}, \bibinfo {author} {\bibfnamefont {Lavi~K.}\
  \bibnamefont {Upreti}}, \bibinfo {author} {\bibfnamefont {Ronny}\
  \bibnamefont {Thomale}}, \ and\ \bibinfo {author} {\bibfnamefont {Igor}\
  \bibnamefont {Boettcher}},\ }\bibfield  {title} {\enquote {\bibinfo {title}
  {Universality of hofstadter butterflies on hyperbolic lattices},}\ }\href
  {\doibase 10.1103/PhysRevLett.128.166402} {\bibfield  {journal} {\bibinfo
  {journal} {Phys. Rev. Lett.}\ }\textbf {\bibinfo {volume} {128}},\ \bibinfo
  {pages} {166402} (\bibinfo {year} {2022})}\BibitemShut {NoStop}%
\bibitem [{\citenamefont {Boettcher}\ \emph {et~al.}(2020)\citenamefont
  {Boettcher}, \citenamefont {Bienias}, \citenamefont {Belyansky},
  \citenamefont {Koll\'ar},\ and\ \citenamefont {Gorshkov}}]{Boettcher:2020}%
  \BibitemOpen
  \bibfield  {author} {\bibinfo {author} {\bibfnamefont {Igor}\ \bibnamefont
  {Boettcher}}, \bibinfo {author} {\bibfnamefont {Przemyslaw}\ \bibnamefont
  {Bienias}}, \bibinfo {author} {\bibfnamefont {Ron}\ \bibnamefont
  {Belyansky}}, \bibinfo {author} {\bibfnamefont {Alicia~J.}\ \bibnamefont
  {Koll\'ar}}, \ and\ \bibinfo {author} {\bibfnamefont {Alexey~V.}\
  \bibnamefont {Gorshkov}},\ }\bibfield  {title} {\enquote {\bibinfo {title}
  {Quantum simulation of hyperbolic space with circuit quantum electrodynamics:
  {F}rom graphs to geometry},}\ }\href {\doibase 10.1103/PhysRevA.102.032208}
  {\bibfield  {journal} {\bibinfo  {journal} {Phys. Rev. A}\ }\textbf {\bibinfo
  {volume} {102}},\ \bibinfo {pages} {032208} (\bibinfo {year}
  {2020})}\BibitemShut {NoStop}%
\bibitem [{\citenamefont {Maciejko}\ and\ \citenamefont
  {Rayan}(2022)}]{Maciejko:2022}%
  \BibitemOpen
  \bibfield  {author} {\bibinfo {author} {\bibfnamefont {Joseph}\ \bibnamefont
  {Maciejko}}\ and\ \bibinfo {author} {\bibfnamefont {Steven}\ \bibnamefont
  {Rayan}},\ }\bibfield  {title} {\enquote {\bibinfo {title} {Automorphic
  {B}loch theorems for hyperbolic lattices},}\ }\href {\doibase
  10.1073/pnas.2116869119} {\bibfield  {journal} {\bibinfo  {journal} {Proc.
  Natl. Acad. Sci. U.S.A.}\ }\textbf {\bibinfo {volume} {119}},\ \bibinfo
  {pages} {e2116869119} (\bibinfo {year} {2022})}\BibitemShut {NoStop}%
\bibitem [{\citenamefont {Zhu}\ \emph {et~al.}(2021)\citenamefont {Zhu},
  \citenamefont {Guo}, \citenamefont {Breuckmann}, \citenamefont {Guo},\ and\
  \citenamefont {Feng}}]{Zhu:2021}%
  \BibitemOpen
  \bibfield  {author} {\bibinfo {author} {\bibfnamefont {Xingchuan}\
  \bibnamefont {Zhu}}, \bibinfo {author} {\bibfnamefont {Jiaojiao}\
  \bibnamefont {Guo}}, \bibinfo {author} {\bibfnamefont {Nikolas~P.}\
  \bibnamefont {Breuckmann}}, \bibinfo {author} {\bibfnamefont {Huaiming}\
  \bibnamefont {Guo}}, \ and\ \bibinfo {author} {\bibfnamefont {Shiping}\
  \bibnamefont {Feng}},\ }\bibfield  {title} {\enquote {\bibinfo {title}
  {Quantum phase transitions of interacting bosons on hyperbolic lattices},}\
  }\href {\doibase 10.1088/1361-648X/ac0a1a} {\bibfield  {journal} {\bibinfo
  {journal} {J. Phys. Condens. Matter}\ }\textbf {\bibinfo {volume} {33}},\
  \bibinfo {pages} {335602} (\bibinfo {year} {2021})}\BibitemShut {NoStop}%
\bibitem [{\citenamefont {Breuckmann}\ and\ \citenamefont
  {Terhal}(2016)}]{Breuckmann:2016}%
  \BibitemOpen
  \bibfield  {author} {\bibinfo {author} {\bibfnamefont {Nikolas~P.}\
  \bibnamefont {Breuckmann}}\ and\ \bibinfo {author} {\bibfnamefont
  {Barbara~M.}\ \bibnamefont {Terhal}},\ }\bibfield  {title} {\enquote
  {\bibinfo {title} {Constructions and {N}oise {T}hreshold of {H}yperbolic
  {S}urface {C}odes},}\ }\href {\doibase 10.1109/TIT.2016.2555700} {\bibfield
  {journal} {\bibinfo  {journal} {IEEE Trans. Inf. Theory}\ }\textbf {\bibinfo
  {volume} {62}},\ \bibinfo {pages} {3731--3744} (\bibinfo {year}
  {2016})}\BibitemShut {NoStop}%
\bibitem [{\citenamefont {Boettcher}\ \emph {et~al.}(2022)\citenamefont
  {Boettcher}, \citenamefont {Gorshkov}, \citenamefont {Koll\'ar},
  \citenamefont {Maciejko}, \citenamefont {Rayan},\ and\ \citenamefont
  {Thomale}}]{Boettcher:2021}%
  \BibitemOpen
  \bibfield  {author} {\bibinfo {author} {\bibfnamefont {Igor}\ \bibnamefont
  {Boettcher}}, \bibinfo {author} {\bibfnamefont {Alexey~V.}\ \bibnamefont
  {Gorshkov}}, \bibinfo {author} {\bibfnamefont {Alicia~J.}\ \bibnamefont
  {Koll\'ar}}, \bibinfo {author} {\bibfnamefont {Joseph}\ \bibnamefont
  {Maciejko}}, \bibinfo {author} {\bibfnamefont {Steven}\ \bibnamefont
  {Rayan}}, \ and\ \bibinfo {author} {\bibfnamefont {Ronny}\ \bibnamefont
  {Thomale}},\ }\bibfield  {title} {\enquote {\bibinfo {title} {Crystallography
  of hyperbolic lattices},}\ }\href {\doibase 10.1103/PhysRevB.105.125118}
  {\bibfield  {journal} {\bibinfo  {journal} {Phys. Rev. B}\ }\textbf {\bibinfo
  {volume} {105}},\ \bibinfo {pages} {125118} (\bibinfo {year}
  {2022})}\BibitemShut {NoStop}%
\bibitem [{\citenamefont {Bienias}\ \emph {et~al.}(2022)\citenamefont
  {Bienias}, \citenamefont {Boettcher}, \citenamefont {Belyansky},
  \citenamefont {Koll\'ar},\ and\ \citenamefont {Gorshkov}}]{Bienias:2021}%
  \BibitemOpen
  \bibfield  {author} {\bibinfo {author} {\bibfnamefont {Przemyslaw}\
  \bibnamefont {Bienias}}, \bibinfo {author} {\bibfnamefont {Igor}\
  \bibnamefont {Boettcher}}, \bibinfo {author} {\bibfnamefont {Ron}\
  \bibnamefont {Belyansky}}, \bibinfo {author} {\bibfnamefont {Alicia~J.}\
  \bibnamefont {Koll\'ar}}, \ and\ \bibinfo {author} {\bibfnamefont
  {Alexey~V.}\ \bibnamefont {Gorshkov}},\ }\bibfield  {title} {\enquote
  {\bibinfo {title} {Circuit quantum electrodynamics in hyperbolic space: From
  photon bound states to frustrated spin models},}\ }\href {\doibase
  10.1103/PhysRevLett.128.013601} {\bibfield  {journal} {\bibinfo  {journal}
  {Phys. Rev. Lett.}\ }\textbf {\bibinfo {volume} {128}},\ \bibinfo {pages}
  {013601} (\bibinfo {year} {2022})}\BibitemShut {NoStop}%
\bibitem [{\citenamefont {Attar}\ and\ \citenamefont
  {Boettcher}(2022)}]{Attar2022}%
  \BibitemOpen
  \bibfield  {author} {\bibinfo {author} {\bibfnamefont {Adil}\ \bibnamefont
  {Attar}}\ and\ \bibinfo {author} {\bibfnamefont {Igor}\ \bibnamefont
  {Boettcher}},\ }\bibfield  {title} {\enquote {\bibinfo {title} {Selberg trace
  formula in hyperbolic band theory},}\ }\href {\doibase
  10.1103/PhysRevE.106.034114} {\bibfield  {journal} {\bibinfo  {journal}
  {Phys. Rev. E}\ }\textbf {\bibinfo {volume} {106}},\ \bibinfo {pages}
  {034114} (\bibinfo {year} {2022})}\BibitemShut {NoStop}%
\bibitem [{\citenamefont {Ruzzene}\ \emph {et~al.}(2021)\citenamefont
  {Ruzzene}, \citenamefont {Prodan},\ and\ \citenamefont
  {Prodan}}]{Ruzzene:2021}%
  \BibitemOpen
  \bibfield  {author} {\bibinfo {author} {\bibfnamefont {Massimo}\ \bibnamefont
  {Ruzzene}}, \bibinfo {author} {\bibfnamefont {Emil}\ \bibnamefont {Prodan}},
  \ and\ \bibinfo {author} {\bibfnamefont {Camelia}\ \bibnamefont {Prodan}},\
  }\bibfield  {title} {\enquote {\bibinfo {title} {Dynamics of elastic
  hyperbolic lattices},}\ }\href {\doibase 10.1016/j.eml.2021.101491}
  {\bibfield  {journal} {\bibinfo  {journal} {Extreme Mech. Lett.}\ }\textbf
  {\bibinfo {volume} {49}},\ \bibinfo {pages} {101491} (\bibinfo {year}
  {2021})}\BibitemShut {NoStop}%
\bibitem [{\citenamefont {Koll\'{a}r}\ \emph {et~al.}(2019)\citenamefont
  {Koll\'{a}r}, \citenamefont {Fitzpatrick}, \citenamefont {Sarnak},\ and\
  \citenamefont {A.}}]{Kollar:2019b}%
  \BibitemOpen
  \bibfield  {author} {\bibinfo {author} {\bibfnamefont {Alicia~J.}\
  \bibnamefont {Koll\'{a}r}}, \bibinfo {author} {\bibfnamefont {Mattias}\
  \bibnamefont {Fitzpatrick}}, \bibinfo {author} {\bibfnamefont {Peter~Sarnak}\
  \bibnamefont {Sarnak}}, \ and\ \bibinfo {author} {\bibfnamefont
  {Houck.~Andrew}\ \bibnamefont {A.}},\ }\bibfield  {title} {\enquote {\bibinfo
  {title} {{L}ine-{G}raph {L}attices: {E}uclidean and {N}on-{E}uclidean {F}lat
  {B}ands, and {I}mplementations in {C}ircuit {Q}uantum {E}lectrodynamics},}\
  }\href {\doibase 10.1007/s00220-019-03645-8} {\bibfield  {journal} {\bibinfo
  {journal} {Commun. Math. Phys.}\ }\textbf {\bibinfo {volume} {376}},\
  \bibinfo {pages} {1909--1956} (\bibinfo {year} {2019})}\BibitemShut {NoStop}%
\bibitem [{\citenamefont {Saa}\ \emph {et~al.}(2021)\citenamefont {Saa},
  \citenamefont {Miranda},\ and\ \citenamefont {Rouxinol}}]{Saa:2021}%
  \BibitemOpen
  \bibfield  {author} {\bibinfo {author} {\bibfnamefont {Alberto}\ \bibnamefont
  {Saa}}, \bibinfo {author} {\bibfnamefont {Eduardo}\ \bibnamefont {Miranda}},
  \ and\ \bibinfo {author} {\bibfnamefont {Francisco}\ \bibnamefont
  {Rouxinol}},\ }\href@noop {} {\enquote {\bibinfo {title} {{Higher-dimensional
  Euclidean and non-Euclidean structures in planar circuit quantum
  electrodynamics}},}\ } (\bibinfo {year} {2021}),\ \Eprint
  {http://arxiv.org/abs/2108.08854} {arXiv:2108.08854 [quant-ph]} \BibitemShut
  {NoStop}%
\bibitem [{\citenamefont {Bzdu\v{s}ek}\ and\ \citenamefont
  {Maciejko}(2022)}]{Bzdusek:2022}%
  \BibitemOpen
  \bibfield  {author} {\bibinfo {author} {\bibfnamefont {Tom\'a\v{s}}\
  \bibnamefont {Bzdu\v{s}ek}}\ and\ \bibinfo {author} {\bibfnamefont {Joseph}\
  \bibnamefont {Maciejko}},\ }\bibfield  {title} {\enquote {\bibinfo {title}
  {Flat bands and band-touching from real-space topology in hyperbolic
  lattices},}\ }\href {\doibase 10.1103/PhysRevB.106.155146} {\bibfield
  {journal} {\bibinfo  {journal} {Phys. Rev. B}\ }\textbf {\bibinfo {volume}
  {106}},\ \bibinfo {pages} {155146} (\bibinfo {year} {2022})}\BibitemShut
  {NoStop}%
\bibitem [{\citenamefont {Mosseri}\ \emph {et~al.}(2022)\citenamefont
  {Mosseri}, \citenamefont {Vogeler},\ and\ \citenamefont
  {Vidal}}]{Mosseri:2022}%
  \BibitemOpen
  \bibfield  {author} {\bibinfo {author} {\bibfnamefont {R\'emy}\ \bibnamefont
  {Mosseri}}, \bibinfo {author} {\bibfnamefont {Roger}\ \bibnamefont
  {Vogeler}}, \ and\ \bibinfo {author} {\bibfnamefont {Julien}\ \bibnamefont
  {Vidal}},\ }\bibfield  {title} {\enquote {\bibinfo {title} {{Aharonov-Bohm
  cages, flat bands, and gap labeling in hyperbolic tilings}},}\ }\href
  {\doibase 10.1103/PhysRevB.106.155120} {\bibfield  {journal} {\bibinfo
  {journal} {Phys. Rev. B}\ }\textbf {\bibinfo {volume} {106}},\ \bibinfo
  {pages} {155120} (\bibinfo {year} {2022})}\BibitemShut {NoStop}%
\bibitem [{\citenamefont {Zhang}\ \emph {et~al.}(2022)\citenamefont {Zhang},
  \citenamefont {Yuan}, \citenamefont {Sun}, \citenamefont {Sun},\ and\
  \citenamefont {Zhang}}]{Zhang:2022}%
  \BibitemOpen
  \bibfield  {author} {\bibinfo {author} {\bibfnamefont {Weixuan}\ \bibnamefont
  {Zhang}}, \bibinfo {author} {\bibfnamefont {Hao}\ \bibnamefont {Yuan}},
  \bibinfo {author} {\bibfnamefont {Na}~\bibnamefont {Sun}}, \bibinfo {author}
  {\bibfnamefont {Houjun}\ \bibnamefont {Sun}}, \ and\ \bibinfo {author}
  {\bibfnamefont {Xiangdong}\ \bibnamefont {Zhang}},\ }\bibfield  {title}
  {\enquote {\bibinfo {title} {Observation of novel topological states in
  hyperbolic lattices},}\ }\href {\doibase 10.1038/s41467-022-30631-x}
  {\bibfield  {journal} {\bibinfo  {journal} {Nat. Commun}\ }\textbf {\bibinfo
  {volume} {13}},\ \bibinfo {pages} {2937} (\bibinfo {year}
  {2022})}\BibitemShut {NoStop}%
\bibitem [{\citenamefont {Liu}\ \emph {et~al.}(2022)\citenamefont {Liu},
  \citenamefont {Hua}, \citenamefont {Peng},\ and\ \citenamefont
  {Zhou}}]{Liu:2022}%
  \BibitemOpen
  \bibfield  {author} {\bibinfo {author} {\bibfnamefont {Zheng-Rong}\
  \bibnamefont {Liu}}, \bibinfo {author} {\bibfnamefont {Chun-Bo}\ \bibnamefont
  {Hua}}, \bibinfo {author} {\bibfnamefont {Tan}\ \bibnamefont {Peng}}, \ and\
  \bibinfo {author} {\bibfnamefont {Bin}\ \bibnamefont {Zhou}},\ }\bibfield
  {title} {\enquote {\bibinfo {title} {Chern insulator in a hyperbolic
  lattice},}\ }\href {\doibase 10.1103/PhysRevB.105.245301} {\bibfield
  {journal} {\bibinfo  {journal} {Phys. Rev. B}\ }\textbf {\bibinfo {volume}
  {105}},\ \bibinfo {pages} {245301} (\bibinfo {year} {2022})}\BibitemShut
  {NoStop}%
\bibitem [{\citenamefont {Katok}(1992)}]{Katok:1992}%
  \BibitemOpen
  \bibfield  {author} {\bibinfo {author} {\bibfnamefont {Svetlana}\
  \bibnamefont {Katok}},\ }\href@noop {} {\emph {\bibinfo {title} {Fuchsian
  groups}}}\ (\bibinfo  {publisher} {University of Chicago press},\ \bibinfo
  {address} {Chicago},\ \bibinfo {year} {1992})\BibitemShut {NoStop}%
\bibitem [{\citenamefont {Kitaev}(2009)}]{Kitaev:2009}%
  \BibitemOpen
  \bibfield  {author} {\bibinfo {author} {\bibfnamefont {Alexei}\ \bibnamefont
  {Kitaev}},\ }\bibfield  {title} {\enquote {\bibinfo {title} {Periodic table
  for topological insulators and superconductors},}\ }\href {\doibase
  10.1063/1.3149495} {\bibfield  {journal} {\bibinfo  {journal} {AIP Conf.
  Proc.}\ }\textbf {\bibinfo {volume} {1134}},\ \bibinfo {pages} {22} (\bibinfo
  {year} {2009})}\BibitemShut {NoStop}%
\bibitem [{\citenamefont {Ryu}\ \emph {et~al.}(2010)\citenamefont {Ryu},
  \citenamefont {Schnyder}, \citenamefont {Furusaki},\ and\ \citenamefont
  {Ludwig}}]{Ryu:2010}%
  \BibitemOpen
  \bibfield  {author} {\bibinfo {author} {\bibfnamefont {Shinsei}\ \bibnamefont
  {Ryu}}, \bibinfo {author} {\bibfnamefont {Andreas~P.}\ \bibnamefont
  {Schnyder}}, \bibinfo {author} {\bibfnamefont {Akira}\ \bibnamefont
  {Furusaki}}, \ and\ \bibinfo {author} {\bibfnamefont {Andreas W.~W.}\
  \bibnamefont {Ludwig}},\ }\bibfield  {title} {\enquote {\bibinfo {title}
  {Topological insulators and superconductors: tenfold way and dimensional
  hierarchy},}\ }\href {\doibase 10.1088/1367-2630/12/6/065010} {\bibfield
  {journal} {\bibinfo  {journal} {New J. Phys.}\ }\textbf {\bibinfo {volume}
  {12}},\ \bibinfo {pages} {065010} (\bibinfo {year} {2010})}\BibitemShut
  {NoStop}%
\bibitem [{\citenamefont {Haldane}(1988)}]{Haldane:1988}%
  \BibitemOpen
  \bibfield  {author} {\bibinfo {author} {\bibfnamefont {F.~D.~M.}\
  \bibnamefont {Haldane}},\ }\bibfield  {title} {\enquote {\bibinfo {title}
  {{Model for a {Q}uantum {H}all {E}ffect without {L}andau {L}evels:
  {C}ondensed-{M}atter {R}ealization of the ``Parity Anomaly''}},}\ }\href
  {\doibase 10.1103/PhysRevLett.61.2015} {\bibfield  {journal} {\bibinfo
  {journal} {Phys. Rev. Lett.}\ }\textbf {\bibinfo {volume} {61}},\ \bibinfo
  {pages} {2015--2018} (\bibinfo {year} {1988})}\BibitemShut {NoStop}%
\bibitem [{\citenamefont {Kane}\ and\ \citenamefont {Mele}(2005)}]{Kane:2005}%
  \BibitemOpen
  \bibfield  {author} {\bibinfo {author} {\bibfnamefont {C.~L.}\ \bibnamefont
  {Kane}}\ and\ \bibinfo {author} {\bibfnamefont {E.~J.}\ \bibnamefont
  {Mele}},\ }\bibfield  {title} {\enquote {\bibinfo {title} {{${Z}_{2}$
  Topological Order and the Quantum Spin Hall Effect}},}\ }\href {\doibase
  10.1103/PhysRevLett.95.146802} {\bibfield  {journal} {\bibinfo  {journal}
  {Phys. Rev. Lett.}\ }\textbf {\bibinfo {volume} {95}},\ \bibinfo {pages}
  {146802} (\bibinfo {year} {2005})}\BibitemShut {NoStop}%
\bibitem [{\citenamefont {Khanikaev}\ \emph {et~al.}(2013)\citenamefont
  {Khanikaev}, \citenamefont {Mousavi}, \citenamefont {Tse}, \citenamefont
  {Kargarian}, \citenamefont {MacDonald},\ and\ \citenamefont
  {Shvets}}]{Khanikaev:2013}%
  \BibitemOpen
  \bibfield  {author} {\bibinfo {author} {\bibfnamefont {Alexander~B}\
  \bibnamefont {Khanikaev}}, \bibinfo {author} {\bibfnamefont {S~Hossein}\
  \bibnamefont {Mousavi}}, \bibinfo {author} {\bibfnamefont {Wang-Kong}\
  \bibnamefont {Tse}}, \bibinfo {author} {\bibfnamefont {Mehdi}\ \bibnamefont
  {Kargarian}}, \bibinfo {author} {\bibfnamefont {Allan~H}\ \bibnamefont
  {MacDonald}}, \ and\ \bibinfo {author} {\bibfnamefont {Gennady}\ \bibnamefont
  {Shvets}},\ }\bibfield  {title} {\enquote {\bibinfo {title} {Photonic
  topological insulators},}\ }\href {\doibase 10.1038/nmat3520} {\bibfield
  {journal} {\bibinfo  {journal} {Nat. Mater.}\ }\textbf {\bibinfo {volume}
  {12}},\ \bibinfo {pages} {233--239} (\bibinfo {year} {2013})}\BibitemShut
  {NoStop}%
\bibitem [{\citenamefont {Jotzu}\ \emph {et~al.}(2014)\citenamefont {Jotzu},
  \citenamefont {Messer}, \citenamefont {Desbuquois}, \citenamefont {Lebrat},
  \citenamefont {Uehlinger}, \citenamefont {Greif},\ and\ \citenamefont
  {Esslinger}}]{Jotzu:2014}%
  \BibitemOpen
  \bibfield  {author} {\bibinfo {author} {\bibfnamefont {Gregor}\ \bibnamefont
  {Jotzu}}, \bibinfo {author} {\bibfnamefont {Michael}\ \bibnamefont {Messer}},
  \bibinfo {author} {\bibfnamefont {R\'{e}mi}\ \bibnamefont {Desbuquois}},
  \bibinfo {author} {\bibfnamefont {Martin}\ \bibnamefont {Lebrat}}, \bibinfo
  {author} {\bibfnamefont {Thomas}\ \bibnamefont {Uehlinger}}, \bibinfo
  {author} {\bibfnamefont {Daniel}\ \bibnamefont {Greif}}, \ and\ \bibinfo
  {author} {\bibfnamefont {Tilman}\ \bibnamefont {Esslinger}},\ }\bibfield
  {title} {\enquote {\bibinfo {title} {Experimental realization of the
  topological {H}aldane model with ultracold fermions},}\ }\href {\doibase
  10.1038/nature13915} {\bibfield  {journal} {\bibinfo  {journal} {Nature}\
  }\textbf {\bibinfo {volume} {515}},\ \bibinfo {pages} {237--240} (\bibinfo
  {year} {2014})}\BibitemShut {NoStop}%
\bibitem [{\citenamefont {Ding}\ \emph {et~al.}(2019)\citenamefont {Ding},
  \citenamefont {Peng}, \citenamefont {Zhu}, \citenamefont {Fan}, \citenamefont
  {Yang}, \citenamefont {Liang}, \citenamefont {Zhu}, \citenamefont {Wan},\
  and\ \citenamefont {Cheng}}]{Ding:2019}%
  \BibitemOpen
  \bibfield  {author} {\bibinfo {author} {\bibfnamefont {Yujiang}\ \bibnamefont
  {Ding}}, \bibinfo {author} {\bibfnamefont {Yugui}\ \bibnamefont {Peng}},
  \bibinfo {author} {\bibfnamefont {Yifan}\ \bibnamefont {Zhu}}, \bibinfo
  {author} {\bibfnamefont {Xudong}\ \bibnamefont {Fan}}, \bibinfo {author}
  {\bibfnamefont {Jing}\ \bibnamefont {Yang}}, \bibinfo {author} {\bibfnamefont
  {Bin}\ \bibnamefont {Liang}}, \bibinfo {author} {\bibfnamefont {Xuefeng}\
  \bibnamefont {Zhu}}, \bibinfo {author} {\bibfnamefont {Xiangang}\
  \bibnamefont {Wan}}, \ and\ \bibinfo {author} {\bibfnamefont {Jianchun}\
  \bibnamefont {Cheng}},\ }\bibfield  {title} {\enquote {\bibinfo {title}
  {{Experimental Demonstration of Acoustic Chern Insulators}},}\ }\href
  {\doibase 10.1103/PhysRevLett.122.014302} {\bibfield  {journal} {\bibinfo
  {journal} {Phys. Rev. Lett.}\ }\textbf {\bibinfo {volume} {122}},\ \bibinfo
  {pages} {014302} (\bibinfo {year} {2019})}\BibitemShut {NoStop}%
\bibitem [{\citenamefont {Imhof}\ \emph {et~al.}(2018)\citenamefont {Imhof},
  \citenamefont {Berger}, \citenamefont {Bayer}, \citenamefont {Brehm},
  \citenamefont {Molenkamp}, \citenamefont {Kiessling}, \citenamefont
  {Schindler}, \citenamefont {Lee}, \citenamefont {Greiter}, \citenamefont
  {Neupert},\ and\ \citenamefont {Thomale}}]{Imhof:2018}%
  \BibitemOpen
  \bibfield  {author} {\bibinfo {author} {\bibfnamefont {Stefan}\ \bibnamefont
  {Imhof}}, \bibinfo {author} {\bibfnamefont {Christian}\ \bibnamefont
  {Berger}}, \bibinfo {author} {\bibfnamefont {Florian}\ \bibnamefont {Bayer}},
  \bibinfo {author} {\bibfnamefont {Johannes}\ \bibnamefont {Brehm}}, \bibinfo
  {author} {\bibfnamefont {Laurens~W}\ \bibnamefont {Molenkamp}}, \bibinfo
  {author} {\bibfnamefont {Tobias}\ \bibnamefont {Kiessling}}, \bibinfo
  {author} {\bibfnamefont {Frank}\ \bibnamefont {Schindler}}, \bibinfo {author}
  {\bibfnamefont {Ching~Hua}\ \bibnamefont {Lee}}, \bibinfo {author}
  {\bibfnamefont {Martin}\ \bibnamefont {Greiter}}, \bibinfo {author}
  {\bibfnamefont {Titus}\ \bibnamefont {Neupert}}, \ and\ \bibinfo {author}
  {\bibfnamefont {Ronny}\ \bibnamefont {Thomale}},\ }\bibfield  {title}
  {\enquote {\bibinfo {title} {Topolectrical-circuit realization of topological
  corner modes},}\ }\href {\doibase 10.1038/s41567-018-0246-1} {\bibfield
  {journal} {\bibinfo  {journal} {Nat. Phys.}\ }\textbf {\bibinfo {volume}
  {14}},\ \bibinfo {pages} {925} (\bibinfo {year} {2018})}\BibitemShut
  {NoStop}%
\bibitem [{\citenamefont {Lee}\ \emph {et~al.}(2018)\citenamefont {Lee},
  \citenamefont {Imhof}, \citenamefont {Berger}, \citenamefont {Bayer},
  \citenamefont {Brehm}, \citenamefont {Molenkamp}, \citenamefont {Kiessling},\
  and\ \citenamefont {Thomale}}]{Lee:2018}%
  \BibitemOpen
  \bibfield  {author} {\bibinfo {author} {\bibfnamefont {Ching~Hua}\
  \bibnamefont {Lee}}, \bibinfo {author} {\bibfnamefont {Stefan}\ \bibnamefont
  {Imhof}}, \bibinfo {author} {\bibfnamefont {Christian}\ \bibnamefont
  {Berger}}, \bibinfo {author} {\bibfnamefont {Florian}\ \bibnamefont {Bayer}},
  \bibinfo {author} {\bibfnamefont {Johannes}\ \bibnamefont {Brehm}}, \bibinfo
  {author} {\bibfnamefont {Laurens~W.}\ \bibnamefont {Molenkamp}}, \bibinfo
  {author} {\bibfnamefont {Tobias}\ \bibnamefont {Kiessling}}, \ and\ \bibinfo
  {author} {\bibfnamefont {Ronny}\ \bibnamefont {Thomale}},\ }\bibfield
  {title} {\enquote {\bibinfo {title} {Topolectrical {C}ircuits},}\ }\href
  {\doibase 10.1038/s42005-018-0035-2} {\bibfield  {journal} {\bibinfo
  {journal} {Commun. Phys.}\ }\textbf {\bibinfo {volume} {1}},\ \bibinfo
  {pages} {39} (\bibinfo {year} {2018})}\BibitemShut {NoStop}%
\bibitem [{\citenamefont {Hofmann}\ \emph {et~al.}(2019)\citenamefont
  {Hofmann}, \citenamefont {Helbig}, \citenamefont {Lee}, \citenamefont
  {Greiter},\ and\ \citenamefont {Thomale}}]{Hofmann:2019}%
  \BibitemOpen
  \bibfield  {author} {\bibinfo {author} {\bibfnamefont {Tobias}\ \bibnamefont
  {Hofmann}}, \bibinfo {author} {\bibfnamefont {Tobias}\ \bibnamefont
  {Helbig}}, \bibinfo {author} {\bibfnamefont {Ching~Hua}\ \bibnamefont {Lee}},
  \bibinfo {author} {\bibfnamefont {Martin}\ \bibnamefont {Greiter}}, \ and\
  \bibinfo {author} {\bibfnamefont {Ronny}\ \bibnamefont {Thomale}},\
  }\bibfield  {title} {\enquote {\bibinfo {title} {{Chiral Voltage Propagation
  and Calibration in a Topolectrical Chern Circuit}},}\ }\href {\doibase
  10.1103/PhysRevLett.122.247702} {\bibfield  {journal} {\bibinfo  {journal}
  {Phys. Rev. Lett.}\ }\textbf {\bibinfo {volume} {122}},\ \bibinfo {pages}
  {247702} (\bibinfo {year} {2019})}\BibitemShut {NoStop}%
\bibitem [{\citenamefont {Bellissard}\ \emph {et~al.}(1994)\citenamefont
  {Bellissard}, \citenamefont {van Elst},\ and\ \citenamefont
  {Schulz‐~Baldes}}]{Bellissard:1994}%
  \BibitemOpen
  \bibfield  {author} {\bibinfo {author} {\bibfnamefont {J.}~\bibnamefont
  {Bellissard}}, \bibinfo {author} {\bibfnamefont {A.}~\bibnamefont {van
  Elst}}, \ and\ \bibinfo {author} {\bibfnamefont {H.}~\bibnamefont
  {Schulz‐~Baldes}},\ }\bibfield  {title} {\enquote {\bibinfo {title} {{The
  noncommutative geometry of the quantum Hall effect}},}\ }\href {\doibase
  10.1063/1.530758} {\bibfield  {journal} {\bibinfo  {journal} {J. Math.
  Phys.}\ }\textbf {\bibinfo {volume} {35}},\ \bibinfo {pages} {5373--5451}
  (\bibinfo {year} {1994})}\BibitemShut {NoStop}%
\bibitem [{\citenamefont {Kitaev}(2006)}]{Kitaev:2006}%
  \BibitemOpen
  \bibfield  {author} {\bibinfo {author} {\bibfnamefont {Alexei}\ \bibnamefont
  {Kitaev}},\ }\bibfield  {title} {\enquote {\bibinfo {title} {Anyons in an
  exactly solved model and beyond},}\ }\href {\doibase
  https://doi.org/10.1016/j.aop.2005.10.005} {\bibfield  {journal} {\bibinfo
  {journal} {Ann. Phys.}\ }\textbf {\bibinfo {volume} {321}},\ \bibinfo {pages}
  {2--111} (\bibinfo {year} {2006})}\BibitemShut {NoStop}%
\bibitem [{\citenamefont {Bianco}\ and\ \citenamefont
  {Resta}(2011)}]{Bianco:2011}%
  \BibitemOpen
  \bibfield  {author} {\bibinfo {author} {\bibfnamefont {Raffaello}\
  \bibnamefont {Bianco}}\ and\ \bibinfo {author} {\bibfnamefont {Raffaele}\
  \bibnamefont {Resta}},\ }\bibfield  {title} {\enquote {\bibinfo {title}
  {Mapping topological order in coordinate space},}\ }\href {\doibase
  10.1103/PhysRevB.84.241106} {\bibfield  {journal} {\bibinfo  {journal} {Phys.
  Rev. B}\ }\textbf {\bibinfo {volume} {84}},\ \bibinfo {pages} {241106(R)}
  (\bibinfo {year} {2011})}\BibitemShut {NoStop}%
\bibitem [{\citenamefont {Prodan}(2011)}]{Prodan:2011}%
  \BibitemOpen
  \bibfield  {author} {\bibinfo {author} {\bibfnamefont {Emil}\ \bibnamefont
  {Prodan}},\ }\bibfield  {title} {\enquote {\bibinfo {title} {Disordered
  topological insulators: a non-commutative geometry perspective},}\ }\href
  {\doibase 10.1088/1751-8113/44/11/113001} {\bibfield  {journal} {\bibinfo
  {journal} {J. Phys. A Math. Theor.}\ }\textbf {\bibinfo {volume} {44}},\
  \bibinfo {pages} {113001} (\bibinfo {year} {2011})}\BibitemShut {NoStop}%
\bibitem [{\citenamefont {Huang}\ and\ \citenamefont {Liu}(2018)}]{Huang:2018}%
  \BibitemOpen
  \bibfield  {author} {\bibinfo {author} {\bibfnamefont {Huaqing}\ \bibnamefont
  {Huang}}\ and\ \bibinfo {author} {\bibfnamefont {Feng}\ \bibnamefont {Liu}},\
  }\bibfield  {title} {\enquote {\bibinfo {title} {{Theory of spin Bott index
  for quantum spin Hall states in nonperiodic systems}},}\ }\href {\doibase
  10.1103/PhysRevB.98.125130} {\bibfield  {journal} {\bibinfo  {journal} {Phys.
  Rev. B}\ }\textbf {\bibinfo {volume} {98}},\ \bibinfo {pages} {125130}
  (\bibinfo {year} {2018})}\BibitemShut {NoStop}%
\bibitem [{\citenamefont {Kazaryan}\ \emph {et~al.}(2019)\citenamefont
  {Kazaryan}, \citenamefont {Lando},\ and\ \citenamefont
  {Prasolov}}]{kazaryan2019}%
  \BibitemOpen
  \bibfield  {author} {\bibinfo {author} {\bibfnamefont {M.~E.}\ \bibnamefont
  {Kazaryan}}, \bibinfo {author} {\bibfnamefont {S.~K.}\ \bibnamefont {Lando}},
  \ and\ \bibinfo {author} {\bibfnamefont {V.~V.}\ \bibnamefont {Prasolov}},\
  }\href {\doibase 10.1007/978-3-030-02943-2} {\emph {\bibinfo {title}
  {{Algebraic Curves: Towards Moduli Spaces}}}},\ \bibinfo {series} {Moscow
  Lectures}, Vol.~\bibinfo {volume} {2}\ (\bibinfo  {publisher} {Springer},\
  \bibinfo {address} {Cham, Switzerland},\ \bibinfo {year} {2019})\BibitemShut
  {NoStop}%
\bibitem [{sup()}]{supp}%
  \BibitemOpen
  \href@noop {} {}\bibinfo {note} {The Supplemental Material contains
  supporting data and detailed information about the methods.}\BibitemShut
  {Stop}%
\bibitem [{\citenamefont {Cheng}\ \emph {et~al.}(2022)\citenamefont {Cheng},
  \citenamefont {Serafin}, \citenamefont {McInerney}, \citenamefont {Rocklin},
  \citenamefont {Sun},\ and\ \citenamefont {Mao}}]{Cheng2022}%
  \BibitemOpen
  \bibfield  {author} {\bibinfo {author} {\bibfnamefont {Nan}\ \bibnamefont
  {Cheng}}, \bibinfo {author} {\bibfnamefont {Francesco}\ \bibnamefont
  {Serafin}}, \bibinfo {author} {\bibfnamefont {James}\ \bibnamefont
  {McInerney}}, \bibinfo {author} {\bibfnamefont {Zeb}\ \bibnamefont
  {Rocklin}}, \bibinfo {author} {\bibfnamefont {Kai}\ \bibnamefont {Sun}}, \
  and\ \bibinfo {author} {\bibfnamefont {Xiaoming}\ \bibnamefont {Mao}},\
  }\bibfield  {title} {\enquote {\bibinfo {title} {{Band Theory and Boundary
  Modes of High-Dimensional Representations of Infinite Hyperbolic
  Lattices}},}\ }\href {\doibase 10.1103/PhysRevLett.129.088002} {\bibfield
  {journal} {\bibinfo  {journal} {Phys. Rev. Lett.}\ }\textbf {\bibinfo
  {volume} {129}},\ \bibinfo {pages} {088002} (\bibinfo {year}
  {2022})}\BibitemShut {NoStop}%
\bibitem [{\citenamefont {Chen}\ \emph {et~al.}(2022)\citenamefont {Chen},
  \citenamefont {Brand}, \citenamefont {Helbig}, \citenamefont {Hofmann},
  \citenamefont {Imhof}, \citenamefont {Fritzsche}, \citenamefont {Kießling},
  \citenamefont {Stegmaier}, \citenamefont {Upreti}, \citenamefont {Neupert},
  \citenamefont {Bzdu\v{s}ek}, \citenamefont {Greiter}, \citenamefont
  {Thomale},\ and\ \citenamefont {Boettcher}}]{Chen2022}%
  \BibitemOpen
  \bibfield  {author} {\bibinfo {author} {\bibfnamefont {Anffany}\ \bibnamefont
  {Chen}}, \bibinfo {author} {\bibfnamefont {Hauke}\ \bibnamefont {Brand}},
  \bibinfo {author} {\bibfnamefont {Tobias}\ \bibnamefont {Helbig}}, \bibinfo
  {author} {\bibfnamefont {Tobias}\ \bibnamefont {Hofmann}}, \bibinfo {author}
  {\bibfnamefont {Stefan}\ \bibnamefont {Imhof}}, \bibinfo {author}
  {\bibfnamefont {Alexander}\ \bibnamefont {Fritzsche}}, \bibinfo {author}
  {\bibfnamefont {Tobias}\ \bibnamefont {Kießling}}, \bibinfo {author}
  {\bibfnamefont {Alexander}\ \bibnamefont {Stegmaier}}, \bibinfo {author}
  {\bibfnamefont {Lavi~K.}\ \bibnamefont {Upreti}}, \bibinfo {author}
  {\bibfnamefont {Titus}\ \bibnamefont {Neupert}}, \bibinfo {author}
  {\bibfnamefont {Tom\'{a}\v{s}}\ \bibnamefont {Bzdu\v{s}ek}}, \bibinfo
  {author} {\bibfnamefont {Martin}\ \bibnamefont {Greiter}}, \bibinfo {author}
  {\bibfnamefont {Ronny}\ \bibnamefont {Thomale}}, \ and\ \bibinfo {author}
  {\bibfnamefont {Igor}\ \bibnamefont {Boettcher}},\ }\href {\doibase
  10.48550/ARXIV.2205.05106} {\enquote {\bibinfo {title} {Hyperbolic matter in
  electrical circuits with tunable complex phases},}\ } (\bibinfo {year}
  {2022}),\ \Eprint {http://arxiv.org/abs/2205.05106} {arXiv:2205.05106}
  \BibitemShut {NoStop}%
\bibitem [{\citenamefont {Altland}\ and\ \citenamefont
  {Zirnbauer}(1997)}]{Altland:1997}%
  \BibitemOpen
  \bibfield  {author} {\bibinfo {author} {\bibfnamefont {Alexander}\
  \bibnamefont {Altland}}\ and\ \bibinfo {author} {\bibfnamefont {Martin~R.}\
  \bibnamefont {Zirnbauer}},\ }\bibfield  {title} {\enquote {\bibinfo {title}
  {Nonstandard symmetry classes in mesoscopic normal-superconducting hybrid
  structures},}\ }\href {\doibase 10.1103/PhysRevB.55.1142} {\bibfield
  {journal} {\bibinfo  {journal} {Phys. Rev. B}\ }\textbf {\bibinfo {volume}
  {55}},\ \bibinfo {pages} {1142--1161} (\bibinfo {year} {1997})}\BibitemShut
  {NoStop}%
\bibitem [{\citenamefont {Nakahara}(1990)}]{Nakahara:1990}%
  \BibitemOpen
  \bibfield  {author} {\bibinfo {author} {\bibfnamefont {Mikio}\ \bibnamefont
  {Nakahara}},\ }\href@noop {} {\emph {\bibinfo {title} {{Geometry, topology
  and physics}}}},\ Graduate student series in physics\ (\bibinfo  {publisher}
  {Hilger},\ \bibinfo {address} {Bristol},\ \bibinfo {year} {1990})\BibitemShut
  {NoStop}%
\bibitem [{\citenamefont {Urwyler}(2021)}]{Urwyler:2021}%
  \BibitemOpen
  \bibfield  {author} {\bibinfo {author} {\bibfnamefont {David~M.}\
  \bibnamefont {Urwyler}},\ }\emph {\bibinfo {title} {{Hyperbolic Topological
  Insulator}}},\ \href {\doibase 10.13140/RG.2.2.34715.34081} {Master's
  thesis},\ \bibinfo  {school} {University of Z\"{u}rich}, \bibinfo {address}
  {Switzerland} (\bibinfo {year} {2021})\BibitemShut {NoStop}%
\bibitem [{\citenamefont {Urwyler}\ \emph
  {et~al.}(2022{\natexlab{a}})\citenamefont {Urwyler}, \citenamefont
  {Lenggenhager}, \citenamefont {Neupert},\ and\ \citenamefont
  {Bzdu\v{s}ek}}]{Urwyler:2022b}%
  \BibitemOpen
  \bibfield  {author} {\bibinfo {author} {\bibfnamefont {David~M.}\
  \bibnamefont {Urwyler}}, \bibinfo {author} {\bibfnamefont {Patrick~M.}\
  \bibnamefont {Lenggenhager}}, \bibinfo {author} {\bibfnamefont {Titus}\
  \bibnamefont {Neupert}}, \ and\ \bibinfo {author} {\bibfnamefont
  {Tom\'{a}\v{s}}\ \bibnamefont {Bzdu\v{s}ek}},\ }\href
  {https://meetings.aps.org/Meeting/MAR22/Session/N66.7} {\enquote {\bibinfo
  {title} {{Topological hyperbolic band insulators}},}\ } (\bibinfo {year}
  {2022}{\natexlab{a}}),\ \bibinfo {note} {{APS March Meeting 2022, Session
  N66}}\BibitemShut {NoStop}%
\bibitem [{\citenamefont {Soluyanov}\ and\ \citenamefont
  {Vanderbilt}(2011)}]{Soluyanov:2011}%
  \BibitemOpen
  \bibfield  {author} {\bibinfo {author} {\bibfnamefont {Alexey~A.}\
  \bibnamefont {Soluyanov}}\ and\ \bibinfo {author} {\bibfnamefont {David}\
  \bibnamefont {Vanderbilt}},\ }\bibfield  {title} {\enquote {\bibinfo {title}
  {Computing topological invariants without inversion symmetry},}\ }\href
  {\doibase 10.1103/PhysRevB.83.235401} {\bibfield  {journal} {\bibinfo
  {journal} {Phys. Rev. B}\ }\textbf {\bibinfo {volume} {83}},\ \bibinfo
  {pages} {235401} (\bibinfo {year} {2011})}\BibitemShut {NoStop}%
\bibitem [{\citenamefont {Gresch}\ \emph {et~al.}(2017)\citenamefont {Gresch},
  \citenamefont {Aut\`es}, \citenamefont {Yazyev}, \citenamefont {Troyer},
  \citenamefont {Vanderbilt}, \citenamefont {Bernevig},\ and\ \citenamefont
  {Soluyanov}}]{Gresch:2017}%
  \BibitemOpen
  \bibfield  {author} {\bibinfo {author} {\bibfnamefont {Dominik}\ \bibnamefont
  {Gresch}}, \bibinfo {author} {\bibfnamefont {Gabriel}\ \bibnamefont
  {Aut\`es}}, \bibinfo {author} {\bibfnamefont {Oleg~V.}\ \bibnamefont
  {Yazyev}}, \bibinfo {author} {\bibfnamefont {Matthias}\ \bibnamefont
  {Troyer}}, \bibinfo {author} {\bibfnamefont {David}\ \bibnamefont
  {Vanderbilt}}, \bibinfo {author} {\bibfnamefont {B.~Andrei}\ \bibnamefont
  {Bernevig}}, \ and\ \bibinfo {author} {\bibfnamefont {Alexey~A.}\
  \bibnamefont {Soluyanov}},\ }\bibfield  {title} {\enquote {\bibinfo {title}
  {{Z2Pack: Numerical implementation of hybrid Wannier centers for identifying
  topological materials}},}\ }\href {\doibase 10.1103/PhysRevB.95.075146}
  {\bibfield  {journal} {\bibinfo  {journal} {Phys. Rev. B}\ }\textbf {\bibinfo
  {volume} {95}},\ \bibinfo {pages} {075146} (\bibinfo {year}
  {2017})}\BibitemShut {NoStop}%
\bibitem [{\citenamefont {Fu}\ \emph {et~al.}(2007)\citenamefont {Fu},
  \citenamefont {Kane},\ and\ \citenamefont {Mele}}]{Fu:2007}%
  \BibitemOpen
  \bibfield  {author} {\bibinfo {author} {\bibfnamefont {Liang}\ \bibnamefont
  {Fu}}, \bibinfo {author} {\bibfnamefont {C.~L.}\ \bibnamefont {Kane}}, \ and\
  \bibinfo {author} {\bibfnamefont {E.~J.}\ \bibnamefont {Mele}},\ }\bibfield
  {title} {\enquote {\bibinfo {title} {{Topological Insulators in Three
  Dimensions}},}\ }\href {\doibase 10.1103/PhysRevLett.98.106803} {\bibfield
  {journal} {\bibinfo  {journal} {Phys. Rev. Lett.}\ }\textbf {\bibinfo
  {volume} {98}},\ \bibinfo {pages} {106803} (\bibinfo {year}
  {2007})}\BibitemShut {NoStop}%
\bibitem [{\citenamefont {Zhang}\ and\ \citenamefont {Hu}(2001)}]{Zhang:2001}%
  \BibitemOpen
  \bibfield  {author} {\bibinfo {author} {\bibfnamefont {Shou-Cheng}\
  \bibnamefont {Zhang}}\ and\ \bibinfo {author} {\bibfnamefont {Jiangping}\
  \bibnamefont {Hu}},\ }\bibfield  {title} {\enquote {\bibinfo {title} {A
  {F}our-{D}imensional {G}eneralization of the {Q}uantum {H}all {E}ffect},}\
  }\href {\doibase 10.1126/science.294.5543.823} {\bibfield  {journal}
  {\bibinfo  {journal} {Science}\ }\textbf {\bibinfo {volume} {294}},\ \bibinfo
  {pages} {823--828} (\bibinfo {year} {2001})}\BibitemShut {NoStop}%
\bibitem [{\citenamefont {Borgnia}\ \emph {et~al.}(2020)\citenamefont
  {Borgnia}, \citenamefont {Kruchkov},\ and\ \citenamefont
  {Slager}}]{Borgnia:2020}%
  \BibitemOpen
  \bibfield  {author} {\bibinfo {author} {\bibfnamefont {Dan~S.}\ \bibnamefont
  {Borgnia}}, \bibinfo {author} {\bibfnamefont {Alex~Jura}\ \bibnamefont
  {Kruchkov}}, \ and\ \bibinfo {author} {\bibfnamefont {Robert-Jan}\
  \bibnamefont {Slager}},\ }\bibfield  {title} {\enquote {\bibinfo {title}
  {Non-hermitian boundary modes and topology},}\ }\href {\doibase
  10.1103/PhysRevLett.124.056802} {\bibfield  {journal} {\bibinfo  {journal}
  {Phys. Rev. Lett.}\ }\textbf {\bibinfo {volume} {124}},\ \bibinfo {pages}
  {056802} (\bibinfo {year} {2020})}\BibitemShut {NoStop}%
\bibitem [{\citenamefont {Okuma}\ \emph {et~al.}(2020)\citenamefont {Okuma},
  \citenamefont {Kawabata}, \citenamefont {Shiozaki},\ and\ \citenamefont
  {Sato}}]{Okuma:2020}%
  \BibitemOpen
  \bibfield  {author} {\bibinfo {author} {\bibfnamefont {Nobuyuki}\
  \bibnamefont {Okuma}}, \bibinfo {author} {\bibfnamefont {Kohei}\ \bibnamefont
  {Kawabata}}, \bibinfo {author} {\bibfnamefont {Ken}\ \bibnamefont
  {Shiozaki}}, \ and\ \bibinfo {author} {\bibfnamefont {Masatoshi}\
  \bibnamefont {Sato}},\ }\bibfield  {title} {\enquote {\bibinfo {title}
  {{Topological Origin of Non-Hermitian Skin Effects}},}\ }\href {\doibase
  10.1103/PhysRevLett.124.086801} {\bibfield  {journal} {\bibinfo  {journal}
  {Phys. Rev. Lett.}\ }\textbf {\bibinfo {volume} {124}},\ \bibinfo {pages}
  {086801} (\bibinfo {year} {2020})}\BibitemShut {NoStop}%
\bibitem [{\citenamefont {Weidemann}\ \emph {et~al.}(2020)\citenamefont
  {Weidemann}, \citenamefont {Kremer}, \citenamefont {Helbig}, \citenamefont
  {Hofmann}, \citenamefont {Stegmaier}, \citenamefont {Greiter}, \citenamefont
  {Thomale},\ and\ \citenamefont {Szameit}}]{Weidemann:2020}%
  \BibitemOpen
  \bibfield  {author} {\bibinfo {author} {\bibfnamefont {Sebastian}\
  \bibnamefont {Weidemann}}, \bibinfo {author} {\bibfnamefont {Mark}\
  \bibnamefont {Kremer}}, \bibinfo {author} {\bibfnamefont {Tobias}\
  \bibnamefont {Helbig}}, \bibinfo {author} {\bibfnamefont {Tobias}\
  \bibnamefont {Hofmann}}, \bibinfo {author} {\bibfnamefont {Alexander}\
  \bibnamefont {Stegmaier}}, \bibinfo {author} {\bibfnamefont {Martin}\
  \bibnamefont {Greiter}}, \bibinfo {author} {\bibfnamefont {Ronny}\
  \bibnamefont {Thomale}}, \ and\ \bibinfo {author} {\bibfnamefont {Alexander}\
  \bibnamefont {Szameit}},\ }\bibfield  {title} {\enquote {\bibinfo {title}
  {Topological funneling of light},}\ }\href {\doibase 10.1126/science.aaz8727}
  {\bibfield  {journal} {\bibinfo  {journal} {Science}\ }\textbf {\bibinfo
  {volume} {368}},\ \bibinfo {pages} {311--314} (\bibinfo {year}
  {2020})}\BibitemShut {NoStop}%
\bibitem [{\citenamefont {Hofmann}\ \emph {et~al.}(2020)\citenamefont
  {Hofmann}, \citenamefont {Helbig}, \citenamefont {Schindler}, \citenamefont
  {Salgo}, \citenamefont {Brzezi\'{n}ska}, \citenamefont {Greiter},
  \citenamefont {Kiessling}, \citenamefont {Wolf}, \citenamefont {Vollhardt},
  \citenamefont {Kaba\v{s}i}, \citenamefont {Lee}, \citenamefont
  {Bilu\v{s}i\'{c}}, \citenamefont {Thomale},\ and\ \citenamefont
  {Neupert}}]{Hofmann:2020}%
  \BibitemOpen
  \bibfield  {author} {\bibinfo {author} {\bibfnamefont {Tobias}\ \bibnamefont
  {Hofmann}}, \bibinfo {author} {\bibfnamefont {Tobias}\ \bibnamefont
  {Helbig}}, \bibinfo {author} {\bibfnamefont {Frank}\ \bibnamefont
  {Schindler}}, \bibinfo {author} {\bibfnamefont {Nora}\ \bibnamefont {Salgo}},
  \bibinfo {author} {\bibfnamefont {Marta}\ \bibnamefont {Brzezi\'{n}ska}},
  \bibinfo {author} {\bibfnamefont {Martin}\ \bibnamefont {Greiter}}, \bibinfo
  {author} {\bibfnamefont {Tobias}\ \bibnamefont {Kiessling}}, \bibinfo
  {author} {\bibfnamefont {David}\ \bibnamefont {Wolf}}, \bibinfo {author}
  {\bibfnamefont {Achim}\ \bibnamefont {Vollhardt}}, \bibinfo {author}
  {\bibfnamefont {Anton}\ \bibnamefont {Kaba\v{s}i}}, \bibinfo {author}
  {\bibfnamefont {Ching~Hua}\ \bibnamefont {Lee}}, \bibinfo {author}
  {\bibfnamefont {Ante}\ \bibnamefont {Bilu\v{s}i\'{c}}}, \bibinfo {author}
  {\bibfnamefont {Ronny}\ \bibnamefont {Thomale}}, \ and\ \bibinfo {author}
  {\bibfnamefont {Titus}\ \bibnamefont {Neupert}},\ }\bibfield  {title}
  {\enquote {\bibinfo {title} {Reciprocal skin effect and its realization in a
  topolectrical circuit},}\ }\href {\doibase 10.1103/PhysRevResearch.2.023265}
  {\bibfield  {journal} {\bibinfo  {journal} {Phys. Rev. Research}\ }\textbf
  {\bibinfo {volume} {2}},\ \bibinfo {pages} {023265} (\bibinfo {year}
  {2020})}\BibitemShut {NoStop}%
\bibitem [{\citenamefont {Helbig}\ \emph {et~al.}(2020)\citenamefont {Helbig},
  \citenamefont {Hofmann}, \citenamefont {Imhof}, \citenamefont {Abdelghany},
  \citenamefont {Kiessling}, \citenamefont {Molenkamp}, \citenamefont {Lee},
  \citenamefont {Szameit}, \citenamefont {Greiter},\ and\ \citenamefont
  {Thomale}}]{Helbig:2020}%
  \BibitemOpen
  \bibfield  {author} {\bibinfo {author} {\bibfnamefont {T.}~\bibnamefont
  {Helbig}}, \bibinfo {author} {\bibfnamefont {T.}~\bibnamefont {Hofmann}},
  \bibinfo {author} {\bibfnamefont {S.}~\bibnamefont {Imhof}}, \bibinfo
  {author} {\bibfnamefont {M.}~\bibnamefont {Abdelghany}}, \bibinfo {author}
  {\bibfnamefont {T.}~\bibnamefont {Kiessling}}, \bibinfo {author}
  {\bibfnamefont {L.~W.}\ \bibnamefont {Molenkamp}}, \bibinfo {author}
  {\bibfnamefont {C.~H.}\ \bibnamefont {Lee}}, \bibinfo {author} {\bibfnamefont
  {A.}~\bibnamefont {Szameit}}, \bibinfo {author} {\bibfnamefont
  {M.}~\bibnamefont {Greiter}}, \ and\ \bibinfo {author} {\bibfnamefont
  {R.}~\bibnamefont {Thomale}},\ }\bibfield  {title} {\enquote {\bibinfo
  {title} {Generalized bulk--boundary correspondence in non-hermitian
  topolectrical circuits},}\ }\href {\doibase 10.1038/s41567-020-0922-9}
  {\bibfield  {journal} {\bibinfo  {journal} {Nat. Phys.}\ }\textbf {\bibinfo
  {volume} {16}},\ \bibinfo {pages} {747--750} (\bibinfo {year}
  {2020})}\BibitemShut {NoStop}%
\bibitem [{\citenamefont {Lv}\ \emph {et~al.}(2022)\citenamefont {Lv},
  \citenamefont {Zhang}, \citenamefont {Zhai},\ and\ \citenamefont
  {Zhou}}]{Lv:2021}%
  \BibitemOpen
  \bibfield  {author} {\bibinfo {author} {\bibfnamefont {Chenwei}\ \bibnamefont
  {Lv}}, \bibinfo {author} {\bibfnamefont {Ren}\ \bibnamefont {Zhang}},
  \bibinfo {author} {\bibfnamefont {Zhengzheng}\ \bibnamefont {Zhai}}, \ and\
  \bibinfo {author} {\bibfnamefont {Qi}~\bibnamefont {Zhou}},\ }\bibfield
  {title} {\enquote {\bibinfo {title} {Curving the space by
  non-{H}ermiticity},}\ }\href {\doibase 10.1038/s41467-022-29774-8} {\bibfield
   {journal} {\bibinfo  {journal} {Nat. Commun.}\ }\textbf {\bibinfo {volume}
  {13}},\ \bibinfo {pages} {2184} (\bibinfo {year} {2022})}\BibitemShut
  {NoStop}%
\bibitem [{\citenamefont {Urwyler}\ \emph
  {et~al.}(2022{\natexlab{b}})\citenamefont {Urwyler}, \citenamefont
  {Lenggenhager}, \citenamefont {Boettcher}, \citenamefont {Thomale},
  \citenamefont {Neupert},\ and\ \citenamefont
  {Bzdu\v{s}ek}}]{Urwyler:2022:SDC}%
  \BibitemOpen
  \bibfield  {author} {\bibinfo {author} {\bibfnamefont {David~M.}\
  \bibnamefont {Urwyler}}, \bibinfo {author} {\bibfnamefont {Patrick~M.}\
  \bibnamefont {Lenggenhager}}, \bibinfo {author} {\bibfnamefont {Igor}\
  \bibnamefont {Boettcher}}, \bibinfo {author} {\bibfnamefont {Ronny}\
  \bibnamefont {Thomale}}, \bibinfo {author} {\bibfnamefont {Titus}\
  \bibnamefont {Neupert}}, \ and\ \bibinfo {author} {\bibfnamefont
  {Tom\'{a}\v{s}}\ \bibnamefont {Bzdu\v{s}ek}},\ }\href {\doibase
  10.5281/zenodo.6380568} {\enquote {\bibinfo {title} {Data and code for:
  {H}yperbolic topological band insulators},}\ } (\bibinfo {year}
  {2022}{\natexlab{b}}),\ \bibinfo {note} {{DOI}:
  \href{https://doi.org/10.5281/zenodo.6380568}{https://doi.org/10.5281/zenodo.6380568}}\BibitemShut
  {NoStop}%
\end{thebibliography}%
    \let\addcontentsline\oldaddcontentsline 
        
    \onecolumngrid  
    
    \newpage 
    
\begin{bibunit}
\onecolumngrid
%\resetlinenumber
\renewcommand\thesection{S\arabic{section}}
\renewcommand\thesubsection{\alph{subsection}}
\renewcommand\theequation{S\arabic{equation}}  
\setcounter{page}{1}
\setcounter{equation}{0}
\setcounter{figure}{0}
\setcounter{table}{0}

\renewcommand\thefigure{S{\arabic{figure}}}
\renewcommand{\theHfigure}{S\arabic{figure}}
\renewcommand\thetable{{S\arabic{table}}}
\renewcommand{\theHtable}{S\arabic{table}}

\titleformat{\section}{\normalfont\sffamily\bfseries\filcenter}{Supplementary Note \thesection:}{1em}{}

\makeatletter
\renewcommand{\fnum@figure}{Supplementary Figure~\thefigure}
\renewcommand{\fnum@table}{Supplementary Table~\thetable}
\makeatother

\title{Supplementary Material to: \texorpdfstring{\\ \medskip}{} Hyperbolic Topological Band Insulators}

\author{David M. Urwyler}
\affiliation{Department of Physics, University of Zurich, Winterthurerstrasse 190, 8057 Zurich, Switzerland}

\author{Patrick M. Lenggenhager\,\orcidlink{0000-0001-6746-1387}}
\affiliation{Department of Physics, University of Zurich, Winterthurerstrasse 190, 8057 Zurich, Switzerland}
\affiliation{Condensed Matter Theory Group, Paul Scherrer Institute, 5232 Villigen PSI, Switzerland}
\affiliation{Institute for Theoretical Physics, ETH Zurich, 8093 Zurich, Switzerland}

\author{Igor Boettcher\,\orcidlink{0000-0002-1634-4022}}
\affiliation{Department of Physics, University of Alberta, Edmonton, Alberta T6G 2E1, Canada}
\affiliation{Theoretical Physics Institute, University of Alberta, Edmonton, Alberta T6G 2E1, Canada}

\author{Ronny Thomale\,\orcidlink{0000-0002-3979-8836}}\affiliation{Institut für Theoretische Physik und Astrophysik, Universität Würzburg, 97074 Würzburg, Germany}

\author{Titus Neupert\,\orcidlink{0000-0003-0604-041X}}
\affiliation{Department of Physics, University of Zurich, Winterthurerstrasse 190, 8057 Zurich, Switzerland}

\author{Tom\'{a}\v{s} Bzdu\v{s}ek\,\orcidlink{0000-0001-6904-5264}}\email[corresponding author: ]{tomas.bzdusek@psi.ch}
\affiliation{Condensed Matter Theory Group, Paul Scherrer Institute, 5232 Villigen PSI, Switzerland}
\affiliation{Department of Physics, University of Zurich, Winterthurerstrasse 190, 8057 Zurich, Switzerland}

\date{\today}

\maketitle

\newpage
\onecolumngrid

\tableofcontents

\section{Geometry of the hyperbolic plane.}
We adopt the Poincar\'{e}-disk representation of the hyperbolic plane, i.e., as the unit disk in the complex plane, $\mathbb{D}=\{z\in\mathbb{C}\;|\,\abs{z}<1\}$, with the hyperbolic metric given by 
\begin{equation}
\de s^2 = (2\kappa)^2 \frac{\de z \, \de \bar{z}}{\left(1-z\bar{z}\right)^2},
\label{eq:distance_element}
\end{equation} 
where $\kappa$ is a unit of length and the bar in $\bar{z}$ indicates complex conjugation. 
With this choice, the Gaussian curvature equals $K=-\kappa^2$. 
In our work, we fix $\kappa=1/2$ (leading to curvature $K=-4$). 
With this convention, 
\begin{equation}
d(z_1,z_2) = \frac{1}{2}\arccosh\left[1+\frac{2\abs{z_1-z_2}^2}{(1-\abs{z_1}^2)(1-\abs{z_2}^2)}\right]\label{eqn:hyperbolic-arc-length}
\end{equation}
determines the geodesic distance for a pair of points $z_{1,2}\in\mathbb{D}$~\cite{Boettcher:2021}. 

We briefly investigate geometric aspects of a disk with radius $0<R<1$ in the complex plane, labelled $\mathbb{D}_R$. Its surface area is computed as
\begin{equation}
S(R) = \int_0^{2\pi} \de \phi \ \int_0^R \de r \frac{r}{(1-r^2)^2} = \frac{\pi R^2}{1-R^2}.
\end{equation}
The hyperbolic distance from the center to the boundary of the disk is 
\begin{equation}
d(0,R)=\int_0^R \frac{\de r}{(1-r^2)} = \frac{1}{2}\arccosh\left(\frac{1+R^2}{1-R^2}\right) = \arctanh R, \label{eqn:hyperbolic-distance}
\end{equation}
and its perimeter is 
\begin{equation}
p(R)= \int_0^{2\pi} \de \phi \frac{R}{(1-R^2)} = \frac{2\pi R}{1-R^2}.\label{eqn:perimeter}
\end{equation}
It is easily verified that $p(R)/d(0,R)>2\pi$ (for perimeter) and $S(R)/d^2(0,R)>\pi$ (for surface area), as expected for a negatively curved space.

\FloatBarrier
\newpage

\section{Geometry of \texorpdfstring{$\{p,q\}$}{(p,q)} tessellations.}
We determine the distance $d_1^{\{p,q\}}$ of nearest-neighbor vertices of the $\{p,q\}$ lattice. 
For that purpose, we consider a triangle $ABC$ where: $A$ is the center of a regular $p$-sided polygon (a `$p$-gon', for short), $B$ is one of the $p$-gon's vertices, and $C$ is a point that lies on the boundary of the $p$-gon at the middle of an edge connecting to $B$. 
Note that the hyperbolic length of the triangle side $\abs{BC}=d_1^{\{p,q\}}/2$. 
The internal angles of triangle $ABC$ at vertices $A$, $B$, and $C$ are identified as $\alpha=\pi/p$, $\beta=\pi/q$, and $\gamma=\pi/2$, respectively. 
According to the hyperbolic law of cosines~\cite{Katok:1992},
\begin{equation}
\cosh\left(\frac{\abs{BC}}{\kappa}\right) = \frac{\cos\alpha+\cos\beta\cos\gamma}{\sin\beta\sin\gamma} = \frac{\cos \tfrac{\pi}{p}}{\sin \tfrac{\pi}{q}},
\end{equation} 
which implies $d_1^{\{8,3\}}=\arccosh\big({\cos\frac{\pi}{8}}/{\sin\frac{\pi}{3}}\big)\approx 0.36352$ for our choice $\kappa=1/2$. 
Next, the distance $\abs{AB}=r^{\{p,q\}}$ from the center of the $p$-gon to its vertex is also determined from the law of cosines,
\begin{equation}
\cosh\left(\frac{\abs{AB}}{\kappa}\right)=\frac{\cos\gamma+\cos\alpha\cos\beta}{\sin\alpha\sin\beta}=\frac{1}{\tan\tfrac{\pi}{p}\tan\tfrac{\pi}{q}},
\end{equation}
leading to $r^{\{8,3\}}=\arccosh\big[1/\big(\tan\frac{\pi}{8}\tan\frac{\pi}{3}\big)\big]/2\approx 0.430353$. 

If the center of the $p$-gon is placed at the center of the Poincar\'{e} disk ($z=0$), then Eq.~(\ref{eqn:hyperbolic-distance}) governs the complex coordinates of the $p$-gon vertices, $|{z_B^{\{p,q\}}}|=\tanh \big(r^{\{p,q\}}\big)$, leading to $|{z_B^{\{8,3\}}}| \approx 0.405616$.
The information obtained thus far is sufficient to find the complex coordinates $\{z_{(\varnothing,a)}\}_{a=1}^{n_\textrm{cell}}$ (where $n_\textrm{cell}=16$) of all sites of the $\{8,3\}$ lattice which reside inside the Bolza cell centered at $z=0$ (listed in Supplementary Table~\ref{tab:8-3-coordinates}). 
For later purposes, we use Eq.~(\ref{eqn:hyperbolic-arc-length}) to also determine the distance of next-nearest-neighbor sites as
\begin{equation}
d_2^{\{p,q\}}=d\left(|{z_B^{\{p,q\}}}|\e^{2\pi \imi/p }\,,\,|{z_B^{\{p,q\}}}|\e^{-2\pi \imi/p}\right),
\end{equation}
which leads to $d_2^{\{8,3\}}=\arccosh\big(1+\frac{2\sqrt{2}}{3}\big)/2\approx 0.641645$.

\begin{table}[hbt!]
    \centering
    \begin{tabular}{|c||c|c|c|c|}
    \hline
    \rowcolor{lightgray!40}
    site label $a$  &  1 & 2 & 3 & 4 \\
    \hline
    coordinate $z_{(\varnothing,a)}$ &
    $\; 0.374741 +0.155223 \,\imi \;$ &
    $\; 0.155223 +0.374741 \,\imi \;$ &
    $\; -0.155223+0.374741 \,\imi \;$ &
    $\; -0.374741+0.155223 \,\imi \;$ \\
    \hline \hline    
    \rowcolor{lightgray!40}
    site label $a$  &  5 & 6 & 7 & 8 \\
    \hline
    coordinate $z_{(\varnothing,a)}$ &
    $\; -0.374741-0.155223 \,\imi \;$ &
    $\; -0.155223-0.374741 \,\imi \;$ &
    $\; 0.155223 -0.374741 \,\imi \;$ &
    $\; 0.374741 -0.155223  \,\imi \;$ \\
    \hline \hline
    \rowcolor{lightgray!40}
    site label $a$  &  9 & 10 & 11 & 12 \\
    \hline
    coordinate $z_{(\varnothing,a)}$ &
    $\; 0.610313 +0.252800 \,\imi \;$ &
    $\; 0.252800 +0.610313 \,\imi \;$ &
    $\; -0.252800+0.610313 \,\imi \;$ &
    $\; -0.610313+0.252800  \,\imi \;$ \\
    \hline \hline    
    \rowcolor{lightgray!40}
    site label $a$  &  13 & 14 & 15 & 16 \\
    \hline
    coordinate $z_{(\varnothing,a)}$ &
    $\; -0.610313-0.252800 \,\imi \;$ &
    $\; -0.252800-0.610313 \,\imi \;$ &
    $\; 0.252800 -0.610313 \,\imi \;$ &
    $\; 0.610313 -0.252800 \,\imi \;$ \\
    \hline
    \end{tabular}
    \caption[Coordinates of the sites in the Bolza cell.]{\textbf{Coordinates of the sites in the Bolza cell.} The table lists the complex coordinates of the $16$ sites of the $\{8,3\}$ lattice which belong to the innermost Bolza cell (centered at $z=0$), represented inside the Poincar\'{e} disk with radius $1$. For derivation of these values, and for the meaning of the subscript ``$(\varnothing,a)$'', see Methods.}
   \label{tab:8-3-coordinates}
\end{table}

The Gauss-Bonnet theorem~\cite{Nakahara:1990} relates surface area $S^{P}$ of a geodesic $p$-sided polygon $P$ in space of constant curvature $K$ to its internal angles $\{\alpha_j\}_{j=1}^p$, 
\begin{equation}
    \sum_{j=1}^p\left(\pi -\alpha_j\right)=2\pi - K S^{P}.
    \label{eqn:Gauss-Bonnet}
\end{equation}
The elementary cell of the $\{p,q\}$ lattice is a regular geodesic $p$-gon with all internal angles of size $\alpha = 2\pi/q$. 
It follows from Eq.~(\ref{eqn:Gauss-Bonnet}) that the surface area of the elementary $p$-gon is \begin{equation}
S^{\{p,q\}} = \pi\frac{(p-2)(q-2)-4}{4q}.
\end{equation}
We obtain $S^{\{8,3\}}=\pi/6$, while the area of the Bolza cell ($p$-gon of $\{8,8\}$ lattice) is six times larger, $S^{\textrm{Bolza}}=\pi$. 
The ratio 
\begin{equation}
\widetilde{N}_\textrm{UC}(R) = \frac{S(R)}{S^{\textrm{Bolza}}} = \frac{R^2}{1-R^2}\label{eqn:Nuc-to-radius}
\end{equation}
gives an approximate number of Bolza cells that fit into a disk with radius $R$ in the complex plane.

\FloatBarrier
\newpage

\section{Reduced Kane-Mele model.}\label{sec:rKM-model}
\begin{figure}[hbt!]
\centering
    \includegraphics{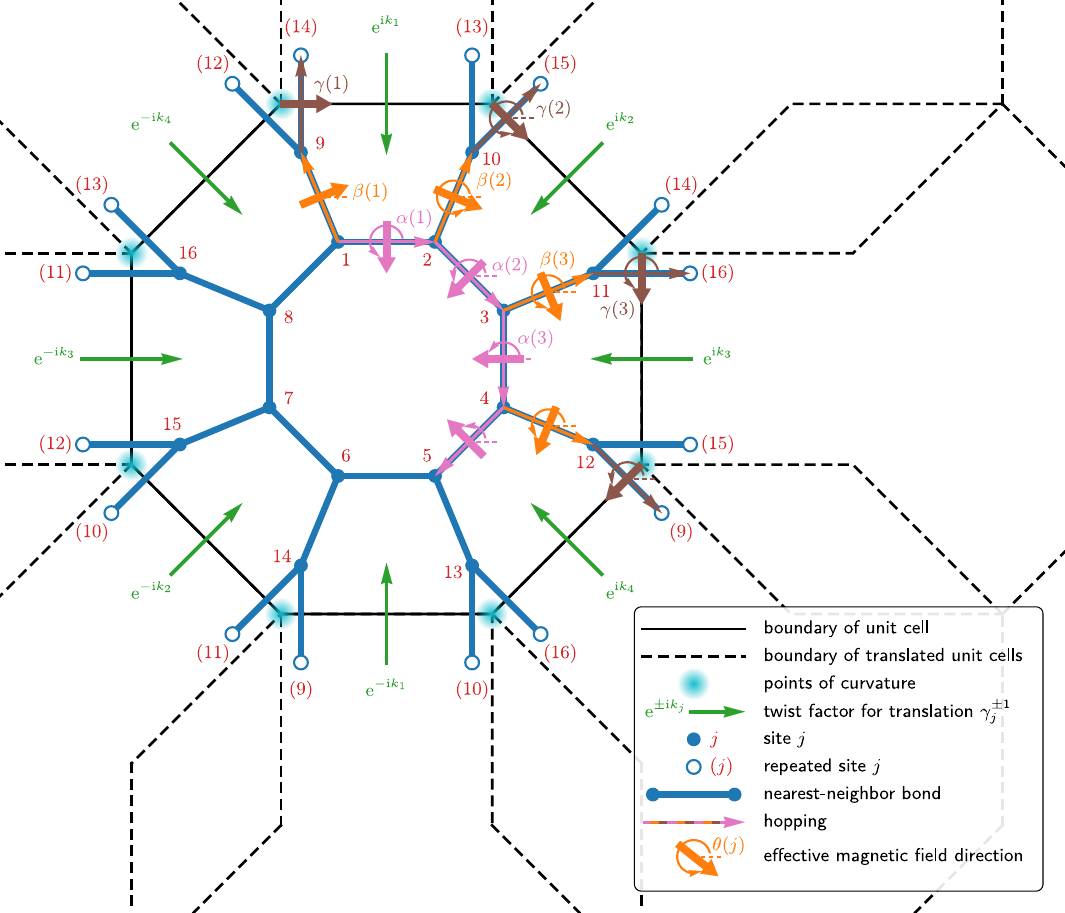}
    \caption[Geometry of the reduced Kane-Mele model.]{\textbf{Geometry of the reduced Kane-Mele model.}
    (For details of the construction, see Methods). In the reduced hyperbolic Kane-Mele model, the lattice is deformed such that all the negative curvature is concentrated at the corners (cyan dots) of the unit cell (the Bolza cell; deformed into the black octagon).
    As a consequence, space is flat everywhere except at those points (octagon corners,) and the neighbouring unit cells (dashed black octagons) seem to overlap in our representation of the lattice.
    The unit cell contains $16$ sites (blue dots labelled by red numbers).
    Assuming an electric field due to the substrate pointing out of the plane, hopping between two sites (pink, orange and brown thin arrows) leads to an effective magnetic field pointing in the direction of the corresponding thick arrow (of the same color).
    This direction is parametrized by an angle $\theta(j)$, where $\theta=\alpha,\beta,\gamma$ depending on whether the hopping is on the inner ring, in the radial direction, or across different Bolza unit cells, respectively, and $j\in\{1,\ldots,8\}$ enumerates those nearest-neighbour bonds. 
    Assuming first that we set the alternating on-site potential to $M=0$, the constructed model preserves time-reversal symmetry $\mcT$, the eightfold rotation symmetry with respect to the center of the Bolza cell (corresponding to symmetry `$R$' in the Supplementary Material to Ref.~\cite{Maciejko:2021}), the mirror symmetry with respect to lines connecting antipodal corners of the unit cell (symmetry `$S$' in Ref.~\cite{Maciejko:2021}), and the mirror symmetry with respect to lines connecting centers of antipodal edges of the unit cell (composition of $R$ and $S$).
    However, the model breaks the threefold rotation symmetry with respect to any site of the $\{8,3\}$ lattice (symmetry `$U$' in Ref.~\cite{Maciejko:2021}), as is apparent from the varying geometrical distortions of the six elementary (blue) octagons in the Bolza unit cell. 
    The on-site $\pm M$ term further breaks $R$ (but not $\mcT$ and $S$) symmetry.
    [Note that in the main text we use symbol `$R$' to indicate a $(\pi/2)$-rotation around the center of the Bolza cell, whereas the same symbol is used by Ref.~\cite{Maciejko:2021} to indicate a $(\pi/4)$-rotation around the same point.] }
\label{fig:reduced-KM-geometry}
\end{figure}

To avoid complications that arise from non-trivial spin holonomy in curved spaces, we construct the \emph{reduced} Kane-Mele (rhKM) model through the following simplification [cf.~Supplementary Fig.~\ref{fig:reduced-KM-geometry}]: instead of taking the negative curvature to be constant inside the Bolza cell, we consider a continuous deformation where all the curvature becomes concentrated at the corners of the cell (cyan dots in Supplementary Fig.~\ref{fig:reduced-KM-geometry}) while the manifold becomes flat everywhere else.
Then, the Bolza cell essentially becomes a Euclidean regular octagon (black outline in Supplementary Fig.~\ref{fig:reduced-KM-geometry}).
[Note that the adjacent Bolza cells (dashed frames in Supplementary Fig.~\ref{fig:reduced-KM-geometry}) seemingly overlap one another, but this is consistent with the $-4\pi$ quantum of curvature at the vertex.]
Then there exists a unique arrangement of the $16$ sites of the $\{8,3\}$ lattice (blue frame in Supplementary Fig.~\ref{fig:reduced-KM-geometry}) within the octagon such that (1)~all pairs of NN sites have the same distance, and (2)~the eightfold rotation and mirror symmetries of the octagon are preserved. 

The advantage of the deformed model is that the NN hopping of electrons occurs within regions of zero curvature, allowing us to construct the Rashba SOC terms using the procedure familiar from the flat Euclidean space. 
Namely, we imagine the presence of a substrate that creates a constant electric field $E_z$ perpendicular to the 2D plane of the system. 
As the spinful electron hops (within the locally flat region) between a pair of NN sites $b \mapsfrom a$ (represented in Supplementary Fig.~\ref{fig:reduced-KM-geometry} with thin pink/orange/brown arrows), characterized by displacement vector  $\bs{r}_{ba} = \bs{r}_{b}-\bs{r}_{a}$ [where the vectors are decomposed into local $(x,y)$ coordinates that run rightwards/upwards inside the illustration in Supplementary Fig.~\ref{fig:reduced-KM-geometry}], the electron perceives in its moving reference frame a magnetic field $\bs{B}_{ba}\propto \bs{r}_{ba} \times \bs{E}$ [where $\bs{E}=(0,0,E_z)$, and $\bs{r}_{ba}$ has been supplemented with vanishing component in the third direction]. 
The obtained magnetic fields are in-plane vectors [indicated for several hopping processes in Supplementary Fig.~\ref{fig:reduced-KM-geometry} with thick pink/orange/brown arrows], which are characterized by their direction $\theta_{ba}$. 
The corresponding Rashba term added to the rhKM Hamiltonian is 
\begin{equation}
\mcH_{ba} = \imi \lambda_\textrm{R}\left[\cos(\theta_{ba})\sigma_x + \sin(\theta_{ba}) \sigma_y \right].\label{eqn:Rashba-term}
\end{equation}
By repeating the above procedure for hopping $a\mapsfrom b$, we find that $\mathcal{H}_{ab} = -\mathcal{H}_{ba}$, which is compatible with time-reversal symmetry $\mcT = \imi \sigma_y \mcK$.

In the corresponding code in the enclosed data repository~\cite{Urwyler:2022:SDC} (see also Supplementary Note~\ref{eqn:HBT-Hamiltonians}), we label the angles $\theta_{ba}$ as one of $\{\alpha(j),\beta(j),\gamma(j)\}_{j=1}^8$ (illustrated in pink/orange/brown in Supplementary Fig.~\ref{fig:reduced-KM-geometry}); the three symbols distinguish respectively the Rashba terms on NN bonds (1)~along the inner elementary (blue) octagon, (2)~directed towards the corners of the large (black) octagon, and (3)~crossing the boundary of the large octagon.

Let us emphasize that the interpretation with the non-homogeneous curvature is only adopted to \emph{construct} the Hamiltonian elements in Eq.~(\ref{eqn:Rashba-term}). 
When discussing the geometry of the system in the subsequent text, we still assume the $16$ sites of the inner Bolza cell to be located at the complex coordinates listed in Supplementary Table~\ref{tab:8-3-coordinates}, while keeping the Hamiltonian elements derived above. 

\FloatBarrier
\newpage

\section{Hyperbolic Bloch Hamiltonian.}
To obtain the hyperbolic Bloch Hamiltonian $\mcH(\bs{k})$ from tight-binding Hamiltonian $\mcH$ in position space, we replace the infinite lattice by a single Bolza cell with identified antipodal boundaries [pairs of colored edges in Fig.~\ref{fig:Bolza-cell}(a)]~\cite{Maciejko:2021}. 
To implement the twisted boundary conditions, tunnelling amplitudes ($t$) for hopping process that cross edges displaced by $\gamma_j^{\pm 1}$ are multiplied by phase factors $\e^{\pm i k_j}$. 
Since there are four phases $\{k_j\}_{j=1}^4\equiv \bs{k}$, it follows that $\mcH(\bs{k})$ is defined over a 4D BZ.
For models on the $\{8,3\}$ lattice, the presence of $n_\textrm{cell}{=}16$ sites per Bol\-za cell implies that $\mcH(\bs{k})$ is a matrix of dimensions $D_\textrm{Bloch} {=} n_\textrm{cell}$ ($D_\textrm{Bloch}=2 n_\textrm{cell}$) in the absence (presence) of the spin degree of freedom~\cite{Urwyler:2021,*Urwyler:2022b}.
The Bloch Hamiltonians for the studied models are constructed in Supplementary Note~\ref{eqn:HBT-Hamiltonians}), and made available as \Mathematica{} notebooks in the data repository~\cite{Urwyler:2022:SDC}.

We further consider the generation of density from HBT. 
Given a hyperbolic Bloch Hamiltonian, we perform random sampling of $N_{\bs{k}}$ momenta over the 4D hypercubic BZ, $\forall j\in\{1,2,3,4\}: k_j \in [-\pi,\pi]$. 
The collected list of eigenvalues $\{\varepsilon_j\}_{j=1}^{N_{\bs{k}}D_\textrm{Bloch}}$ is converted into a continuous DoS function via
\begin{equation}
\rho^\textrm{HBT}(E)=\frac{1}{N_{\bs{k}} D_\textrm{Bloch}}\sum_{j=1}^{N_{\bs{k}}D _\textrm{Bloch}} f_\eta(E-\varepsilon_j)  \label{eqn:HBT-DoS} 
\end{equation}
where $f_\eta(\varepsilon)=\tfrac{1}{\eta\sqrt{2\pi}}\exp\big[-\frac{\varepsilon^2}{2\eta^2}\big]$ is a Gaussian smearing function. 
The factor in front of the summation symbol is chosen such that $\rho^\textrm{HBT}(E)$ integrates to $1$.
We use $\eta=0.063$ and $N_{\bs{k}}= \num{2e4}$ throughout the manuscript.

\FloatBarrier
\newpage

\section{Flake Hamiltonian in position space.}
Two inputs are used to algorithmically construct the lattice Hamiltonian $\mcH^\textrm{flake}$ on a circular-flake in position space: the hyperbolic Bloch Hamiltonian $\mcH(\bs{k})$, and the target number $\widetilde{N}_\textrm{UC}$ of unit cells. 
The Bloch Hamiltonian has components $\mcH(\bs{k})=\left\{h_{ab}(\bs{k})\right\}_{a,b=1}^{n_\textrm{cell}}$
For the models presented in this work, one finds that $h_{ab}(\bs{k}) = t_{ab} \e^{\imi \bs{k}\cdot \bs{\varrho}_{ab}}$ with a unique value of the 4-component vector $\bs{\varrho}_{ab}$ and of the hopping amplitude $t_{ab}$ for each  $a,b\in\{1,\ldots,n_\textrm{cell}\}$. (Note that here we treat $t_{ab}$ as a single complex number for spinless models, and as a complex-valued $2\times 2$ matrix for spinful models.) 
The algorithm, which is implemented in the data repository~\cite{Urwyler:2022:SDC}, proceeds in the following steps.

First, we invert Eq.~(\ref{eqn:Nuc-to-radius}) to determine an estimated radius $R(\widetilde{N}_\textrm{UC})=\big[\widetilde{N}_\textrm{UC}/\big(1+\widetilde{N}_\textrm{UC}\big)\big]^{1/2}$ of a disk in the complex plane that contains $\widetilde{N}_\textrm{UC}$ Bolza cells. 
We center one of the Bolza cells at $z=0$, and we use the known action $\rho_{\gamma_j}$ of generators $\gamma_j$ on the complex coordinates in $\mathbb{D}$ [Eq.~(45) in Ref.~\onlinecite{Boettcher:2020}] to identify the centers of all cells of the $\{8,8\}$ (Bolza) lattice that lie within the distance $R(\widetilde{N}_\textrm{UC})$ from the origin. 
(Note that the centers of the Bolza cells do not coincide with any site of the $\{8,3\}$ lattice.)
This results in a \emph{list of Bolza cells}, $L_\textrm{cells}=\{w_\ell\}_{\ell=1}^{N_\textrm{UC}}$; here, each $w_\ell$ is a sequence of generators and their inverses (a \enquote{word}) that translates $z=0$ to $\rho_{w_\ell}(0) = z_j$ within the disk of radius $R$ (i.e., $|{z_j}|<R$), and $N_\textrm{UC}$ is the total number of selected Bolza cells.
The words $w_\ell$ are elements of the hyperbolic translation group, and act on complex coordinates inside the Poincar\'{e} disk by $\rho_{w_\ell}$, which is the corresponding composition of $\rho_{\gamma_j}$ (and of their inverses). 
The word `$\varnothing$' of length zero (which corresponds to the identity of the translation group) is always present in $L_\textrm{cells}$; it encodes the Bolza cell centered at $z=0$.
We then apply $L_\textrm{cells}$ to generate a \emph{list of sites} of the $\{8,3\}$ lattice, $\widetilde{L}_\textrm{sites}=\{(w_\ell,a)\}$, where $1\leq a\leq n_\textrm{cell}$ labels the individual sites within each Bolza cell $w_\ell\in L_\textrm{cells}$. The complex coordinates of the sites are obtained as $z_{(w_\ell,a)} = \rho_{w_\ell}(z_{(\varnothing,a)})$.

We next use the list $\widetilde{L}_\textrm{sites}$ and the amplitudes $t_{ab}$ to construct the real-space Hamiltonian $\widetilde{\mcH}^\textrm{flake}$.
The Hamiltonian has $\widetilde{n}_{\textrm{flake}}\times \widetilde{n}_{\textrm{flake}}$ components, where $\widetilde{n}_{\textrm{flake}} = n_\textrm{cell} N_\textrm{UC}$ is the length of list $\widetilde{L}_\textrm{sites}$.
Since the presently considered tight-binding Hamiltonians contain only NN and NNN terms, we run the following routine. 
For each pair $(x,y)$ with $x=(w_\ell,a)$ and $y=(w_{m},b)$ in $\widetilde{L}_\textrm{sites}$, compute the distance $d(z_x,z_y)$. 
If the distance is equal to the NN distance $d_1^{\{8,3\}}$ or to the NNN distance $d_2^{\{8,3\}}$, we set $\big(\widetilde{\mcH}^\textrm{flake}\big)_{x,y}=t_{ab}$. 
In the very last step, we smooth the boundary of the system by identifying sites that have only a single NN. 
We drop the corresponding rows and columns of $\widetilde{\mcH}^\textrm{flake}$, which results in the final Hamiltonian $\mcH^\textrm{flake}$ that has a slightly decreased number of components $n_\textrm{flake} \times n_\textrm{flake}$ (we similarly define the corresponding shortened list of sites $L_\textrm{sites}$ of length $n_\textrm{flake}$). 
Note that for spinful models, the counted Hamiltonian components are $2\times 2$ blocks, i.e., the actual Hamiltonian dimension is $D_\textrm{flake}=n_\textrm{flake}$ ($D_\textrm{flake}=2n_\textrm{flake}$) for spinless (spinful) models.

\FloatBarrier
\newpage

\section{Considered system sizes in the flake geometry.} 
The assumed system size varies between the figures. We encode the size for each calculation with the triplet ``$\textsc{size}=(\widetilde{N}_\textrm{UC},N_\textrm{UC},n_\textrm{flake})$'' (where $N_\textrm{UC}$ and $n_\textrm{flake}$ are selected automatically by the the above-outlined algorithm upon inputting $\widetilde{N}_\textrm{UC}$). 

\begin{figure}[hbt!]
\centering
    \includegraphics[width=\linewidth]{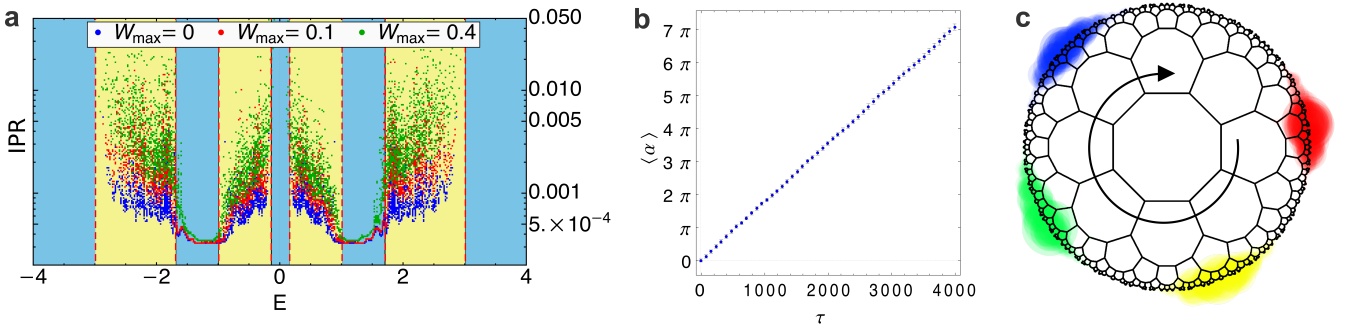}
    \caption[Robustness against Anderson disorder in the hyperbolic Haldane model.]{
    \textbf{Robustness against Anderson disorder in the hyperbolic Haldane model.}
    The disorder chosen for the hyperbolic Haldane model corresponds to the addition of a random on-site potential drawn from a box distribution in range $[-W_\textrm{max},W_\textrm{max}]$. 
    \textbf{a,} IPR of the individual eigenstates for various choices of disorder strength $W_\textrm{max}$. Note that the domain of low values of $\textrm{IPR}$ has reduced to a narrower range of energies for the green data ($W_\textrm{max}=0.4$), indicating the shrinking of the bulk energy gap; nevertheless, the robustness of edge states with low IPR is manifest. 
    \textbf{b--c,} Propagation of chiral edge states in disordered hH model.
    The calculation is analogous to Fig.~\ref{fig:edge-states}(b,c) of the main text, but here we assume the addition of a strong random on-site potential with disorder strength $W_\textrm{max}=0.4$ (green in panel \textbf{a}).
    The wave packet continues to propagate around the flake boundary with nearly uniform angular velocity.
    The parameters of the Gaussian projector are $(\mu,\sigma) = (1.3,0.025)$, and the data in red/yellow/green/blue color consecutively correspond to the wave packet at times $\tau = (0, 240,480, 720)$, $(\mu, \sigma) = (1.3,0.025)$.
    }
\label{fig:HaldaneAnderson}
\end{figure}

For the hH model, the sizes are set up as follows: 
\begin{itemize}
\item  Fig.~\ref{fig:Bolza-cell}(b,c) and Fig.~\ref{fig:Haldane}(c): $\textsc{size}=(700,761,8496)$,
\item  Supplementary Fig.~\ref{fig:topology}(a,c) and Supplementary Fig.~\ref{fig:HaldaneAnderson}(a): $\textsc{size}=(400,409,4520)$,
\item Fig.~\ref{fig:edge-states}(a) and Supplementary Fig.~\ref{fig:edgemode-dispersion}(b): $\textsc{size} = (500,569,6344 \mapsto 896)$,
\item Fig.~\ref{fig:edge-states}(b,c) and Supplementary Fig.~\ref{fig:HaldaneAnderson}(b,c): $\textsc{size} = (200,169,1864)$.
\end{itemize}

\begin{figure}[h!]
\centering
    \includegraphics[width=\linewidth]{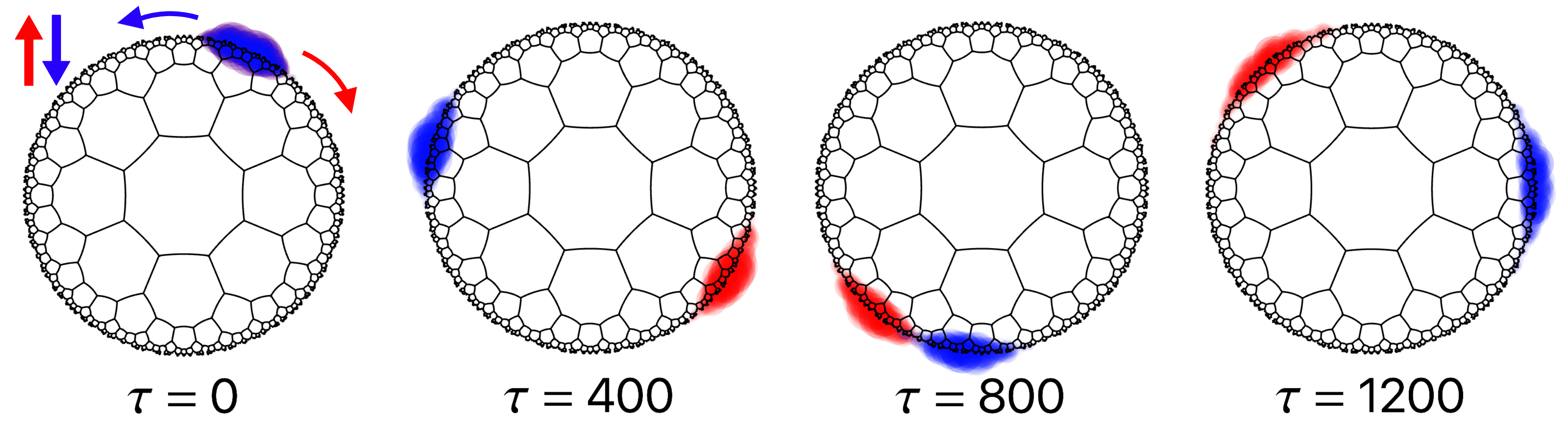} \caption[Propagation of edge states in the rhKM model.]{
    \textbf{Propagation of edge states in the rhKM model.} 
    We compute and plot the propagation of helical edge states around the boundary of a flake supporting the rhKM model.
    The time evolution is computed according to the description in section `\emph{Bulk-boundary correspondence}' of the main text, and we use the Gaussian projector $\mathbb{P}_{(\mu,\sigma)}$ with $(\mu,\sigma)=(1.3,0.025)$.
    The indicated time $\tau$ is counted in multiples of $1/t_1$. The initial single-site localized state $|\varphi_\textrm{site}\rangle$ was chosen to have the form $\frac{1}{\sqrt{2}}(1,1)^\top$ in the spin degree of freedom. 
    We observe that time-evolution splits $\mathbb{P}_{(\mu,\sigma)} |\varphi_\textrm{site}\rangle$ into a pair of counter-propagating wave packets that during the time evolution pass through one another without scattering.
    }
\label{fig:KaneMeleEdgeState}
\end{figure}

For the rhKM model, we chose the following sizes: 
\begin{itemize}
\item Supplementary Fig.~\ref{fig:topology}(a,d) and Supplementary Fig.~\ref{fig:KaneMeleEdgeState}: $\textsc{size}=(200,169,1864)$,
\item Fig.~\ref{fig:Kane-Mele}(b) and Fig.~\ref{fig:Anderson}: $\textsc{size}=(300,297,3304)$,
\item Supplementary Fig.~\ref{fig:edgemode-dispersion}(b): $\textsc{size} = (500,569,6344 \mapsto 896)$.
\end{itemize}
Note that in the calculations leading to Fig.~\ref{fig:edge-states}(a) and to Supplementary Fig.~\ref{fig:edgemode-dispersion}(a,b), an additional level of removing sites at the boundary is applied, namely, starting with the list $\widetilde{L}_\textrm{sites}$, we remove all sites with less than three nearest neighbors. Thus, instead of arriving at a system with $6344$ sites, we obtain a much smaller system with $896$ sites, cf.~Supplementary Note~\ref{sec:edge-mode-dispersion}.

\FloatBarrier
\newpage

\section{Bulk density of states and integrated boundary density of states for the flake Hamiltonian.}\label{sec:DoS-def}
Given a lattice Hamiltonian $\mcH^\textrm{flake}$ on a circular flake, we perform ED to find its eigensystem $\Lambda^\textrm{flake}=\big\{(E_j,|\phi_j\rangle)\big\}_{j=1}^{D_\textrm{flake}}$, where $|\phi_j\rangle$ is an eigenvector with eigenvalue $E_j$.
Each eigenstate is a list of amplitudes, namely $|\phi_j\rangle  = \{\phi_{j,x}\}_{x\in L_\textrm{sites}}$ for spinless models ($|\phi_j\rangle  = \{\phi_{j,x,\sigma}\}_{x\in L_\textrm{sites},\sigma\in\{\uparrow,\downarrow\}}$ for spinful models).
For each state we define the \emph{bulk weight} $\varpi_j$ as the probability that the particle in such a state is located on the innermost unit cell (i.e., sites with $w_\ell=\varnothing$),
\begin{equation}
\varpi_j^\textrm{bulk}=\sum_{a=1}^{16}\abs{\phi_{j,(\varnothing,a)}}^2 \qquad \textrm{and} \qquad \varpi_j^\textrm{bulk}=\sum_{a=1}^{16}\sum_{\sigma\in\{\uparrow,\downarrow\}}\abs{\phi_{j,(\varnothing,a),\sigma}}^2 \label{eqn:bulk-weight}
\end{equation}
for spinless and spinful models, respectively. We then convert the information about the eigenvalues and eigenvectors into a continuous DoS function via
\begin{equation}
\rho^\textrm{ED}_\textrm{bulk}(E)=\frac{1}{D_\textrm{Bloch}}\sum_{j=1}^{D_\textrm{flake}}\varpi_j^\textrm{bulk} f_\eta(E - E_j)\label{eqn:bulk-DoS-def}
\end{equation}
with the smearing function $f_\eta$ defined below Eq.~(\ref{eqn:HBT-DoS}). 
The prefactor $1/D_\textrm{Bloch}$ guarantees that $\rho^\textrm{ED}_\textrm{bulk}(E)$ integrates to $1$. To achieve a simple comparison with the HBT data, we use $\eta=0.063$ throughout the manuscript.

We next define the integrated DoS at the boundary in analogy with Eq.~(\ref{eqn:bulk-DoS-def}), where the only difference is the replacement $\varpi_j^\textrm{bulk} \mapsto \varpi_j^\textrm{bound.}$, with
\begin{equation}
\varpi_j^\textrm{bound.}=\sum_{x \in L_\textrm{bound.}}\!\!\!\abs{\phi_{j,x}}^2\qquad \textrm{and} \quad \varpi_j^\textrm{bound.}=\sum_{x\in L_\textrm{bound.}}\sum_{\sigma\in\{\uparrow,\downarrow\}}\abs{\phi_{j,x,\sigma}}^2
\end{equation}
for spinfless and spinful models, respectively, where $L_\textrm{bound.}$ is a sublist of $L_\textrm{sites}$ that selects sites with two nearest neighbors.

\FloatBarrier
\newpage

\section{Real-space invariants.}\label{sec:RSIs} 
\begin{figure}[h!]
    \centering
        \includegraphics[width=0.55\linewidth]{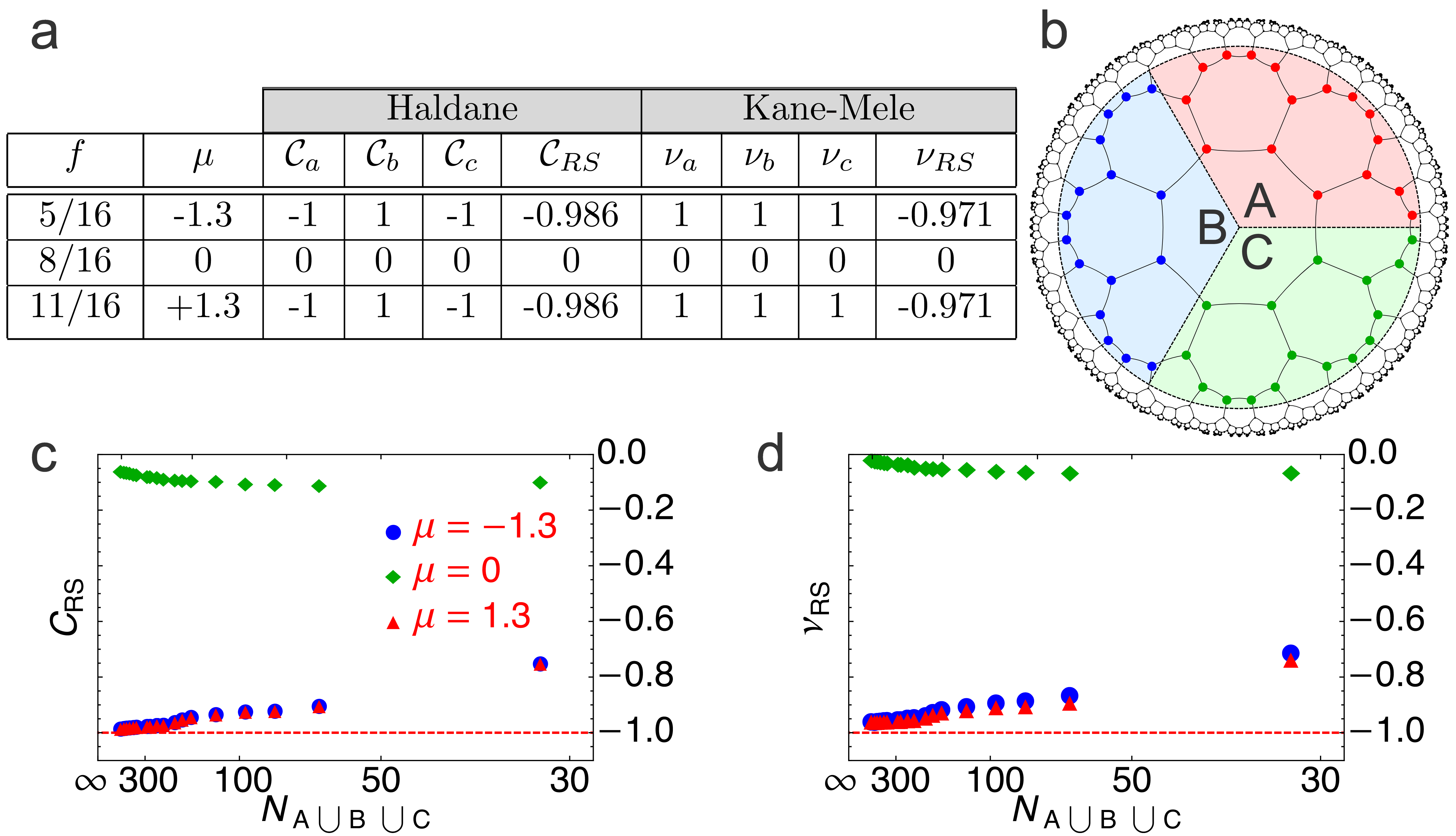} 
        \caption[]{
        \textbf{Topological invariants.}
        \textbf{a.} Summary of topological invariants for the hyperbolic Haldane and reduced hyperbolic KM models for their respective three energy gaps, labelled by filling fraction $f$ and chemical potential $\mu$. We show both momentum-space band invariants (subscripts $a$ and $b$) and  real-space topological markers (subscript $\textrm{RS}$)
        \textbf{b.} Regions $\mathsf{A},\mathsf{B},\mathsf{C}$ involved in the computation of the real-space invariants.
        \textbf{c,~d.} Convergence of the computed real-space Chern number (\textbf{c}) and real-space spin Chern number (\textbf{d}) for the three energy gaps (labelled by $\mu$) upon increasing the size of the regions $\mathsf{A},\mathsf{B},\mathsf{C}$. Here $N_{\mathsf{A}\cup \mathsf{B} \cup \mathsf{C}}$ is the total number of sites in the three regions.
        }
    \label{fig:topology}
    \end{figure}

For the energy gap at chemical potential $\mu$, we compute the \emph{real-space Chern number} using the formula from Ref.~\onlinecite{Kitaev:2006}, 
\begin{equation}
\mathcal{C}_\textrm{RS}(\mu) = 12 \pi \imi \sum_{j\in A} \sum_{k \in B} \sum_{\ell \in C}\left(\mathbb{P}_{jk}^\mu \mathbb{P}_{k\ell}^\mu \mathbb{P}_{\ell j}^\mu - \mathbb{P}_{j\ell}^\mu \mathbb{P}_{\ell k}^\mu \mathbb{P}_{kj}^\mu \right)   \label{eqn:RS-Chern}
\end{equation}
where $\mathbb{P}^\mu$ is the projector onto the subspace of occupied single-particle states at chemical potential $\mu$, and $A,B,C$ are three regions in the bulk of the systems that (\emph{i})~\emph{do not} extend all the way to the boundary, and which (\emph{ii})~are arranged counter-clockwise around the center of the system [cf.~Fig.~\ref{fig:topology}(b)].

In the presence of time-reversal symmetry, we compute the \emph{real-space spin Chern number} $\nu_\textrm{RS}$ following the ideas of Refs.~\onlinecite{Prodan:2011,Huang:2018}. Namely, we first construct the \emph{projected spin operator} $\mathbb{P}_z^\mu = \mathbb{P}^\mu \sigma_z \mathbb{P}^\mu$. Next, we perform spectral decomposition of the projected spin operator into eigenstates ($|\varsigma_j\rangle$) and eigenvalues ($S_{\!\!j}$). 
As long as the spin-mixing Rashba term is weak, the eigenvalues $S_{\!\!j}$ remain close to $\pm 1$, allowing us to define the index sets $\mathfrak{S}_\pm$; in addition, there are unphysical zero eigenvalues, which correspond to the part of the Hilbert space that is projected out by $\mathbb{P}^\mu$. 
This allows us to define projectors $\mathbb{P}_\pm = \sum_{j \in \mathfrak{S}_\pm} |\varsigma_j\rangle\langle \varsigma_j|$.
The integer-valued real-space spin Chern number is then obtained as
\begin{equation}
\nu_\textrm{RS}(\mu) = \frac{1}{2}\left[\nu^+_\textrm{RS}(\mu) - \nu^-_\textrm{RS}(\mu)\right],\label{eqn:RS-Kane-Mele}
\end{equation}
where $\nu^\pm_\textrm{RS}(\mu)$ are computed per Eq.~(\ref{eqn:RS-Chern}) with replaced $\mathbb{P}^\mu \mapsto \mathbb{P}^\mu_{\pm}$. 
Note that for strong SOC, the eigenstates of $\mathbb{P}_z$ no longer exhibit a clear spectral gap between the two sets $\mathfrak{S}_\pm$, and the formula in Eq.~(\ref{eqn:RS-Kane-Mele}) ceases to be applicable. We verified that this issue does not arise for our selected model parameters.

\begin{table}[h!]
 \small
 \centering
 \begin{tabular}{|c||c|c|c|c|c|c|c|c|c|c| } 
 \hline
 \multicolumn{11}{|c|}{\cellcolor{lightgray!40} Haldane/ Kane-Mele} \\
 \hline \hline
 \cellcolor{lightgray!40} $N_{A \cup B \cup C}$ & \cellcolor{lightgray!40}16 & \cellcolor{lightgray!40} 32 & \cellcolor{lightgray!40}64 & \cellcolor{lightgray!40}80 & \cellcolor{lightgray!40}96 & \cellcolor{lightgray!40}120 & \cellcolor{lightgray!40}152 & \cellcolor{lightgray!40}168 &\cellcolor{lightgray!40} 184 &\cellcolor{lightgray!40} 216\\ \hline
 $\mathcal{C_\textrm{RS}} \;\; (\mu = \pm 1.3)$ &-0.661 & -0.751 &  -0.904 & -0.921 & -0.924 & -0.934 & -0.944 & -0.953 & -0.962 & -0.972\\ \hline
 $\mathcal{\nu_\textrm{RS}} \;\; (\mu = -1.3)$ &-0.638 & -0.723 & -0.876 & -0.895 & -0.903 & -0.916 & -0.926 & -0.937 & -0.948 & -0.955   \\ \hline
 $\mathcal{\nu_\textrm{RS}} \;\; (\mu = +1.3)$ & -0.654 & -0.747 & -0.902 & -0.915 & -0.918 & -0.929 & -0.937 & -0.945 & -0.955 & -0.964 \\ \hline 
 
 \rowcolor{lightgray!40}
 $N_{A \cup B \cup C}$ & 240 & 272 & 288 & 368 & 400 & 448 & 496 & 544 & 624 \\ \cline{1-10}
 $\mathcal{C_\textrm{RS}} \;\; (\mu = \pm 1.3)$ & -0.973 & -0.976 & -0.976 & -0.979 & -0.980 & -0.981 & -0.982 & -0.984 & -0.986\\ \cline{1-10}
 $\mathcal{\nu_\textrm{RS}} \;\; (\mu = -1.3)$ & -0.957 & -0.963 & -0.963 & -0.966 & -0.966 & -0.968 & -0.969 & -0.971 & -0.970 \\ \cline{1-10}
 $\mathcal{\nu_\textrm{RS}} \;\; (\mu = +1.3)$ & -0.965 & -0.967 & -0.968 & -0.969 &-0.969  & -0.970 & -0.971 & -0.971 & -0.970 \\  \cline{1-10}  \cline{1-10}
\end{tabular}
\caption[Tabulated values of data in Fig.~\ref{fig:topology}(c,d).]{
    \textbf{Tabulated values of data in Fig.~\ref{fig:topology}(c,d).} The values indicate the convergence of the topological markers in positions space to values $\pm 1$ as the number of sites $N_{A{\cup}B{\cup}C}$ in the summation region is increased.
   }
   \label{Tab:RealSpaceTopInvTable}
\end{table}

\newpage
\FloatBarrier
\clearpage[4]

\section{Gaussian projector operator.}\label{sec:Gaussian}
Given the eigensystem of a flake Hamiltonian, $\Lambda^\textrm{flake} = \{(E_j,|\phi_j\rangle)\}_{j=1}^{D_\textrm{flake}}$, we construct for an energy range $[\mu-\sigma,\mu+\sigma]$ within a topological energy gap the operator
\begin{equation}
\mathbb{P}_{(\mu,\sigma)} = \sum_{j=1}^{D_\textrm{flake}} \exp\left[{-\frac{(E_j-\mu)^2}{2\sigma^2}}\right] | \phi_j\rangle \langle \phi_j|.\label{eqn:mu-sigma-projector}
\end{equation}
Note that $\mathbb{P}_{(\mu,\sigma)}$ is not a projector in the traditional sense, since its eigenvalues are arbitrary numbers in range $[0,1]$. 
The idea behind the exponential weight factor in Eq.~(\ref{eqn:mu-sigma-projector}) is that, per the approximately linear dispersion of the edge state visible in Fig.~\ref{fig:edge-states}(a), we construct a Gaussian function in angular momentum $\ell$. 
Therefore, the wave function $\mathbb{P}_{(\mu,\sigma)}\ket{\varphi_\textrm{site}}$ has approximately Gaussian coefficients in its decomposition to angular momenta, implying it constitutes a Gaussian wave packet in the angular coordinate~$\alpha$.

\FloatBarrier
\newpage

\section{Models of disorder.}\label{sec:disorder}
For the hH Hamiltonian $\mathcal{H}^\textrm{flake}_\textrm{hH}$, we consider the addition of random on-site potential, i.e., for each site $a$ we add to $\big(\mathcal{H}^\textrm{flake}_\textrm{hH}\big)_{aa}$ a random value drawn from the box distribution bounded by $\pm W_\textrm{max}$.

For the rhKM Hamiltonian $\mathcal{H}^\textrm{flake}_\textrm{hH}$, we consider the addition of a random spin-dependent term to each pair $(a,b)$ of NN and NNN sites. We specifically consider terms that are off-diagonal in the spin-degree of freedom; namely, for each pair $(a,b)$ we draw random values $\alpha_{x,y}\in[-W_\textrm{max},+W_\textrm{max}]$, and increase the $2\times 2$ Hamiltonian blocks as follows:
\begin{eqnarray}
\textrm{TR-symmetric disorder}:\;\;\big(\mathcal{H}^\textrm{flake}_\textrm{rhKM}\big)_{ab} \,&\,{{+}{=}}\,&\, \imi(\alpha_x\sigma_x + \alpha_y\sigma_y),\label{eqn:TRS-spin-disorder}\\
\textrm{TR-breaking disorder}:\;\;\big(\mathcal{H}^\textrm{flake}_\textrm{rhKM}\big)_{ab} \,&\,{{+}{=}}\,&\, \phantom{\imi}(\alpha_x \sigma_x + \alpha_y \sigma_y),\label{eqn:TRB-spin-disorder}
\end{eqnarray}
with $\big(\mathcal{H}^\textrm{flake}_\textrm{rhKM}\big)_{ba} = \big(\mathcal{H}^\textrm{flake}_\textrm{rhKM}\big)_{ba}^\dagger$ for both cases. The disorder in Eq.~(\ref{eqn:TRS-spin-disorder}) is interpretable as random Rashba SOC, while the one in Eq.~(\ref{eqn:TRB-spin-disorder}) corresponds to random in-plane magnetic fields along the trajectory connecting sites $(a,b)$.

The localization of a normalized eigenstate $|\phi_j\rangle $ is quantified by the inverse participation ratio, defined for spinless and spinful systems as 
\begin{equation}
\textrm{IPR}_j = \sum_{a=1}^{n_\textrm{flake}} \abs{\phi_{j,a}}^4   \quad \textrm{resp.} \quad \textrm{IPR}_j = \sum_{a=1}^{n_\textrm{flake}} \left(\abs{\phi_{j,a,\uparrow}}^2 + \abs{\phi_{j,a,\downarrow}}^2 \right)^2.
\end{equation}
One easily verifies that if $|\phi_j\rangle$ were homogeneously distributed over $N$ sites, then $\textrm{IPR}_j = 1/N$. This implies the interpretation that an eigenstate characterized by $\textrm{IPR}_j$ as being distributed over approximately $1/\textrm{IPR}_j$ sites.

\FloatBarrier
\newpage

\section{Hyperbolic Bloch Hamiltonians of the studied models}\label{eqn:HBT-Hamiltonians}

Since the hyperbolic Bloch Hamiltonians $\mcH(\bs{k})$ studied in this work are $16\times 16$ and $32 \times 32$ matrices, we do not write them explicitly. 
Instead, we include here the \Mathematica{} code that generates these Hamiltonians when compiled (the corresponding \Mathematica{} notebooks are included in the data repository~\cite{Urwyler:2022:SDC}).

We first specify the following:
\begin{lstlisting}[backgroundcolor = \color{verylightgray},
                    columns=fullflexible,
                   language = Mathematica,
                   xleftmargin = 2cm,
                   framexleftmargin = 1em]
$Assumptions = {
   k1 \[Element] Reals, k2 \[Element] Reals, 
   k3 \[Element] Reals, k4 \[Element] Reals,
   t1 \[Element] Reals, t2 \[Element] Reals, 
   flux \[Element] Reals, M \[Element] Reals, LR \[Element] Reals
   };
   
kList = {k1, k2, k3, k4, -k1, -k2, -k3, -k4};
\end{lstlisting}

The hyperbolic Bloch Hamiltonian of the NN model on the $\{8,3\}$ lattice is obtained as \texttt{Hnn} with the following code:
\begin{lstlisting}[backgroundcolor = \color{verylightgray},
                    columns=fullflexible,
                   language = Mathematica,
                   xleftmargin = 2cm,
                   framexleftmargin = 1em]

Hnn = ConstantArray[0, {16, 16}];

For[a = 1, a < 9, a++,
  Hnn[[Mod[a, 8] + 1, a]] = t1;
  Hnn[[a + 8, a]] = t1;
  Hnn[[8 + Mod[a + 4, 8] + 1, 8 + a]] = t1*Exp[-I*kList[[a]]];
  ];
  
Hnn = FullSimplify[Hnn + ConjugateTranspose[Hnn]];
\end{lstlisting}

After including the mass term, we obtain the hyperbolic Bloch Hamiltonian $\mcH_{\{8,3\}}(\bs{k})$ as \texttt{H83} with the following code:
\begin{lstlisting}[backgroundcolor = \color{verylightgray},
                    columns=fullflexible,
                   language = Mathematica,
                   xleftmargin = 2cm,
                   framexleftmargin = 1em]
Hmass = ConstantArray[0, {16, 16}];
For[a = 1, a < 9, a++,
  Hmass[[a, a]] += M*Power[-1, a + 1];
  Hmass[[a + 8, a + 8]] += -M*Power[-1, a + 1];
  ];
  
H83 = Hnn + Hmass;
\end{lstlisting}

To obtain the hyperbolic Bloch Hamiltonian $\mcH_\textrm{hH}(\bs{k})$ for the hyperbolic Haldane model, encoded as \texttt{Hh} in the code below, we further define a matrix \texttt{Hflux} of NNN terms. 
\begin{lstlisting}[backgroundcolor = \color{verylightgray},
                    columns=fullflexible,
                   language = Mathematica,
                   xleftmargin = 2cm,
                   framexleftmargin = 1em]
Hflux = ConstantArray[0, {16, 16}];

For[a = 1, a < 9, a++,
  Hflux[[Mod[a + 1, 8] + 1, a]] = t2*f;
  
  Hflux[[a, 8 + Mod[a, 8] + 1]] = t2*f;
  Hflux[[8 + Mod[a - 2, 8] + 1, a]] = t2*f;
  
  Hflux[[8 + Mod[a + 4, 8] + 1, a]] = t2*f*Exp[-I*kList[[a]]];
  Hflux[[a, 8 + Mod[a + 2, 8] + 1]] = 
   t2*f*Exp[I*kList[[Mod[a - 2, 8] + 1]]];
  
  Hflux[[8 + Mod[a + 2, 8] + 1, 8 + Mod[a + 4, 8] + 1]] = 
   t2*f*Exp[I*(kList[[a]] - kList[[Mod[a - 2, 8] + 1]])];
  ];

Hflux = FullSimplify[Hflux + ConjugateTranspose[Hflux]];

Hh = FullSimplify[H83 + Hflux /. f \[Rule] Exp[I*flux]];
\end{lstlisting}

We next proceed to construct the hyperbolic Bloch Hamiltonian $\mcH_\textrm{rhKM}(\bs{k})$ of the reduced hyperbolic Kane-Mele model. To that end, we first double the Haldane model into a `quantum spin Hall' Hamiltonian \texttt{Hqsh}:
\begin{lstlisting}[backgroundcolor = \color{verylightgray},
                    columns=fullflexible,
                   language = Mathematica,
                   xleftmargin = 2cm,
                   framexleftmargin = 1em]
Hqsh = ArrayFlatten[
     {
      {((H83 + Hflux) /. f \[Rule] Exp[I*flux]), 0},
      {0, ((H83 + Hflux) /. f \[Rule] Exp[\[Minus]I*flux])}
      }
     ];
\end{lstlisting}

The spin-orbit-coupled rhKM model is finally obtained as \texttt{Hhkm} with the following code:
\begin{lstlisting}[backgroundcolor = \color{verylightgray},
                    columns=fullflexible,
                   language = Mathematica,
                   xleftmargin = 2cm,
                   framexleftmargin = 1em]
Hsoc = ConstantArray[0, {16, 16}];

For[a = 1, a < 9, a++,
  alpha = -Pi/2 - (a - 1)*(2 Pi/8);
  Hsoc[[Mod[a, 8] + 1, a]] = I*(Cos[alpha]*PauliMatrix[1] + Sin[alpha]*PauliMatrix[2]);
  Hsoc[[a, Mod[a, 8] + 1]] = -I*(Cos[alpha]*PauliMatrix[1] + Sin[alpha]*PauliMatrix[2]);
  
  beta = Pi/8 - (a - 1)*(2 Pi/8);
  Hsoc[[a + 8, a]] = I*(Cos[beta]*PauliMatrix[1] + Sin[beta]*PauliMatrix[2]);
  Hsoc[[a, a + 8]] = -I*(Cos[beta]*PauliMatrix[1] + Sin[beta]*PauliMatrix[2]);
  
  gamma = 0 - (a - 1)*(2 Pi/8);
  Hsoc[[8 + Mod[a + 4, 8] + 1, 8 + a]] =
  I*(Cos[gamma]*PauliMatrix[1] + Sin[gamma]*PauliMatrix[2])*Exp[-I*kList[[a]]];
  Hsoc[[8 + a, 8 + Mod[a + 4, 8] + 1]] =
  -I*(Cos[gamma]*PauliMatrix[1] + Sin[gamma]*PauliMatrix[2])*Exp[I*kList[[a]]];
  ];

For[a = 1, a < 17, a++,
  For[b = 1, b < 17, b++,
    If[Hsoc[[a, b]] == 0,
      Hsoc[[a, b]] = ConstantArray[0, {2, 2}];
      ];
    ];
  ];

Hsoc = ArrayFlatten@Transpose[Hsoc, {3, 4, 1, 2}];

Hhkm = (Hqsh /. flux \[Rule] Pi/2) + LR*Hsoc;
\end{lstlisting}

\FloatBarrier
\newpage

\section{Wilson-loop extraction of topological band invariants}\label{eqn:Wilson-spectra}

We extract topological band invariants of hyperbolic Bloch Hamiltonians on two-dimensional planes in the 4D BZ using the Wilson-loop technique~\cite{Soluyanov:2011,Gresch:2017}. 
The utilized code is shared in the data repository~\cite{Urwyler:2022:SDC}.\bigskip

\begin{figure}[hbt!]
\centering
    \includegraphics[width=\linewidth]{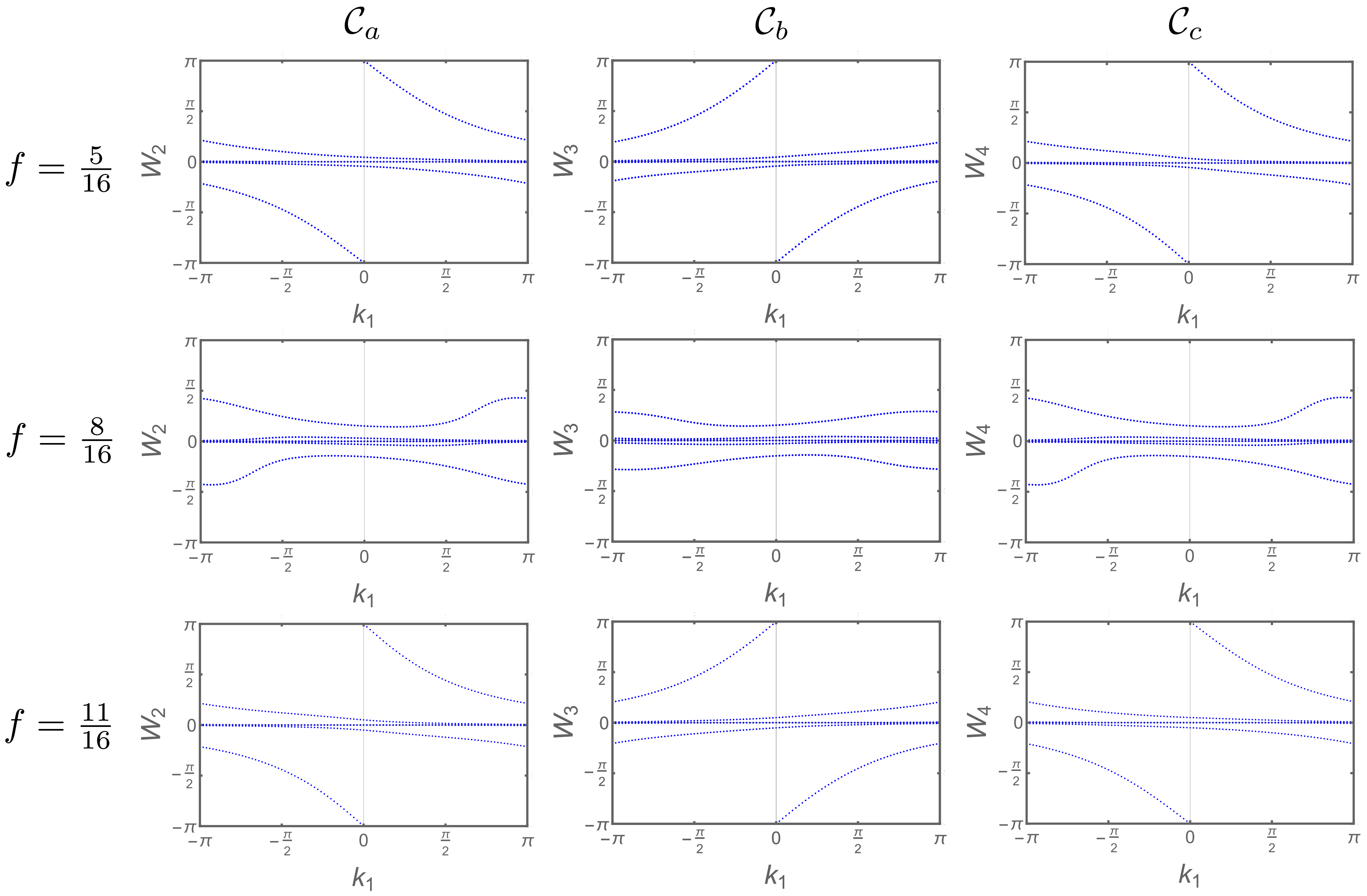} \caption[]{
    \textbf{Wilson-loop spectra for the hH model.} 
    Rows represent the three energy gaps of the model at the indicated filling fractions $f$, while columns correspond to the three pairs of planes that we do not relate by symmetry in the discussion in the main text.
    The Wilson spectra reveal the values of the Chern numbers tabulated in Fig.~\ref{fig:topology}(a). 
    Note that the number of plotted Wilson bands matches the number of filled energy bands (numerators of $f$); however, some of the bands are not resolved since they coincide close to zero value.
    }
\label{fig:Wilson-hH}
\end{figure}

First, we determine the Chern numbers $\mathcal{C}_a,\mathcal{C}_b,\mathcal{C}_c\in\mathbb{Z}$ of the hyperbolic Haldane model $\mcH_\textrm{hH}(\bs{k})$ by computing the Wilson loop in the $k_2$-, $k_3$- resp.~$k_4$-direction (labelled as $W_2$, $W_3$, resp.~$W_4$) as a function of $k_1$.
The results of our analysis are shown in Fig.~\ref{fig:Wilson-hH}, and tabulated in Fig.~\ref{fig:topology}(a). 

\begin{figure}[hbt!]
\centering
    \includegraphics[width=\linewidth]{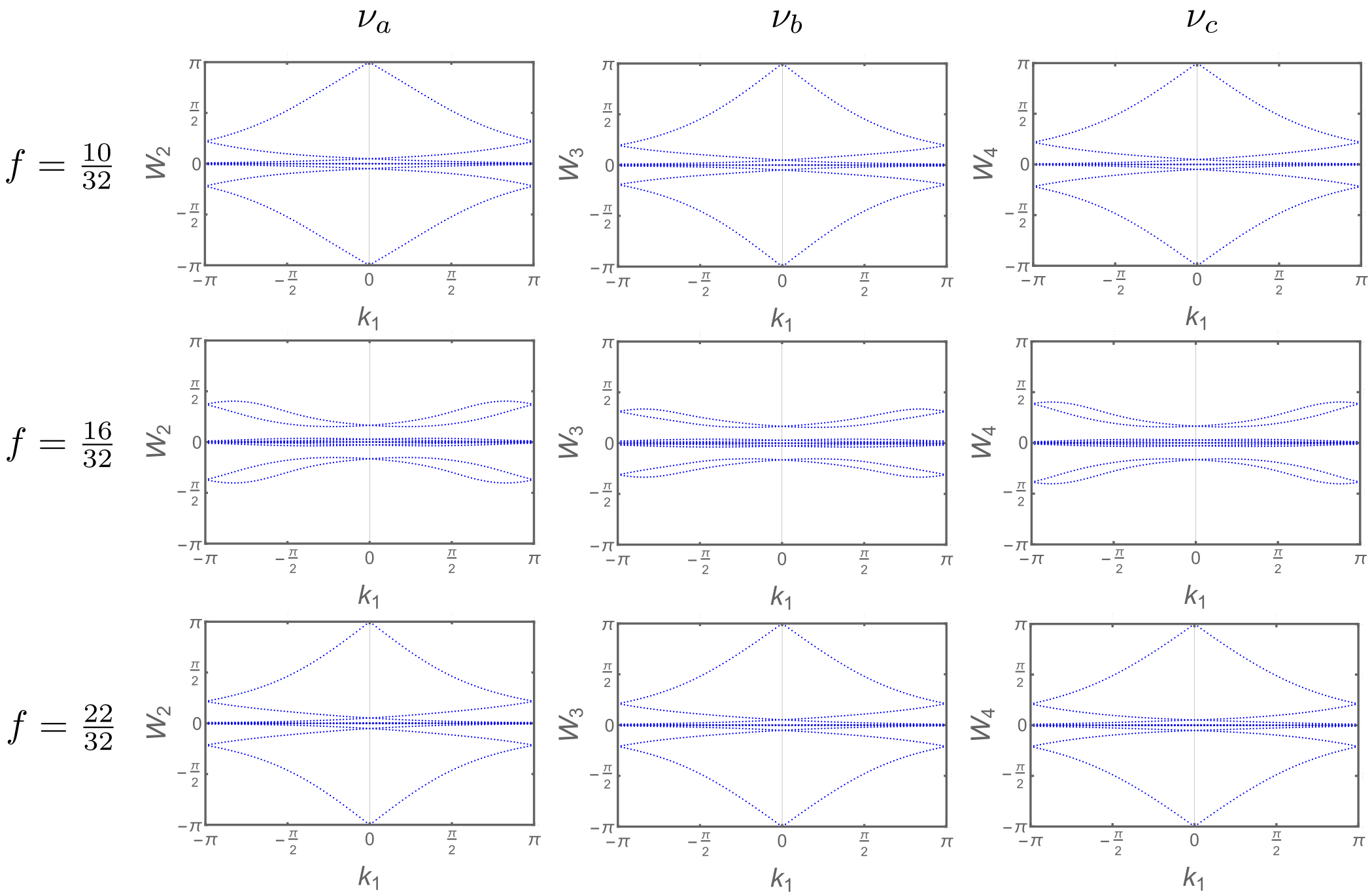} \caption[]{
    \textbf{Wilson-loop spectra for the rhKM model.} 
    Rows represent the three energy gaps of the model at the indicated filling fractions $f$, while columns correspond to the three pairs of planes that we do not relate by symmetry in the discussion in the main text.
    The Wilson spectra reveal the values of the Kane-Mele invariants tabulated in Fig.~\ref{fig:topology}(a). 
    The number of plotted Wilson bands matches the number of filled energy bands (numerators of $f$); however, some of the bands are not resolved since they coincide close to zero value.
    }
\label{fig:Wilson-rhKM}
\end{figure}

\newpage
We next determine the Kane-Mele invariants $\nu_a,\nu_b,\nu_c \in\mathbb{Z}_2$ for the reduced hyperbolic Kane-Mele model $\mcH_\textrm{rhKM}(\bs{k})$ by computing the Wilson loop in the $k_2$-, $k_3$- resp.~$k_4$-direction (labelled again as $W_2$, $W_3$, resp.~$W_4$) as a function of $k_1$.
The results of our analysis are shown in Fig.~\ref{fig:Wilson-rhKM}. 

\FloatBarrier
\newpage

\section{Extraction of the edge mode dispersion}\label{sec:edge-mode-dispersion}

In this supplementary note we describe a method we developed to determine the angular momentum of a given eigenstate of the Hamiltonian defined on a hyperbolic lattice. This method is used to generate the data for the dispersion of the chiral edge state of the hH model plotted in Fig.~\ref{fig:edge-states}a of the main text.
\begin{figure}[hbt!]
\centering
    \includegraphics[width=0.9\textwidth]{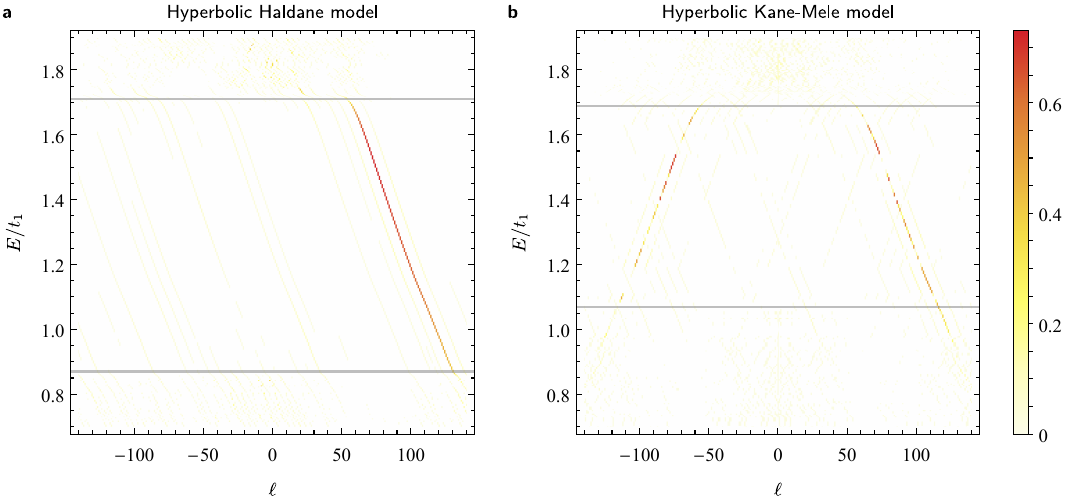}
    \caption[Angular momentum dispersion of edge modes.]{\textbf{Angular momentum dispersion of edge modes.}
        Contribution $\phi_\ell$ [cf.~\cref{SI:eq:phi-l} in Supplementary Note \ref{sec:edge-mode-dispersion}] of different angular momenta to the states lying in the upper energy gap for \textbf{a,} the Haldane model, and \textbf{b,} the reduced Kane-Mele model defined on the flake of the $\{8,3\}$ lattice shown in Supplementary Fig.~\ref{fig:83-lattice}.
        The horizontal grey lines indicate the band edges below and above the gap, and were determined from the density of states $\rho^\textrm{ED}_\textrm{bulk}$ for the here considered system size.
        \textbf{a, } In the Haldane model, there is a single (chiral) propagating edge mode with a very sharp angular momentum dispersion lying exactly in the gap.
        \textbf{b, } In the case of the Kane-Mele model, we find in the gap two (helical) counter-propagating edge modes.
        The two edge-state branches are characterized by opposite sign of the angular momentum ($\ell$) as well as of the angular momentum dispersion ($\mathrm{d}E/\mathrm{d}\ell$).
    }
    \label{fig:edgemode-dispersion}
\end{figure}

The method relies on a decomposition of functions defined on the disk $\mathbb{D}_R=\{z\in\mathbb{C}\;|\;\abs{z}\leq R\}$ of radius $0<R<1$ with the hyperbolic metric given by \cref{eq:distance_element} into eigenmodes of the Laplace-Beltrami operator
\begin{equation}
    \Delta = \left(1-\abs{z}^2\right)^2\left(\pdv[2]{}{x}+\pdv[2]{}{y}\right),
    \label{SI:eq:Laplace-Beltrami}
\end{equation}
where $z=x+\i y\in\mathbb{C}$.
The solutions to the Dirichlet problem
\begin{equation}
	(\Delta+\lambda)u(z) = 0,\qquad \left.u(z)\right|_{(z)\in\partial\mathbb{D}_{R}}=0
	\label{SI:eq:Dirichlet-problem}
\end{equation}
form an orthonormal basis for functions on $\mathbb{D}_R$ and they are given~\cite{Boettcher:2020,Lenggenhager:2021d} by
\begin{equation}
    u_{n,\ell}(z) = \frac{g_{k_n,\ell}(\abs{z})}{\norm{g_{k_n,\ell}}}\e^{\i\ell\arg(z)},
    \label{SI:eq:unl}
\end{equation}
where
\begin{equation}
    g_{k,\ell}(r) =
	\begin{cases}
		P_{\frac{1}{2}\left(-1+\i k\right)}^0\left(\frac{1+r^2}{1-r^2}\right),&\ell=0\\
		\left(\prod_{m=0}^{\ell-1}\left(-\frac{1}{2}-m+\i k\right)\right)^{-1}P_{\frac{1}{2}\left(-1+\i k\right)}^\ell\left(\frac{1+r^2}{1-r^2}\right),&\ell>0\\
		(-1)^\ell g_{k,\abs{\ell}}(r),&\ell<0
	\end{cases},
\end{equation}
$P_q^\ell(s)$ are the associated Legendre functions, $\norm{g}=\sqrt{\cip{g}{g}}$ is the norm induced by the inner product on $\mathbb{D}_R$
\begin{equation}
	\cip{v}{w} = \int_{\abs{z}\leq R}\frac{\dd[2]{z}}{(1-\abs{z}^2)^2}\cconj{v(z)}w(z),
\end{equation}
$k_{n,\ell}$ is the $n$-th zero of
\begin{equation}
	k\mapsto P_{\frac{1}{2}\left(-1+\i k\right)}^\ell\left(\frac{1+R^2}{1-R^2}\right),
	\label{SI:eq:boundary_condition}
\end{equation}
and $\ell\in\mathbb{Z}$.
The solutions to \cref{SI:eq:boundary_condition} correspond to zeroes of $g_{k,\ell}(R)$, cf.~Supplementary Fig.~\ref{fig:LB-eigenvalues}a.

For the flake of the $\{8,3\}$ lattice shown in Supplementary Fig.~\ref{fig:83-lattice} that we define our models on, all the eigenvalues $\lambda_{n,\ell}$ are shown in Supplementary Fig.~\ref{fig:LB-eigenvalues}b as functions of $n$ and $\ell$ and the five solutions to \cref{SI:eq:Dirichlet-problem} with smallest $\lambda_{n,\ell}$ are plotted in the top row of Supplementary Fig.~\ref{fig:LB-eigenmodes}.

\begin{figure}[hbt!]
\centering
    \includegraphics{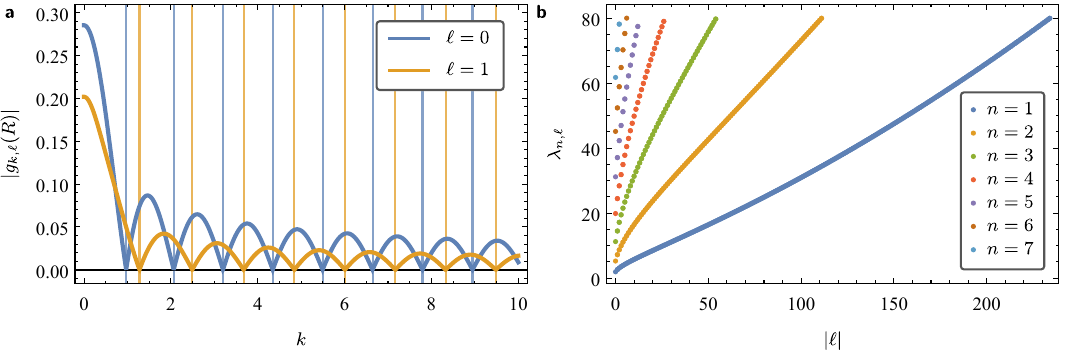}
    \caption[]{\textbf{Eigenvalues of the Laplace-Beltrami operator.}
        Illustration of how to find $k_{n,\ell}$ and the resulting eigenvalues $\lambda_{n,\ell}$ with $896$ sites of the $\{8,3\}$ lattice shown in \cref{fig:83-lattice}a.
        \textbf{a,} The function $k\mapsto g_{k,\ell}(R)$ for $\ell=0$ (blue) and $\ell=1$ (orange) and with $R=0.991437$ chosen as illustrated in \cref{fig:83-lattice}b. The first few zeroes $k_{n,\ell}$ are marked by vertical lines.
        \textbf{b,} The first $896$ eigenvalues $\lambda_{n,\ell}$ as a function of $\abs{\ell}$ with the different branches corresponding to different $n$.
    }
    \label{fig:LB-eigenvalues}
\end{figure}

\begin{figure}[hbt!]
\centering
    \includegraphics{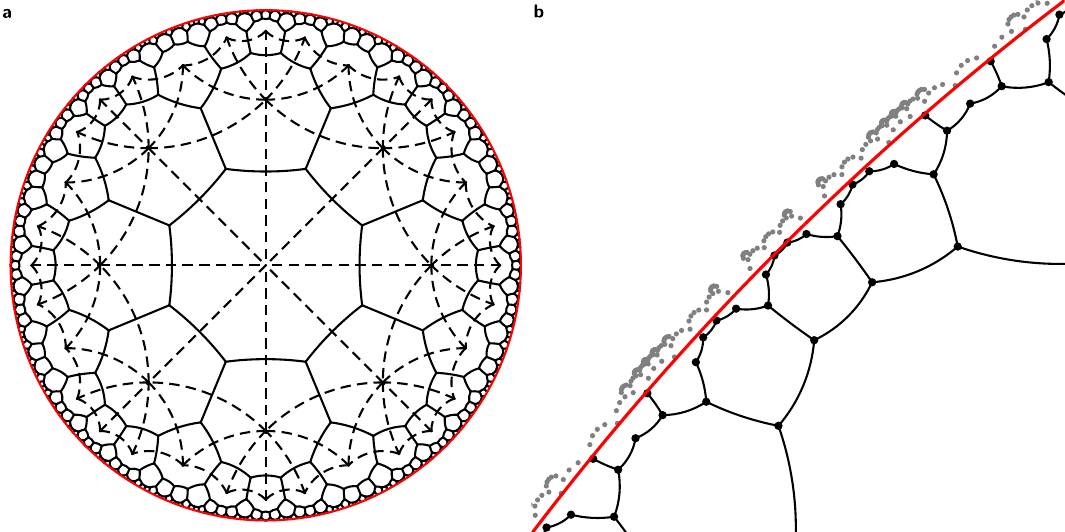}
    \caption[]{\textbf{Flake of the $\{8,3\}$ lattice used to extract the edge mode dispersion.}
        \textbf{a,} The flake with $896$ sites with nearest-neighbour bonds (black lines).
        Some of the Wigner-Seitz unit cells are indicated by black dashed lines. 
        The red circle bounds the disk of radius $R$ which determines the Dirichlet boundary conditions for $u(z)$.
        \textbf{b,} The radius $R$ is defined as the maximal radius $r$ such that additional sites of the infinite hyperbolic lattice (gray points) not included in the flake (whose sites sites are shown with black points) lie outside the disk $\mathbb{D}_r$.
    }
    \label{fig:83-lattice}
\end{figure}

A function $v(z)$ on $\mathbb{D}_R$ can then be decomposed into the eigenfunctions $u_{n,\ell}(z)$:
\begin{subequations}
    \begin{align}
        v(z) &= \sum_{n>0}\sum_{\ell\in\mathbb{Z}}v_{n,\ell}u_{n,\ell}(z),\\
    	v_{n\ell} &= \cip{u_{n,\ell}}{v} = \int_{\abs{z}\leq R}\frac{\dd[2]{z}}{(1-\abs{z}^2)^2}\cconj{u_{n,\ell}(z)}v(z).
    \end{align}
    \label{SI:eq:LBdecomp}
\end{subequations}
In \cref{SI:eq:unl} we recognize that $\ell$ can be interpreted as angular momentum.
If $v(z)$ is normalizable, $\norm{v}<\infty$,
\begin{equation}
    v_\ell = \sum_{n>0}\abs{v_{n,\ell}}^2
\end{equation}
gives the contributions of different values of angular momentum $\ell$ to the function $v(z)$.

This can be used to find the contributions of angular momenta to quantities (vectors) defined on the lattice, e.g., the eigenstates of the Hamiltonian $\mcH^\textrm{flake}_{\{8,3\}}$ defined on a flake of the $\{8,3\}$ lattice.
A normalized vector $\ket{\phi}=(\phi_1,\dotsc,\phi_N)^\top$ defined on the lattice given by the finite set of sites $\{z_i\}_{i=1}^N$ induces the following function on $\mathbb{D}_R$
\begin{equation}
    \phi^{(\alpha)}(z) = \sum_{i=1}^N\phi_i\eta_i^{(\alpha)}(z),
\end{equation}
where $\eta_i(z)$ is non-vanishing only in the Wigner-Seitz cell of the lattice site $i$ (the Wigner-Seitz cell is defined as the region of $\mathbb{D}_R$ that has shorter hyperbolic distance to site $i$ than to any other site of the flake), and satisfies
\begin{equation}
    \int_{\abs{z}\leq R}\frac{\dd[2]{z}}{(1-\abs{z}^2)^2}\abs{\eta_i^{(\alpha)}(\vec{r})}^2 = 1.
\end{equation}
Various choices for $\eta_i(z)$ are possible.
Here we consider two options: 
\begin{itemize}[leftmargin=3cm]
    \item[Option (1):$\quad$] $\eta_i^{(1)}(z)=\Theta_i(z)/\sqrt{A_\mathrm{WS}}$, and 
    \item[Option (2):$\quad$] $\eta_i^{(2)}(z)=\sqrt{A_\mathrm{WS}}(1-\abs{z_i}^2)^2\delta^{(2)}(z-z_i)$.
\end{itemize}
In both of the above, $A_\mathrm{WS}$ is the (hyperbolic) area of a Wigner-Seitz unit cell (cf.~\cref{fig:83-lattice}); and $\Theta_i(z)$ in the first expression is a \enquote{region function} that is equal to one inside (and to zero outside) the Wigner-Seitz cell of site $i$.
For both of the above options,
\begin{equation}
	\cip{\eta_i^{(\alpha)}}{\phi^{(\alpha)}} = \sum_j \phi_j\int_{\abs{z}\leq R}\frac{\dd[2]{z}}{(1-\abs{z}^2)^2}\cconj{\left[\eta_i^{(\alpha)}(z)\right]}\eta_j^{(\alpha)}(z) = \phi_i
\end{equation}
allows us get back the $i^\textrm{th}$ component of vector $\ket{\phi}$.

The extension $\phi^{(\alpha)}(z)$ of $\ket{\phi}$ to the full disk $\mathbb{D}_R$ allows us to apply the decomposition into eigenmodes of the Laplace-Beltrami operator given in \cref{SI:eq:LBdecomp} to the vector $\ket{\phi}$:
\begin{equation}
	\phi_{n,\ell}^{(\alpha)} = \cip{u_{n,\ell}}{\phi^{(\alpha)}} = \sum_i\phi_i\cip{u_{n,\ell}}{\eta_i^{(\alpha)}}.
\end{equation}
For choice (1) this becomes
\begin{equation}
	\phi_{n,\ell}^{(1)} = \sqrt{A_\mathrm{WS}}\sum_i\phi_i\cconj{\overline{u_{n,\ell}}(z_i)}
\end{equation}
where $\overline{u_{n,\ell}}(z_i)$ is the average over the $i^\textrm{th}$ Wigner-Seitz cell $\mathrm{WS}_i$:
\begin{equation}
    \overline{u_{n,\ell}}(z_i) = \int_{z\in \mathrm{WS}_i}\int_{\abs{z}\leq R}\frac{\dd[2]{z}}{(1-\abs{z}^2)^2}u_{n,\ell}(z).
\end{equation}
Choice (2) results in a much simpler expression only involving $u_{n,\ell}(z)$ evaluated at the lattice sites:
\begin{equation}
	\phi_{n,\ell}^{(2)} = \sqrt{A_\mathrm{WS}}\sum_i\phi_i\cconj{u_{n,\ell}(z_i)}.
\end{equation}
For the two choices we define 
\begin{itemize}[leftmargin=3cm]
    \item[(1):$\quad$] $\ket*{{\psi}_{n,\ell}^{(1)}}=\sqrt{A_\mathrm{WS}}(\overline{u_{n,\ell}}(z_1),\dotsc,\overline{u_{n,\ell}}(z_N))^\top$, resp.
    \item[(2):$\quad$] $\ket*{\psi_{n,\ell}^{(2)}}=\sqrt{A_\mathrm{WS}}(u_{n,\ell}(z_1),\dotsc,u_{n,\ell}(z_N))^\top$,
\end{itemize}
 allowing us to express the coefficients $\phi_{n,\ell}$ compactly as
\begin{equation}
    \phi_{n,\ell}^{(\alpha)} = \ip{\psi_{n,\ell}^{(\alpha)}}{\phi}.
\end{equation}
Some examples of $\ket*{\psi_{n,\ell}^{(1)}}$ are shown in the bottom row of the Supplementary Fig.~\ref{fig:LB-eigenmodes}.
Finally, we define the angular-momentum components
\begin{equation}
    \phi_\ell^{(\alpha)} = \sum_{n>0}\abs{\phi_{n,\ell}^{(\alpha)}}^2.
    \label{SI:eq:phi-l}
\end{equation}

\begin{figure}[hbt!]
\centering
    \includegraphics[width=18cm]{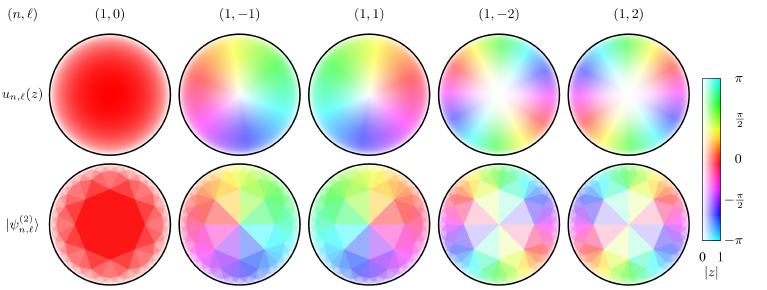}
    \caption[]{\textbf{Eigenmodes of the Laplace-Beltrami operator.}
        Examples of eigenmodes $u_{n,\ell}(z)$ (top row) of the Laplace-Beltrami operator with Dirichlet boundary conditions for $\abs{z}=R=0.991437$.
        The bottom row shows the discretized versions $\ket*{\psi_{n,\ell}^{(2)}}$, i.e., $u_{n,\ell}$ evaluated at the lattice sites of the flake shown in Supplementary Fig.~\ref{fig:83-lattice}.
        In both rows the absolute value is encoded in the intensity and the argument in the color (see legend).
        The header gives the quantum numbers $(n,\ell)$ for each eigenmode.
    }
    \label{fig:LB-eigenmodes}
\end{figure}

Note that the discretized approximations $\ket*{\psi_{n,\ell}^{(\alpha)}}$ of the Laplace-Beltrami eigenmodes $u_{n,\ell}(z)$ are \emph{neither orthogonal nor normalized}, and therefore subsets with $N$ elements generally do not form a basis of $\mathbb{C}^N$.
In general, a large number of $\ket*{\psi_{n,\ell}}$ are required to characterize an arbitrary state $\ket{\phi}$.
Owing to the choice of $t_1=1>0$, eigenstates $\ket{\phi}$ of the flake Hamiltonian, $\mcH^\textrm{flake}_{\{8,3\}}\ket{\phi}=E\ket{\phi}$, with energy $E$ lying towards the upper end of the energy spectrum have larger contributions $\phi_{n,\ell}$ associated to small values of $\lambda_{n,\ell}$, i.e., slowly oscillating eigenfunctions $u_{n,\ell}(z)$.
It is therefore easier to determine $\phi_\ell$ for those states, while states with smaller energy $E$ are highly oscillatory and require larger $\abs{\ell}$ as well as $n$.

In practice, our algorithm for calculating the $\phi_{\ell}$ for all states $\ket{\phi}$ is set up as follows. 
To avoid a computationally heavy numerical integration of the individual Wigner-Seitz cell, we choose option (2) discussed above.
The vectors $\{\ket*{\psi_{n,\ell}^{(2)}}\}_{n>0,\ell\in\mathbb{Z}}$ only depend on the finite lattice, i.e., the flake, and not the Hamiltonian matrix defined on it.
Given a flake of the $\{8,3\}$ lattice (and a compatible choice of bounding radius $R$), a subset of $\{\ket*{\psi_{n,\ell}}\}_{n>0,\ell\in\mathbb{Z}}$ can be precomputed and stored.
To do that, we first need to find solutions of \cref{SI:eq:boundary_condition} for the chosen range of angular momentum $\ell\in[\ell_\mathrm{min},\ell_\mathrm{max}]$; this is done by a root search in a predefined interval $k\in(0,k_\mathrm{max})$.
Note that there is some freedom in choosing $R$ due to the discretization; it must lie beyond the outermost site appearing on our disk-shaped flake (i.e., inside the restricted list $L_\textrm{sites}$), but closer than the nearest site of the $\{8,3\}$ lattice not included in the flake (i.e., not appearing in the slightly larger list $\widetilde{L}_\textrm{sites}$).
For the system size considered here, we choose $R=0.991437$.
The resulting values $k_{n,\ell}$ allow us to define the corresponding eigenfunctions $u_{n,\ell}(z)$ via \cref{SI:eq:unl} and consequently compute $\ket*{\psi_{n,\ell}^{(2)}}$.
Later, the overlaps $\phi_{n,\ell}^{(2)} = \ip*{\psi_{n,\ell}^{(2)}}{\phi}$ can be efficiently computed for all $\ket{\phi}$, resulting in the energy vs.~angular momentum spectrum. 

The results of applying the outlined algorithm to states in (and near) the upper energy gap of the hH model and of the rhKM model are shown, respectively, in the two panels of Fig.~\ref{fig:edgemode-dispersion}.
Let us remark that in the main text Fig.~\ref{fig:edge-states}(a) we plot essentially the same data as in Supplementary Fig.~\ref{fig:edgemode-dispersion}(a). 
However, as shown in the version of the plot in the supplementary figure, the extracted data are very sharp (one pixel-in-$\ell$ wide), which would make them hard to see in the small figure panel in the main text.
For this reason, we opt in the main text to coarse grain the signal in angular momentum over $(2n_\textrm{max}+1)$ values of angular momenta as 
\begin{equation}
    \phi_\ell^{(\alpha)} \mapsto \sum_{a=-n_\textrm{max}}^{+n_\textrm{max}} \phi_{\ell+a}^{(\alpha)}.
    \label{SI:eq:smear}
\end{equation}
to improve the visibility. The result of this coarse graining for $n_\textrm{max}\in\{0,1,2\}$ (i.e., over 1, 3, resp.~5 adjacent values of $\ell$) is shown in Supplementary Fig.~\ref{fig:smearing}. The data in main text Fig.~\ref{fig:edge-states}(a) correspond to $n_\textrm{max}=2$.

\begin{figure}[hbt!]
\centering
    \includegraphics[width=0.3\linewidth]{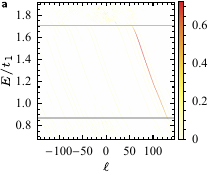}\hspace{0.045\linewidth}
    \includegraphics[width=0.3\linewidth]{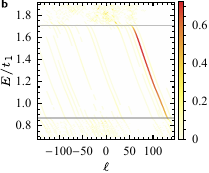}\hspace{0.045\linewidth}
    \includegraphics[width=0.3\linewidth]{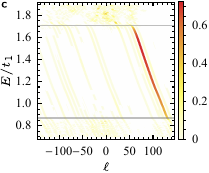}
    \caption[]{
    \textbf{Smearing of chiral edge states of the hH model.} \textbf{a,} The extracted values of $\phi_\ell$ for the chiral edge states of the hH model. \textbf{b,} The result of smearing the values shown in panel a according to \cref{SI:eq:smear} with $n_\textrm{max}=1$ and \textbf{c} with $n_\textrm{max}=2$.
    }
\label{fig:smearing}
\end{figure}

\FloatBarrier
\newpage

\section{Group velocity of the chiral edge states in hyperbolic Haldane model}\label{sec:group-velocity}

In this supplementary note, we show how the data in Fig.~\ref{fig:edge-states}(a) and~in Fig.~\ref{fig:edge-states}(b) provide two independent ways to extract the velocity with which the wave packets of topological edge states in the hH model propagate around the flake boundary, allowing for a consistency check of the numerical modelling. 
We remark that the two figure panels are generated for \emph{different system sizes}; therefore, one should be careful to compare absolute (rather than angular) velocities. 
Furthermore, the focus is not on quantitative rigor but on qualitative comparison; therefore, we approximate most of the discussed quantities to two significant digits.

We begin with Fig.~\ref{fig:edge-states}(a), which is computed for a system with $n_\textrm{sites}= 896$, such that the number of Bolza cells is approximately $N_\textrm{UC} = 896/16 = 56$. 
From Eq.~(\ref{eqn:Nuc-to-radius}) (where we approximate the numerator on the right-hand side by $1$) we obtain for the radius $R$ that $1-R^2 \approx 1/56$. 
Next, from Eq.~(\ref{eqn:perimeter}) we obtain the perimeter $p \approx 112 \pi \approx 350$. 
We further estimate the angular group velocity as $\omega_\textrm{group} = \Delta E/ \Delta \ell$. 
We read from the data in Fig.~\ref{fig:edge-states}(a) that across the energy gap $\Delta E\approx 0.83$ and $\Delta\ell \approx 65 $, leading to $ \omega_\textrm{group} \approx 0.0128$.
Multiplying by the perimeter, we obtain the absolute group velocity $v_\textrm{group}\approx  \omega_\textrm{group} p \approx 4.5$. (Here, units of length are such that the Gaussian curvature is $K=-4$, cf.~Methods. Time is measured in units of $\hbar/t_1$; in numerical modelling we set both $\hbar$ and $t_1$ to $1$.)

On the other hand, the data in Fig.~\ref{fig:edge-states}(b) are obtained for a system with $n_\textrm{sites}=1864$. 
Repeating analogous geometric considerations as above, we find that $N_\textrm{UC} = 116.5$, and $1-R^2 \approx 2/233$. 
The perimeter of the corresponding system is estimated as $p \approx 233 \pi \approx 732$. 
We read from the data in Fig.~\ref{fig:edge-states}(b) that the wave packet traverses angular distance $\Delta \alpha \approx 7.5 \pi $ in time $\Delta \tau = 4000$, implying angular velocity $\omega = \Delta\alpha/\Delta \tau \approx 0.0059$. Multiplying with the perimeter, we obtain the absolute speed of the wave packet propagation $v = \omega p \approx 4.3$.

We find that the two extracted values of the velocity with which the wave packets propagate along the boundary differ by ${\sim}5\%$. 
This is acceptable agreement within our margin of error, given that several of the discussed quantities (proper choice of $R$, as well as intervals $\Delta \ell$ and $\Delta \alpha$) can only be extracted up to a few-percent confidence interval.

\FloatBarrier
\newpage

\section{Phase diagram of the hyperbolic Haldane Bloch Hamiltonian at half-filling and \texorpdfstring{$\Phi=\pi/2$}{Phi = pi/2}}\label{sec:Haldane-gap}

Recall that for the original Haldane model on the Euclidean honeycomb lattice~\cite{Haldane:1988}, the inclusion of $M$ drives a trivial energy gap while it is the inclusion of $t_2$ (at finite flux) that drives the topological gap. 
The boundary between the trivial and the topological insulating phases is given by the analytic formula $\abs{M/t_2} = 3\sqrt{3}\abs{\sin\Phi}$. 
In particular, the topological phase of the Euclidean Haldane model persists when $M$ is set to zero.

In this Supplementary Note, we briefly investigate whether a similar competition between the $M$-driven trivial gap and the $t_2$-driven topological gap also occurs at the half-filling for the hyperbolic Haldane Hamiltonian.
To that end, we numerically determine~\cite{Urwyler:2022:SDC} the energy gap at half-filling as a function of $(M,t_2)$ for fixed values $t_1 =1$ and $\Phi=\pi/2$. The result of this analysis is plotted in Supplementary Fig.~\ref{fig:Haldane-gap}(a), where the red dot indicates the value of model parameters considered throughout the the manuscript. 

\begin{figure}[hbt!]
\centering
    \includegraphics[width=\linewidth]{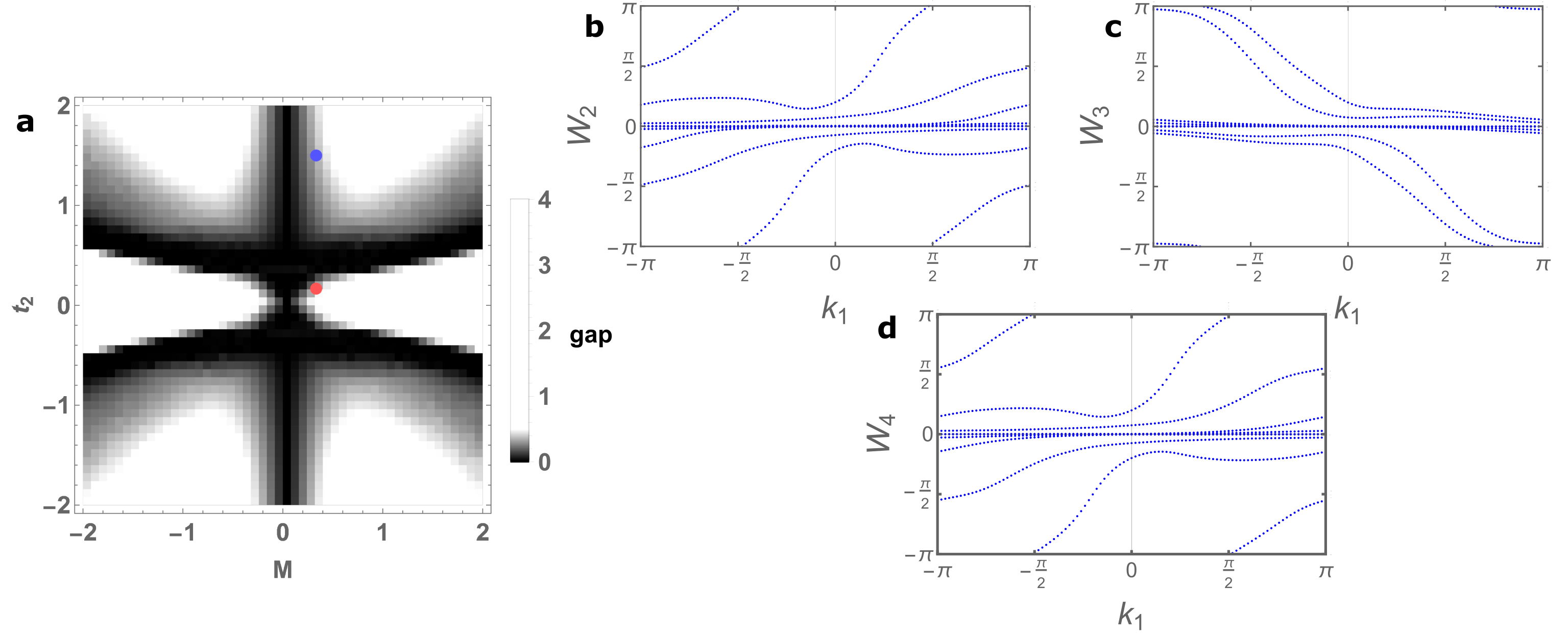} \caption[]{
    \textbf{Energy gap of the hyperbolic Haldane Bloch Hamiltonian at half-filling.} 
    \textbf{a} We set $t_1 =1$ and $\Phi = \pi/2$, while keeping the parameters $M$ and $t_2$ variable.
    Shades of gray indicate the bulk energy gap of the hyperbolic Haldane model as determined by the hyperbolic band theory. Bright tones indicate large values of the gap (expressed in multiples of $t_1 = 1$), while saturated black corresponds to gap closing. Red dot at $M=1/3$ and $t_2 = 1/6$ corresponds to the choice of parameters considered throughout the manuscript, where the energy gap at half-filling is trivial. The blue dot at $M=1/3$ and $t_2 = 3/2$ corresponds to the parameters briefly discussed in Supplementary Note \ref{sec:Haldane-gap}, where the energy gap at half-filling is associated with non-vanishing Chern numbers $\mathcal{C}_a,\mathcal{C}_b,\mathcal{C}_c$.
    \textbf{b--d} Wilson-loop spectra for $M=1/3$ and $t_2 = 3/2$ (blue dot in panel \textbf{a}), which indicate $\mathcal{C}_a = +2$, $\mathcal{C}_b = -2$, and $\mathcal{C}_c=+2$.
    }
\label{fig:Haldane-gap}
\end{figure}

The first striking feature we observe in Supplementary Fig.~\ref{fig:Haldane-gap}(a) is that the hyperbolic Haldane model at half-filling, in contrast with the Euclidean one~\cite{Haldane:1988}, is \emph{gapless} for $M=0$. 
We further observe, in resemblance with the Euclidean case, that besides the insulating phase at small values of $\abs{t_2/M}$ there are additional gapped regions occurring at large values of $\abs{t_2/M}$. 
To determine the band topology of these additional insulating phases, we fix $M=1/3$ and $t_2 = 3/2$, which correspond to the blue dot in Supplementary Fig.~\ref{fig:Haldane-gap}(a). 
We apply the Wilson-loop technique to compute the values of Chern numbers $\mathcal{C}_{a,b,c}$ in the insulating phase that occurs at large and \emph{positive} $t_2/M$.
The results of our analysis, plotted in Supplementary Fig.~\ref{fig:Haldane-gap}(b--d), imply $\mathcal{C}_a = +2$, $\mathcal{C}_b=-2$, and $\mathcal{C}_c=+2$ i.e., the energy gap that occurs at half filling for large ${t_2/M}$ is topologically non-trivial. 
Note, however, that the even value of the invariant implies that the corresponding rhKM model for this choice of parameters exhibits trivial values of the $\mathbb{Z}_2$-valued invariants $\nu_a$ and $\nu_b$. 
For large and \emph{negative} $t_2/M$, the signs of the Chern numbers $\mathcal{C}_{a,b,c}$ are flipped.
We also verified that the second Chern number for these insulating regions is trivial.

\begin{figure}[hbt!]
\centering
    \includegraphics[width=0.36\linewidth]{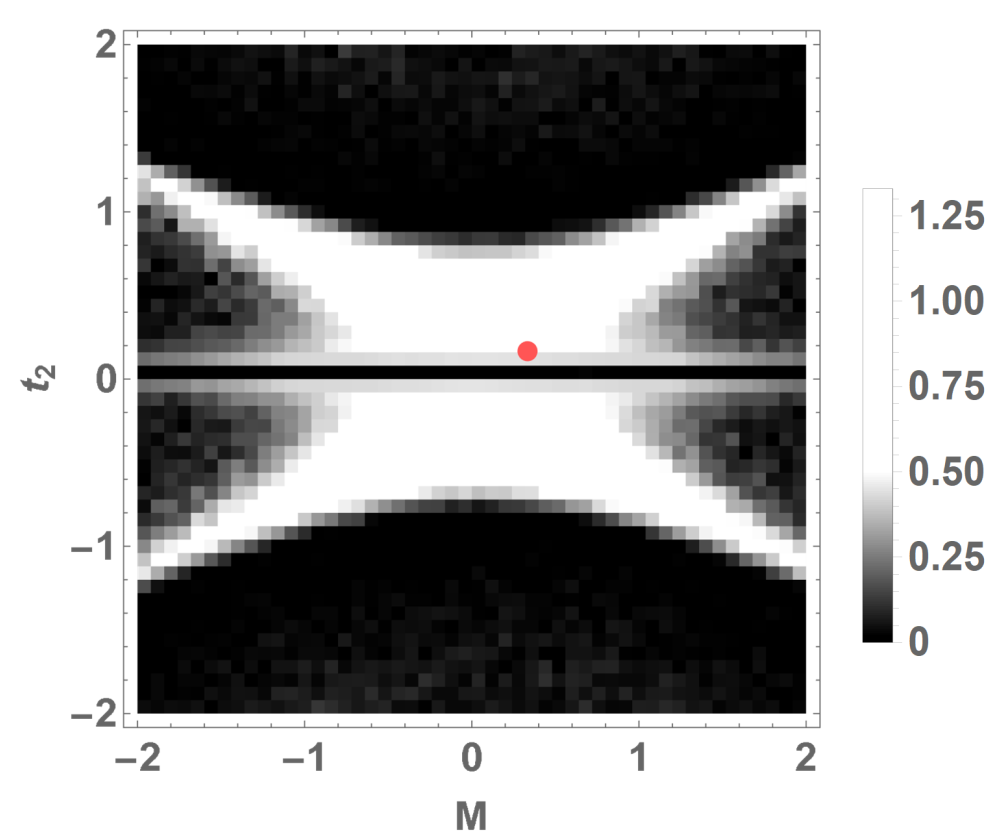} \caption[]{
    \textbf{Energy gap of the hyperbolic Haldane Bloch Hamiltonian at filling $\boldsymbol{f=5/16}$.} 
    Two gapped regions can be identified. These regions are related by a sign flip $t_2 \mapsto -t_2$. Since the same change can be interpreted as the action of time-reversal (complex conjugation flips $\Phi \mapsto -\Phi$, which at $\Phi=\pi/2$ corresponds to a sign flip of the purely imaginary $t_2$ term), the two gapped regions must exhibit opposite sign of all Chern numbers. The red dot indicates the parameter values $t_2 = \tfrac{M}{2} = \tfrac{1}{3}$ which are assumed throughout the main text. 
    Since the gapped phase extends to $M=0$, it follows from an additional $(\pi/4)$-rotation symmetry that necessarily $\mathcal{C}_a = \mathcal{C}_c$, cf.~Eq.~(\ref{eqn:extra-Chern-relation}).
    }
\label{fig:Haldane-gap-fill5}
\end{figure}

Finally, we check that, in contrast to the half-filled case, the bulk energy gap at fillings $f\in\{\tfrac{5}{16},\tfrac{11}{16}\}$ (which correspond to the Chern insulating phases studied in the main text) do not close for $M=0$. This is illustrated explicitly for $f=\tfrac{5}{16}$ in Fig.~\ref{fig:Haldane-gap-fill5} (with the data for $f=\tfrac{11}{16}$ looking essentially identical). 
Note that at $M=0$ the hyperbolic Haldane model acquires an additional symmetry, namely rotation by $\pi/4$ around the center of the Bolza cell, which permutes momenta as $k_1\mapsto k_2\mapsto k_3 \mapsto k_4 \mapsto -k_1$. 
This symmetry implies that at $M=0$ (and also for all gapped phases at finite $M$ that extend to $M=0$) we have
\begin{equation}
\mathcal{C}_a := C_{12} = C_{23} = C_{34} = \mathcal{C}_{14} =:\mathcal{C}_c,\label{eqn:extra-Chern-relation}
\end{equation}
reducing the number of independent Chern numbers to two. 
[Let us remark that the remaining two Chern numbers could potentially be related by the three-fold rotation around a vertex of the $\{8,3\}$ lattice. However, as this symmetry is known to act non-orthogonally on the four momentum components~\cite{Maciejko:2021}, we leave a careful investigation of this symmetry for a future study.]

    \end{bibunit}
    \end{document}